\documentclass{emulateapj}
\usepackage{graphicx}
\usepackage{natbib}
\newcommand\kms{\,km\thinspace s$^{-1}$}  

\bibpunct{(}{)}{;}{a}{}{,} 
\shorttitle{Near-IR properties of LSI galaxies}
\shortauthors{Kirby et al.}
\usepackage{lscape}
\usepackage{hyperref}
\begin{document}

\title{Deep Near-IR Surface Photometry of 57 Galaxies in the Local Sphere of Influence}

\author{Emma M. Kirby\altaffilmark{1,2}\altaffiltext{2}{Affiliated with the Australia Telescope National Facility, CSIRO.} and Helmut Jerjen\altaffilmark{1}}
\affil{Research School of Astronomy and Astrophysics, \\
Mt Stromlo Observatory, Australian National University, \\
Cotter Road, Weston ACT 2611, Australia}
\email{emma@mso.anu.edu.au, jerjen@mso.anu.edu.au}

\and

\author{Stuart D. Ryder\altaffilmark{3}}
\affil{Anglo-Australian Observatory, \\
P.O. Box 296, Epping, NSW 1710, Australia}
\email{sdr@aao.gov.au}

\and

\author{Simon P. Driver\altaffilmark{4}}
\affil{SUPA, School of Physics and Astronomy,\\
University of St. Andrews}
\email{spd3@st-and.ac.uk }

\begin{abstract}

We present $H$-band (1.65$\mu$m) surface photometry of 57 galaxies drawn from the Local Sphere of Influence (LSI) with distances of less than 10\,Mpc from the Milky Way. The images with a typical surface brightness limit 4 mag fainter than 2MASS ($24.5$\,mag arcsec${}^{-2}< \mu_{lim}<26$\,mag arcsec${}^{-2}$ ) have been obtained with IRIS2 on the 3.9~m Anglo-Australian Telescope. A total of 22 galaxies that remained previously undetected in the near-IR and potentially could have been genuinely young galaxies were found to have an old stellar population with a star density $1-2$ magnitudes below the 2MASS detection threshold. The cleaned near-IR images reveal the morphology and extent of many of the galaxies for the first time. For all program galaxies, we derive radial luminosity profiles, ellipticities, and position angles, together with global parameters such as  total magnitude, mean effective surface brightness and half-light radius. Our results show that 2MASS underestimates the total magnitude of galaxies with $\langle\mu_H\rangle_{eff}$ between $18-21$\,mag arcsec${}^{-2}$ by up to 2.5\,mag. The S\'ersic parameters best describing the observed surface brightness profiles are also presented. Adopting accurate galaxy distances and a $H$-band mass-to-light ratio of  $\Upsilon_{\ast}^H=1.0\pm 0.4$, the LSI galaxies are found to cover a stellar mass range of $5.6<\log_{10}(\mathcal{M}_{\rm stars})<11.1$. The results are discussed along with previously obtained optical data. Our sample of low luminosity galaxies is found to follow closely the optical-infrared $B$ versus $H$ luminosity relation defined by brighter galaxies with a slope of $1.14 \pm 0.02$ and scatter of $0.3$\,magnitudes. Finally we analyse the luminosity -- surface brightness relation to determine an empirical mass-to-light ratio of $\Upsilon_{\ast}^H=0.78\pm0.08$ for late-type galaxies in the $H$-band.

\end{abstract}

\maketitle

\section{Introduction}

The observational properties of nearby galaxies such as fluxes, colours, morphologies and sizes reflect their underlying physical properties (stellar/baryonic and dark matter content, star formation rates, formation history and angular momenta). Exactly how these observational and physical properties are related is still poorly understood. By technical necessity, the observational quantities are mainly based on the optical $B$-band ($390 - 480$\,nm). However galaxies evolving in low density environments with little external stimulation for star formation often contain significant quantities of dust (eg.  see \citealt{driver07}) which can attenuate and distort their optical light profiles.  In contrast, dust attenuation is vastly reduced at near-IR wavelengths and hence  the near-IR provides a spectral regime where a more accurate, unaltered representation of a galaxy's underlying stellar distribution can be obtained \citep{gavazzi96}.  Furthermore, the stellar mass of most galaxies is dominated by the quiescent old stellar component whose energy output peaks at near-IR wavelengths.  Even in the extreme case of Blue Compact Dwarf (BCD) galaxies, previously thought to be primeval galaxies forming their first stars at the present epoch \citep{thuan97}, the analysis of their resolved stellar populations has revealed the presence of stars at least a few Gyrs of age (\citealt[eg. the BCD galaxies: VII Zw 403, Mrk 178 and I Zw 36 as discussed by][respectively]{schulteladbeck98,schulteladbeck00, schulteladbeck01}; SBS 1415+437 discussed by \citealt{aloisi05}; I Zw 18 by \citealt{aloisi07}; and CGCG 269-049 by \citealt{corbin08}).

In order to obtain a deeper understanding of the connection between the light and matter distribution in galaxies, a representative sample of nearby stellar systems needs to be studied in detail. The Local Sphere of Influence (LSI, $D< 10\,$Mpc) contains large numbers of early (dE) and late-type (dIrr) dwarf galaxies that make up about 85\% of the local galaxy population~\citep{kraan79, schmidt92, karachentsev04}. Dwarf galaxies contribute about 4\% to the local luminosity density and about 10-15\% to the local H\,{\sc i} mass density \citep{karachentsev04}. Due to their proximity to the Milky Way, LSI galaxies are ideal for a near-IR study which includes significant numbers of dwarf systems.

Previous near-IR surveys include the Two Micron All Sky Survey \citep[2MASS,][]{skrutskie06} as well as deeper targeted galaxy surveys \citep{gavazzi96b, gavazzi96a, gavazzi00, boselli00}.  2MASS photometry for galaxies suffers from  a number of important drawbacks that are becoming more evident as the samples of independently investigated galaxies become larger. The short integration time of 2MASS observations resulted in most of the low surface brightness (LSB) dwarfs in the LSI remaining undetected, and if they were detected, 2MASS underestimated the fluxes by as much as 70\% \citep{andreon02}. The targeted $H$-band observations of \cite{gavazzi96b, gavazzi96a, gavazzi00} and  \cite{boselli00} were inherently deeper however the samples included few LSB dwarfs. This serious limitation demands a deeper and higher resolution study to investigate those galaxies that were beyond the reach of photometric near-IR studies to date. 

A reference atlas of images needs to have the necessary spatial resolution to probe the morphological fine-structure of these nearby galaxies and contain a significant number of dwarf galaxies that are generally overlooked. A LSI sample has the additional advantage that an increasingly large number of nearby galaxies have accurately known distances. \cite{karachentsev06} report that 214 out of 451 LSI galaxies have  distance estimates (with less than 10\% uncertainty) by means of the tip magnitude of the red giant branch (TRGB), the Tully-Fisher relation, and the surface brightness fluctuations (SBF) method (see  for example \citealt{jerjen98, jerjen01} and \citealt{karachentsev04}). The remaining galaxies have rough distance estimates from the luminosity of their brightest stars, radial velocities or their suspected membership to a known galaxy group.

The purpose of this paper is to present a near-IR $H$-band (1.65$\mu$m) atlas of 57 LSI galaxies, probing to flux levels approximately 4 mag arcsec${}^{-2}$ or 40 times fainter than 2MASS. The majority of the galaxies presented here are much fainter than those in previous targeted surveys. We derive photometric parameters for  each object such as the total magnitude, the effective radius and effective surface brightness, S\'ersic fitting parameters, etc. Using the best distances currently available in the literature allows us to derive physical parameters such as their  luminosities and  stellar masses.  

The paper is organised as follows: we describe the sample selection  in \S \ref{s:sample}. In \S\ref{s:obs} and \S\ref{s:photometry} we discuss the observing strategies,  the data reduction, and the photometric calibration of the images.   The 11 galaxies in the sample which remained undetected at our faint detection limit or had images which could not be usefully analysed are discussed in \S\ref{s:nondetect}. The new data is compared to  2MASS photometry and optical ($B$-band) data in \S\ref{s:results} and the luminosity - surface brightness relation discussed. Interesting properties of individual galaxies are described in \S\ref{s:interesting}. Finally,  the results are summarised in \S\ref{s:summary}.

~\linebreak

\section{Sample selection}\label{s:sample}
\begin{figure*}[!hbt]
\centering
\includegraphics[width=0.8\linewidth]{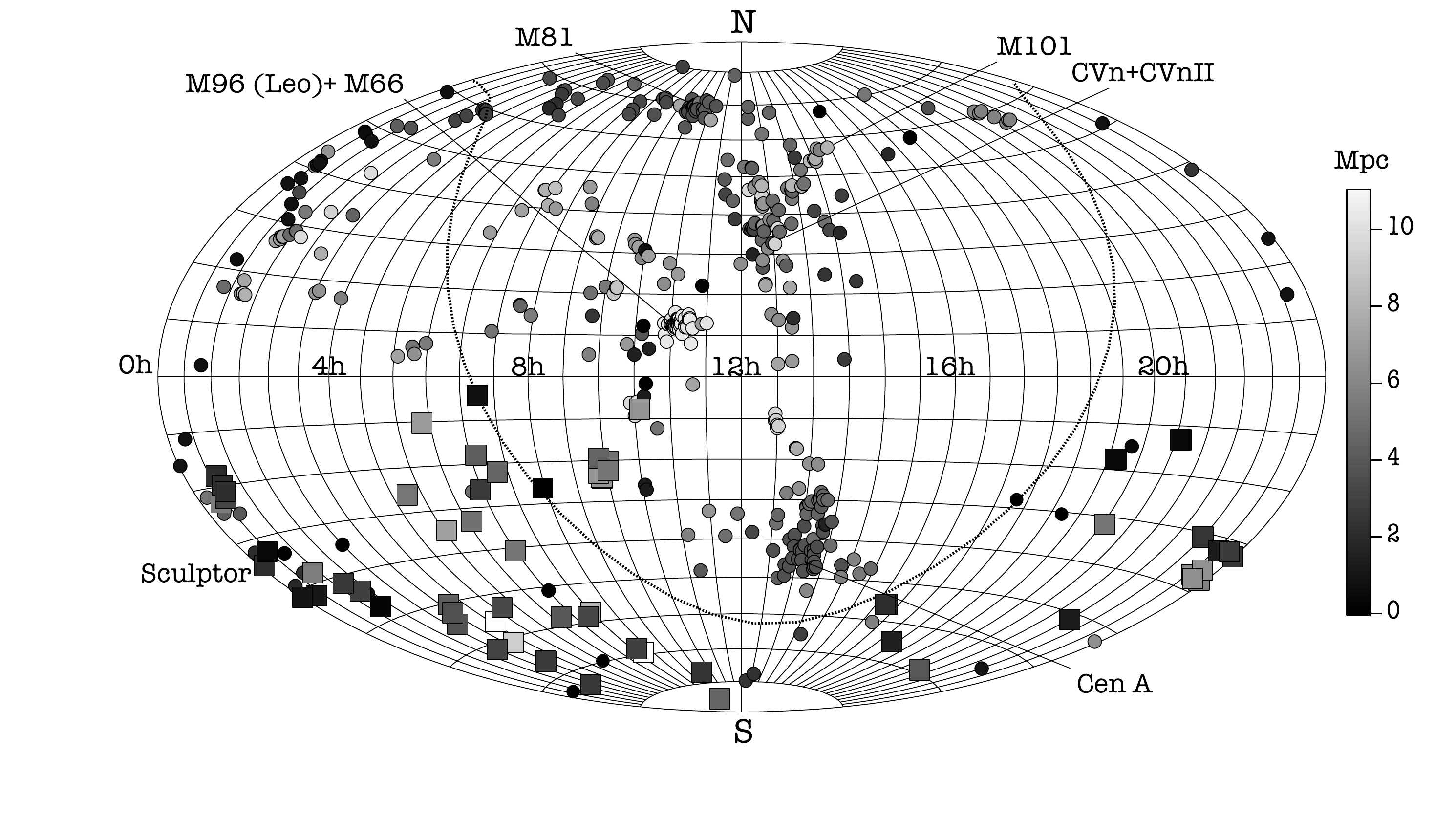}
\caption{{\small  The distribution of galaxies within the Local Sphere of Influence. The circles represent the 451 galaxies listed in \cite{karachentsev04}  plus 19 from other catalogues, and the squares are the  68 galaxies  investigated in our study. The greyscale reflects a galaxy's distance. The curved line shows the location of the Galactic plane.}} 
\label{fig:distrib}
\end{figure*}

We have compiled a list of 470 galaxies with estimated distances less than $10$\,Mpc from the Milky Way\footnote{Note that five galaxies have had their distance estimates revised since their inclusion in the sample and have a distance greater than $10$\,Mpc.}, from the catalogues of~\cite{schmidt92}, \cite{cote97}, \cite{jerjen00}, and \cite{karachentsev04}. Approximately 70\% of LSI galaxies are members of seven nearby galaxy groups including  the Local Group (LG). Each group contains one or more massive spiral or  elliptical galaxies accompanied by a population of dwarf satellites which tend to be dwarf  ellipticals (dE). The southern hemisphere contains 174 LSI galaxies, 113 of which are members of a nearby group.  We randomly selected 68 program galaxies with a  range of total apparent $B$-band magnitude (between $m_B = 9$ and 18\,mag, as well as several with no optical detection to date) and morphology (Hubble types E3 through to Sc, including many irregular and dwarf galaxies), 19 of which were members of a nearby group. Therefore, our sample contains 80\% ($=\frac{68-19}{174-113}$) of southern hemisphere field galaxies and 17\% ($=19/113$)  of  group members.  The distribution of  these Local Sphere of Influence galaxies is shown in Figure~\ref{fig:distrib}.

The selected galaxies further provide a complementary data set to the Local Volume H\,{\sc i} Survey~\citep[LVHIS;][]{koribalski07} which is a H\,{\sc i} imaging survey of all LSI galaxies south of  declination $\delta=-30^{\circ}$ that were detected by the H\,{\sc i} Parkes All-Sky Survey ~\citep[HIPASS,][]{barnes01}. The former H\,{\sc i} survey is currently been carried out at the Australia Telescope Compact Array.  The basic properties of our sample galaxies have been listed in Table~\ref{basicdata} which is organised as follows: 
\begin{itemize} 
	\item[] \emph{Column} (1). - Galaxy name.
	\item[] \emph{Column} (2). -  Morphological type in 
	the \cite{hubble36}, \cite{sandage61}, and \cite{sandage84}
	classification scheme.
	\item[] \emph{Columns} (3) and (4). -  Equatorial coordinates for the epoch J2000.
	\item[] \emph{Column} (5). -   Total $B$-band magnitude and its source.  When the uncertainty associated with this value is not provided, an error of 0.2\,mag has been adopted.
	\item[] \emph{Columns} (6) and (7)  - Distance to the galaxy 
	(from \citealt{karachentsev04, karachentsev06, seth05, carrasco01})
	with an indication of the method used: (TRGB) tip magnitude of the red giant
	branch; (SBF) surface brightness fluctuations; (MEM) 
	group membership; (H)  Hubble flow distance $D=v_{LG}/H_0$ where $H_0=73$\kms Mpc$^{-1}$ is adopted \citep[WMAP,][]{spergel07}.
	\item[] \emph{Column} (8).  - Heliocentric radial velocity, $v_{\odot}$, from the NASA Extragalactic Database (NED). 
	\item[] \emph{Column} (9).  - Local Group velocity, $v_{LG}$, from NED.
	\item[] \emph{Column} (10). - Reddening estimate, $E(B-V)$ from \cite{schlegel98}. The associated error  is 16\%.
	\item[] \emph{Columns} (11), (12), (13) and (14). - The observing log. The observation date, strategy, total exposure time and seeing is listed.
	\end{itemize}

\section{Observations and Reduction}\label{s:obs}

Near-infrared $H$-band images were obtained for the 68 program galaxies during five observing runs between October 2004 and September 2006,
using the Infrared Imager and Spectrograph~2 (IRIS2;
\citealt{tinney04}) at the 3.9~m Anglo-Australian Telescope (AAT).  Table~\ref{basicdata} lists the observing log of the observations. Atmospheric conditions
were clear if not always photometric, and the seeing ranged from
$1\farcs0$ to $2\farcs9$ with a mean of $1\farcs3$. The IRIS2 detector
is a $1024 \times 1024$ Rockwell HAWAII--1 HgCdTe array with a pixel
scale of $0\farcs45$~pixel$^{-1}$, resulting in an instantaneous
field-of-view (FoV) of $7\farcm7 \times 7\farcm7$.

Two different observing strategies were employed, depending on the
anticipated angular extent of the target compared to the IRIS2 FoV:

\begin{itemize}
\item {\sc Jitter Self Flat} (JSF) -- The majority of our sample have optical
diameters $<2^{\prime}$ and, given the sky is typically $7-8$~mag\,arcsec${}^{-2}$
brighter in the $H$-band than the $B$-band, we anticipate these
objects filling barely $10-20$\% of the array in the infrared. These
targets were observed in a $3 \times 3$ grid pattern with a spacing of
$90^{\prime\prime}$, resulting in a $4\farcm7 \times 4\farcm7$ overlap
region common to all pointings that encompasses not just the target
galaxy but also a substantial amount of the background sky. A maximum
of $\sim$30~seconds was spent on any one pointing, consisting of
multiple 5--10~s integrations (depending on the sky brightness at the
time while aiming to keep the combined object $+$ sky counts well
within the linear regime) which were then averaged before being
stored.  This 9-point jitter pattern was repeated up to 6~times,
leading to a total on-source exposure time of just under half an hour
per galaxy. This method was also employed on larger, but well-resolved
targets such as the Argo Dwarf Irregular galaxy.

\item {\sc Chop Sky Jitter} (CSJ) -- In accordance with the recommendations
of \citet{vad04}, objects filling $\gtrsim$40\% of the array FoV
require matching observations of adjacent blank sky to track changes
in the background level and illumination pattern. Five jittered
observations ($10^{\prime\prime}$ offsets) of the target galaxy were
bracketed and interleaved with six jittered observations of the
(relatively blank) sky $10^{\prime}$ north or south. At each object or
sky jitter position, $3 \times 10$~s or $6 \times 5$~s integrations
were averaged. This pattern was repeated between 5 and 12~times, for a
total on-source exposure time of up to half an hour per galaxy.

\end{itemize}

The data reduction was carried out using the
ORAC-DR\footnote{http://www.oracdr.org/} pipeline within the {\sc
starlink} package. Observations made with the JSF method employed the
JITTER\_SELF\_FLAT recipe, while those with the CSJ method used the
CHOP\_SKY\_JITTER recipe. Pre-processing of all raw frames included
subtraction of a matching dark frame; linearity and inter-quadrant
crosstalk correction; and bad pixel masking.

Considerable care was taken to ensure accurate flat-fielding over the 
entire field of the array. For JSF observations, an interim flatfield is created by taking the
median at each pixel of the  nine normalised object frames, then each of
the nine images is divided by this interim flatfield. Extended sources
within these flatfielded object frames are automatically detected and
masked, and an improved flatfield created from masked versions of the
nine normalised object frames. A correction for astrometric distortion
internal to IRIS2 is applied by resampling the properly flatfielded
images, then spatial additive offsets between images are computed using point
sources common to all images. The  nine images are mosaiced together by
applying offsets in intensity to the registered images to produce the
most consistent sky value possible in the overlap regions. A new
flatfield and mosaic is constructed for each set of nine jittered frames,
then all the mosaics registered and co-added to form a master
mosaic. Occasionally, significant variations in the level and/or
structure of the background sky on a temporal or spatial scale smaller
than that sampled by the array within the $\sim$5\,minute period of the
nine jittered frames resulted in noticeable residual structure in the
ensuing mosaic, forcing us to exclude that mosaic from the master
mosaic. The total on-source exposure time, after discarding such
data, is shown in Table~\ref{basicdata}.

For CSJ observations, the six sky frames are first offset in intensity
to a common modal value, then a flatfield formed from the median value
at each pixel. All six sky frames, and five object frames are flatfielded,
then the modal pixel values of the two sky frames bracketing each object
frame are averaged and subtracted from that object frame. Image
registration and mosaicing is then performed on each set of
five sky-subtracted object frames just as for the JSF observations.
These mosaics are registered and co-added to form a master mosaic,
with the exception of any showing residual sky structure as described
above, yielding the on-source exposure times shown in Table~\ref{basicdata}.

\begin{figure*}
\centering
\begin{tabular}{ccc}
  \mbox{\scalebox{0.75}{\includegraphics[width=0.38\linewidth]{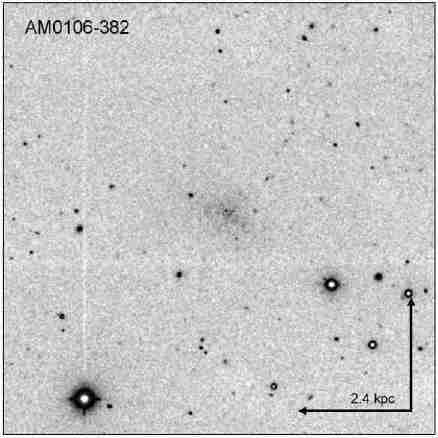}}}&
  \mbox{\scalebox{0.75}{\includegraphics[width=0.38\linewidth]{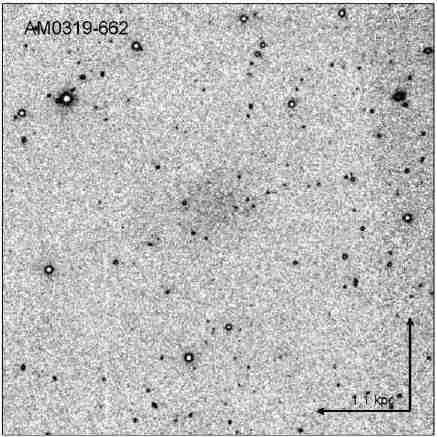}}}&
  \mbox{\scalebox{0.75}{\includegraphics[width=0.38\linewidth]{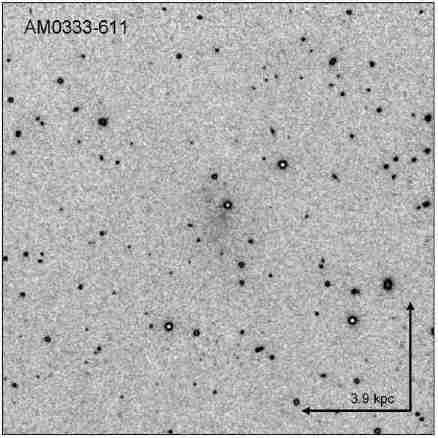}}}\\
  \mbox{\scalebox{0.75}{\includegraphics[width=0.38\linewidth]{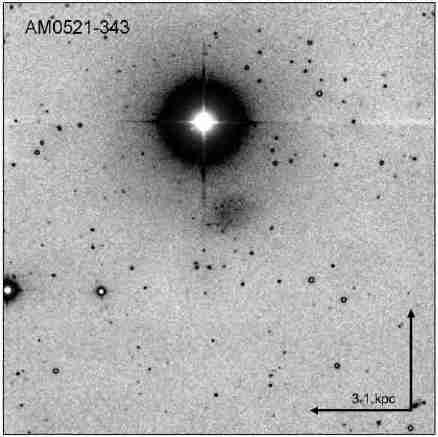}}}&
  \mbox{\scalebox{0.75}{\includegraphics[width=0.38\linewidth]{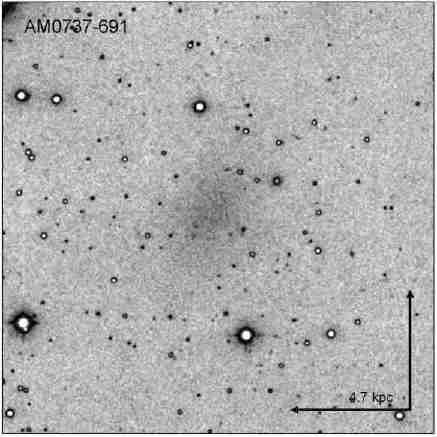}}}&
  \mbox{\scalebox{0.75}{\includegraphics[width=0.38\linewidth]{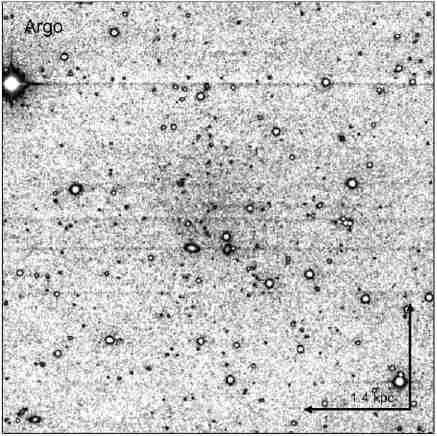}}}\\
  \mbox{\scalebox{0.75}{\includegraphics[width=0.38\linewidth]{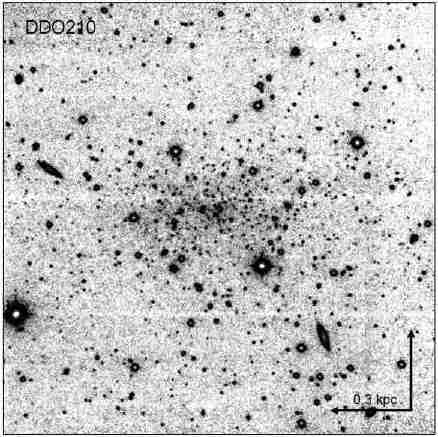}}}&
  \mbox{\scalebox{0.75}{\includegraphics[width=0.38\linewidth]{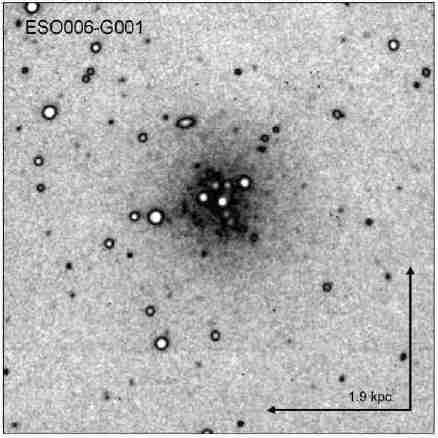}}}&
  \mbox{\scalebox{0.75}{\includegraphics[width=0.38\linewidth]{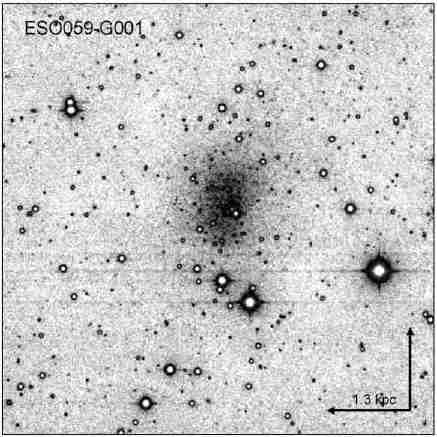}}}\\
  \mbox{\scalebox{0.75}{\includegraphics[width=0.38\linewidth]{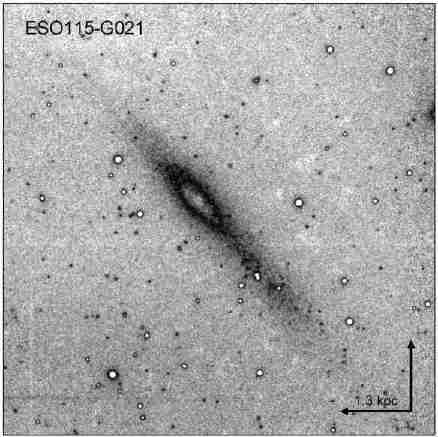}}}&
  \mbox{\scalebox{0.75}{\includegraphics[width=0.38\linewidth]{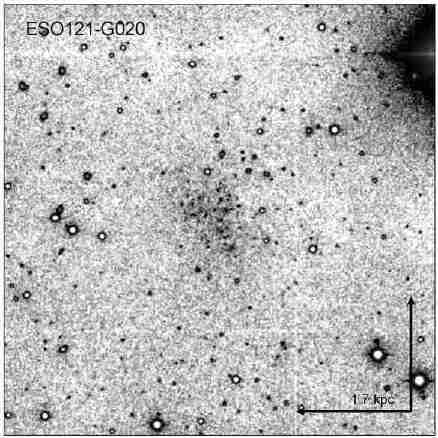}}}&
  \mbox{\scalebox{0.75}{\includegraphics[width=0.38\linewidth]{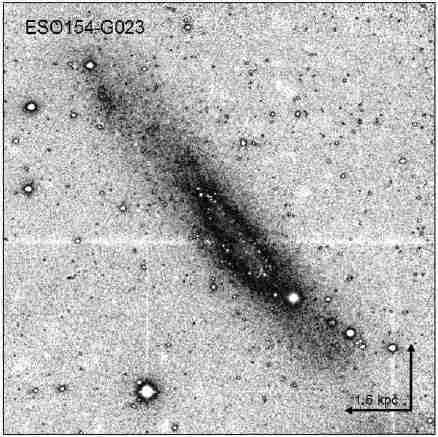}}}\\
  \end{tabular}
\caption{{LSI deep $H$-band images from the 3.9m Anglo-Australian Telescope. Here the scale represents 1 arcmin. The corresponding linear scale is also indicated. North is up and East is to the left. The intensity is represented by a greyscale which goes from white  (low intensity) to black (medium intensity) and then  back to white (high intensity). Higher resolution version available at  \url{http://www.mso.anu.edu.au/~emma/KirbyHband.pdf}}} 
\label{fig:galimages1}
\end{figure*}

\begin{figure*}
\centering
\begin{tabular}{ccc}
\mbox{\scalebox{0.75}{\includegraphics[width=0.38\linewidth]{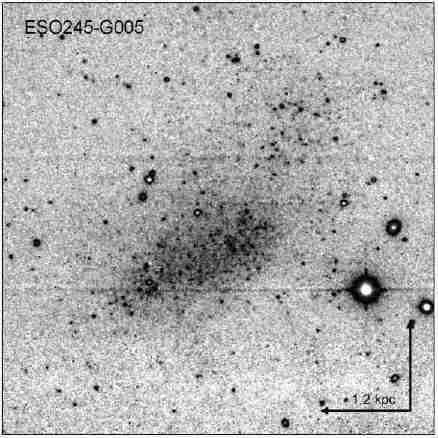}}}&
\mbox{\scalebox{0.75}{\includegraphics[width=0.38\linewidth]{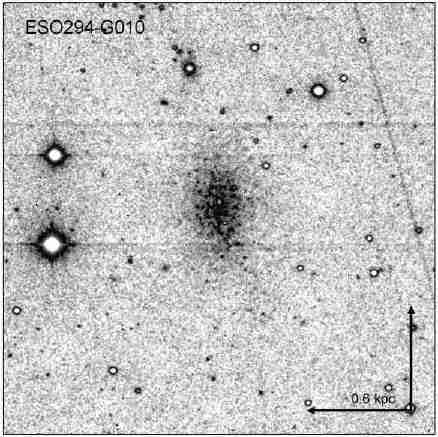}}}&
\mbox{\scalebox{0.75}{\includegraphics[width=0.38\linewidth]{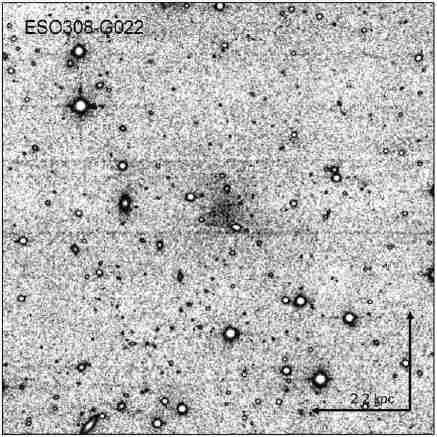}}}\\
\mbox{\scalebox{0.75}{\includegraphics[width=0.38\linewidth]{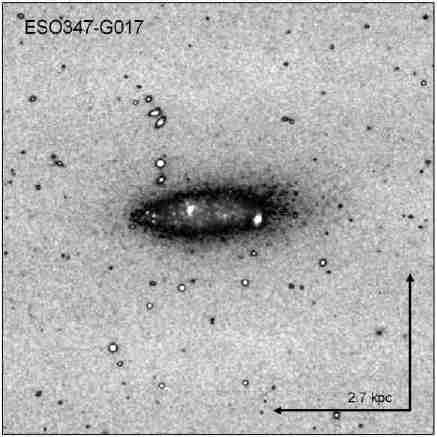}}}&
\mbox{\scalebox{0.75}{\includegraphics[width=0.38\linewidth]{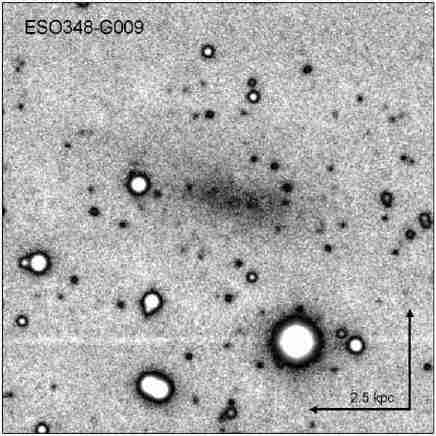}}}&
\mbox{\scalebox{0.75}{\includegraphics[width=0.38\linewidth]{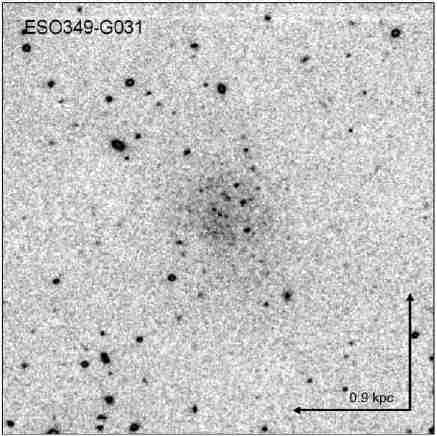}}}\\
\mbox{\scalebox{0.75}{\includegraphics[width=0.38\linewidth]{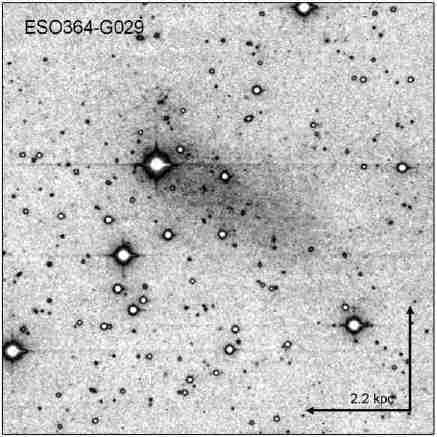}}}&
\mbox{\scalebox{0.75}{\includegraphics[width=0.38\linewidth]{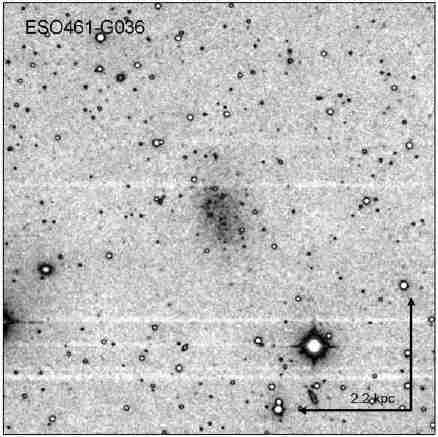}}}&
\mbox{\scalebox{0.75}{\includegraphics[width=0.38\linewidth]{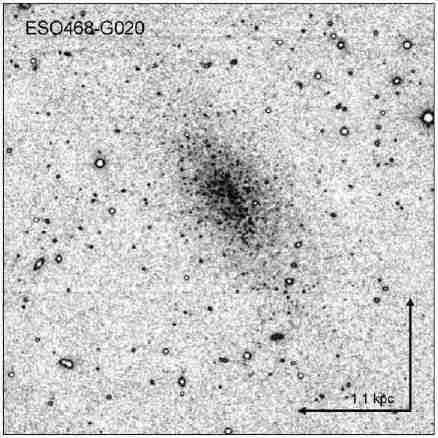}}}\\
\mbox{\scalebox{0.75}{\includegraphics[width=0.38\linewidth]{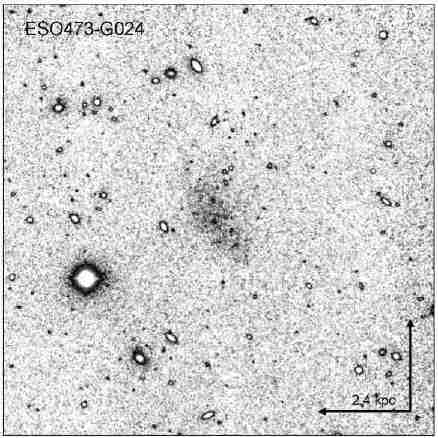}}}&
\mbox{\scalebox{0.75}{\includegraphics[width=0.38\linewidth]{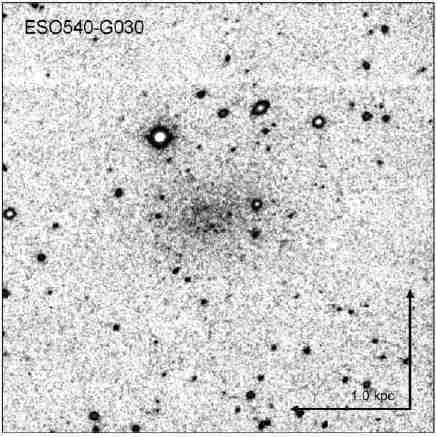}}}&
\mbox{\scalebox{0.75}{\includegraphics[width=0.38\linewidth]{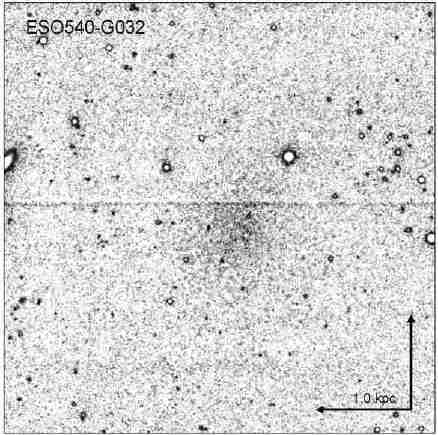}}}\\
  \end{tabular}
\caption{{LSI deep $H$-band images from the 3.9m Anglo-Australian Telescope. Here the scale represents 1 arcmin. The corresponding linear scale is also indicated. North is up and East is to the left. The intensity is represented by a greyscale which goes from white  (low intensity) to black (medium intensity) and then  back to white (high intensity). Higher resolution version available at  \url{http://www.mso.anu.edu.au/~emma/KirbyHband.pdf}}} 
\label{fig:galimages2}
\end{figure*}

\begin{figure*} 
\centering
\begin{tabular}{ccc}
\mbox{\scalebox{0.75}{\includegraphics[width=0.38\linewidth]{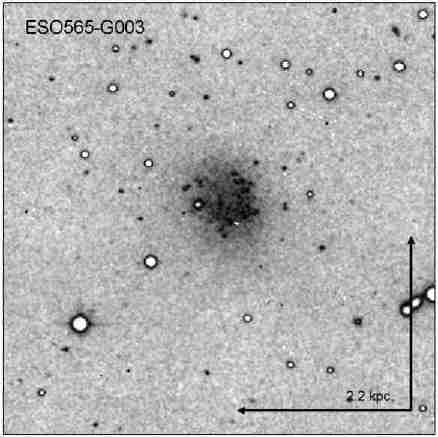}}}&
\mbox{\scalebox{0.75}{\includegraphics[width=0.38\linewidth]{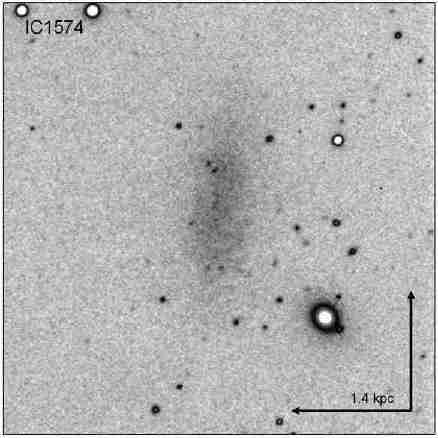}}}&
\mbox{\scalebox{0.75}{\includegraphics[width=0.38\linewidth]{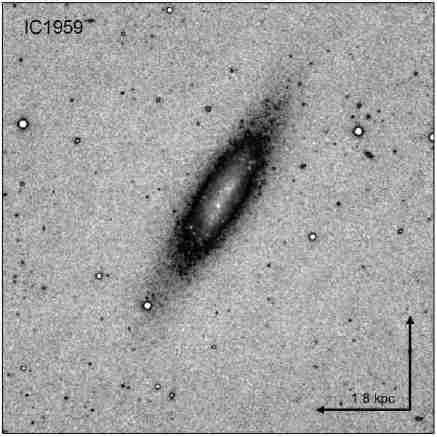}}}\\
\mbox{\scalebox{0.75}{\includegraphics[width=0.38\linewidth]{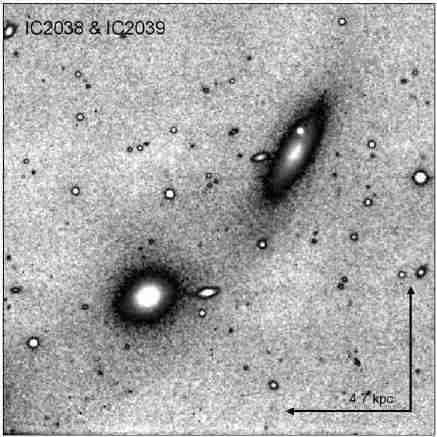}}}&
\mbox{\scalebox{0.75}{\includegraphics[width=0.38\linewidth]{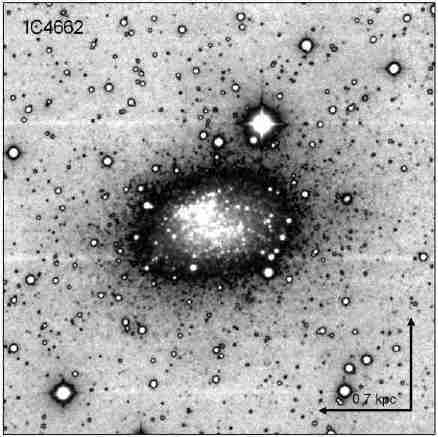}}}&
\mbox{\scalebox{0.75}{\includegraphics[width=0.38\linewidth]{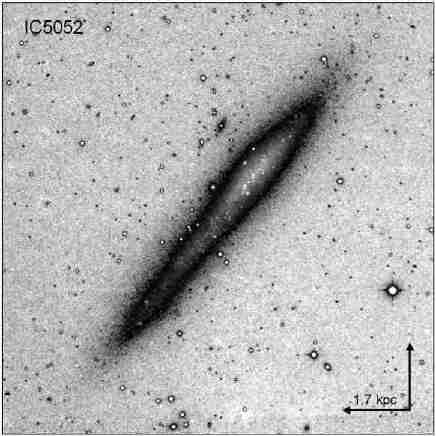}}}\\
\mbox{\scalebox{0.75}{\includegraphics[width=0.38\linewidth]{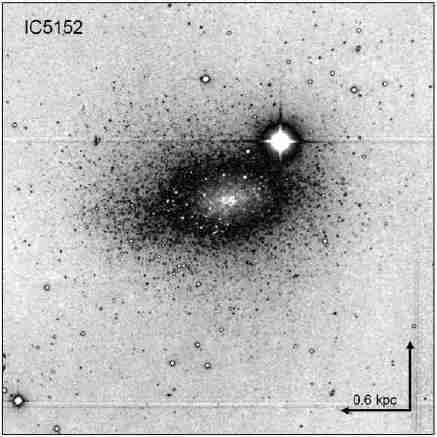}}}&
\mbox{\scalebox{0.75}{\includegraphics[width=0.38\linewidth]{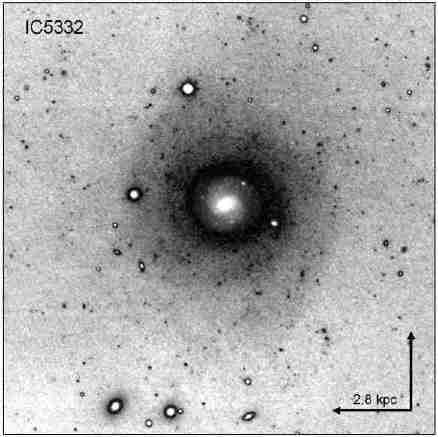}}}&
\mbox{\scalebox{0.75}{\includegraphics[width=0.38\linewidth]{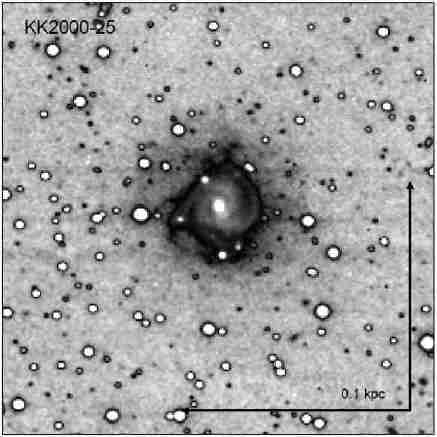}}}\\
\mbox{\scalebox{0.75}{\includegraphics[width=0.38\linewidth]{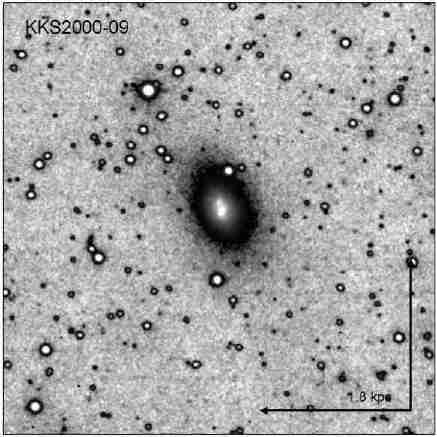}}}&
\mbox{\scalebox{0.75}{\includegraphics[width=0.38\linewidth]{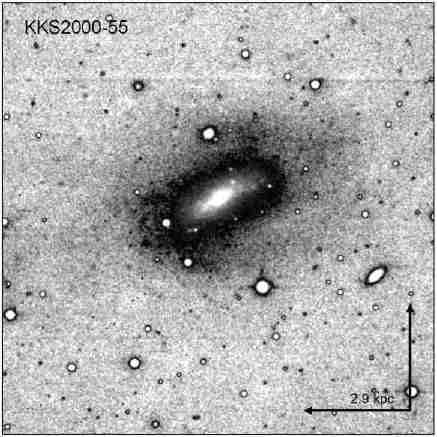}}}&
\mbox{\scalebox{0.75}{\includegraphics[width=0.38\linewidth]{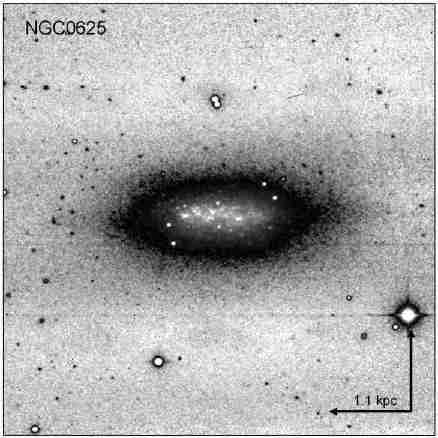}}}\\
  \end{tabular}
\caption{{LSI deep $H$-band images from the 3.9m Anglo-Australian Telescope. Here the scale represents 1 arcmin. The corresponding linear scale is also indicated. North is up and East is to the left. The intensity is represented by a greyscale which goes from white  (low intensity) to black (medium intensity) and then  back to white (high intensity). Higher resolution version available at  \url{http://www.mso.anu.edu.au/~emma/KirbyHband.pdf}}} 
\label{fig:galimages3}
\end{figure*}

\begin{figure*} 
\centering
\begin{tabular}{ccc}
\mbox{\scalebox{0.75}{\includegraphics[width=0.38\linewidth]{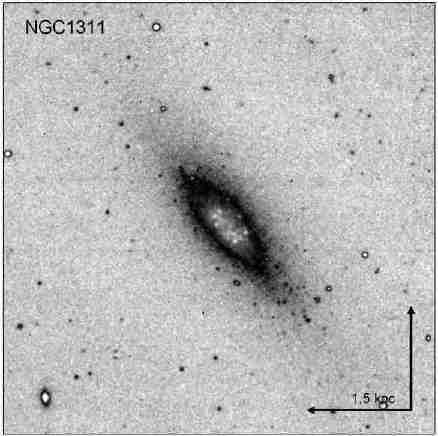}}}&
\mbox{\scalebox{0.75}{\includegraphics[width=0.38\linewidth]{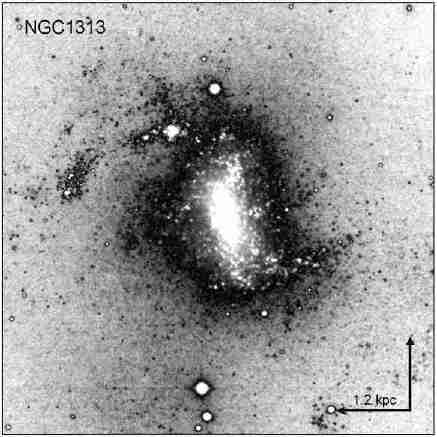}}}&
\mbox{\scalebox{0.75}{\includegraphics[width=0.38\linewidth]{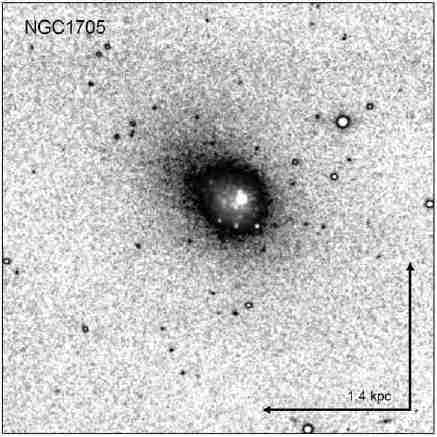}}}\\
\mbox{\scalebox{0.75}{\includegraphics[width=0.38\linewidth]{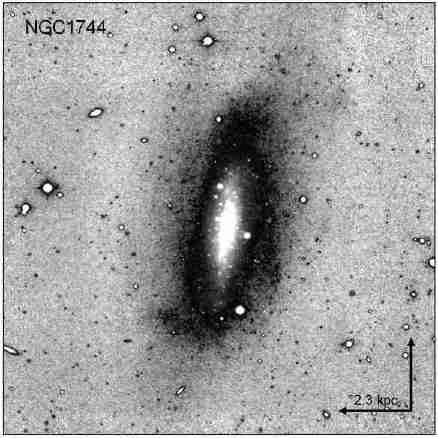}}}&
\mbox{\scalebox{0.75}{\includegraphics[width=0.38\linewidth]{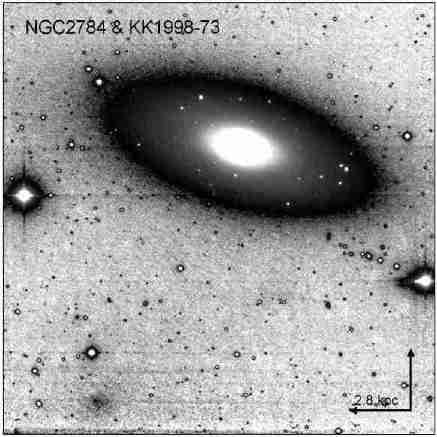}}}&
\mbox{\scalebox{0.75}{\includegraphics[width=0.38\linewidth]{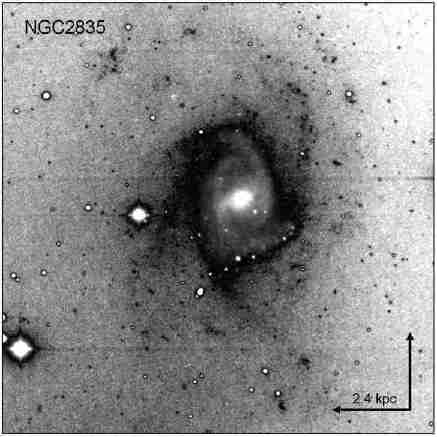}}}\\
\mbox{\scalebox{0.75}{\includegraphics[width=0.38\linewidth]{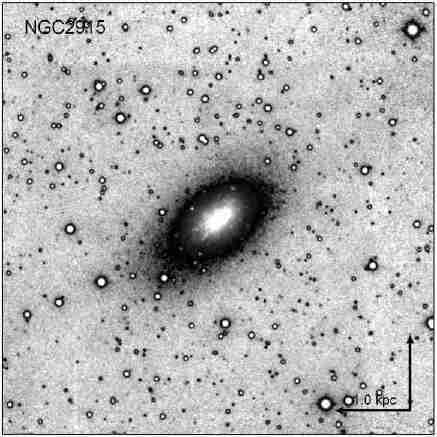}}}&
\mbox{\scalebox{0.75}{\includegraphics[width=0.38\linewidth]{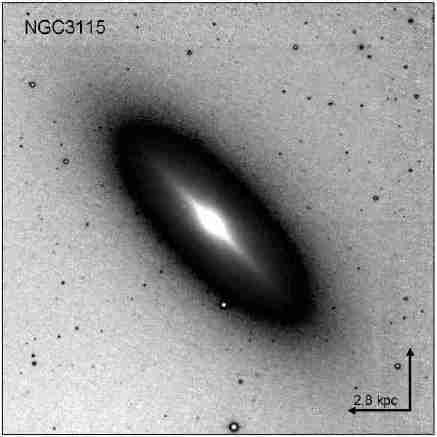}}}&
\mbox{\scalebox{0.75}{\includegraphics[width=0.38\linewidth]{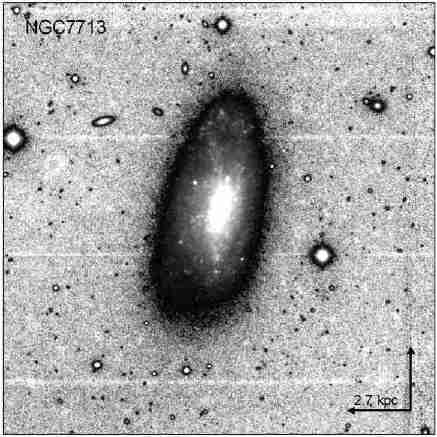}}}\\
\mbox{\scalebox{0.75}{\includegraphics[width=0.38\linewidth]{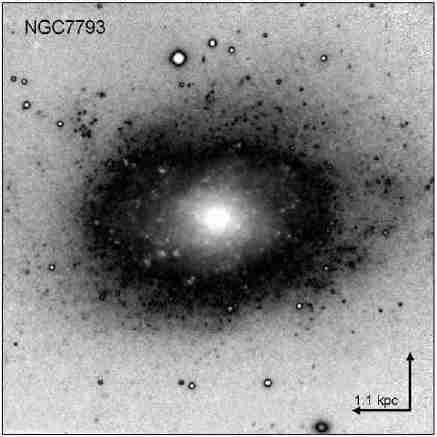}}}&
\mbox{\scalebox{0.75}{\includegraphics[width=0.38\linewidth]{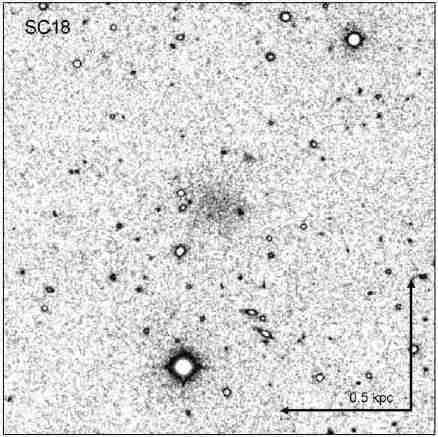}}}&
\mbox{\scalebox{0.75}{\includegraphics[width=0.38\linewidth]{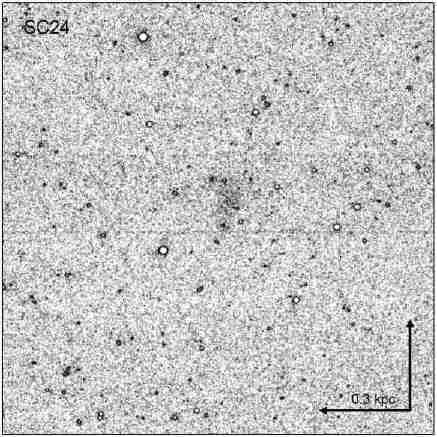}}}\\
  \end{tabular}
\caption{{LSI deep $H$-band images from the 3.9m Anglo-Australian Telescope. Here the scale represents 1 arcmin. The corresponding linear scale is also indicated. North is up and East is to the left. The intensity is represented by a greyscale which goes from white  (low intensity) to black (medium intensity) and then  back to white (high intensity). Higher resolution version available at  \url{http://www.mso.anu.edu.au/~emma/KirbyHband.pdf}}} 
\label{fig:galimages4}
\end{figure*}

\begin{figure*} 
\centering
\begin{tabular}{ccc}
\mbox{\scalebox{0.75}{\includegraphics[width=0.38\linewidth]{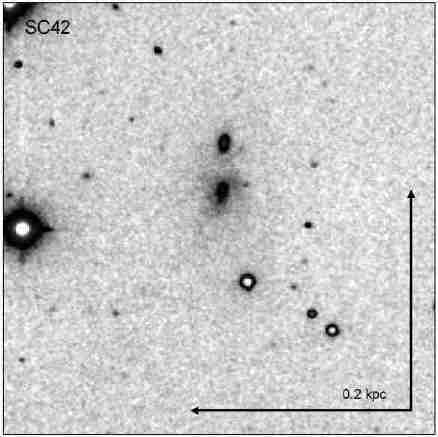}}}&
\mbox{\scalebox{0.75}{\includegraphics[width=0.38\linewidth]{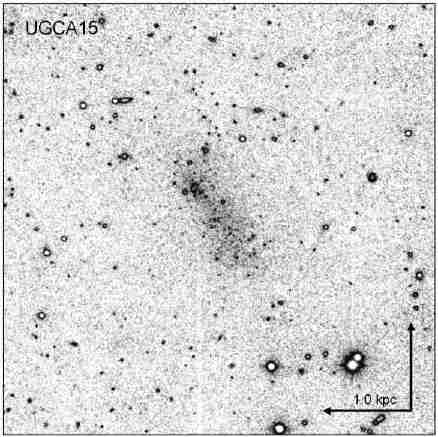}}}&
\mbox{\scalebox{0.75}{\includegraphics[width=0.38\linewidth]{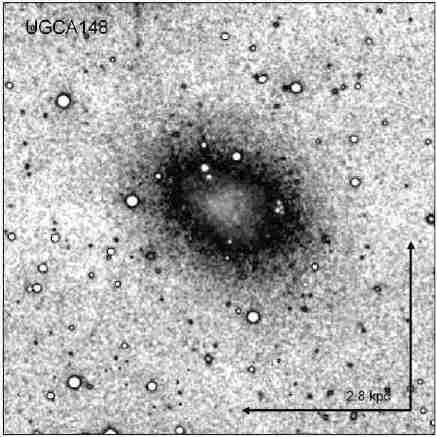}}}\\
\mbox{\scalebox{0.75}{\includegraphics[width=0.38\linewidth]{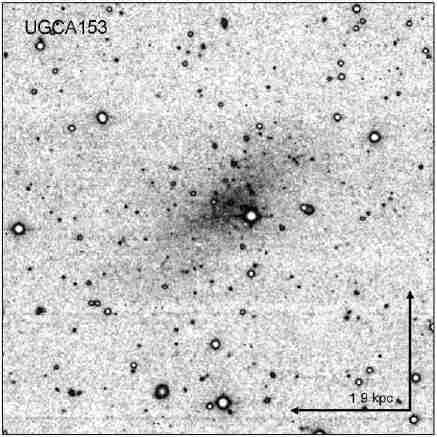}}}&
\mbox{\scalebox{0.75}{\includegraphics[width=0.38\linewidth]{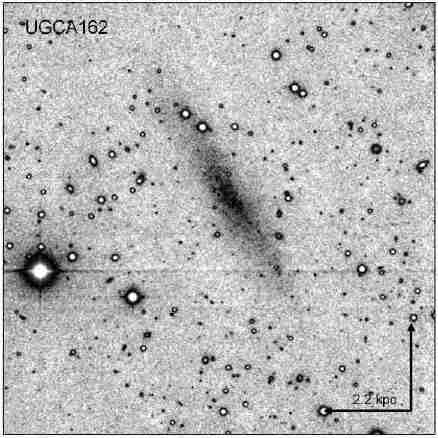}}}&
\mbox{\scalebox{0.75}{\includegraphics[width=0.38\linewidth]{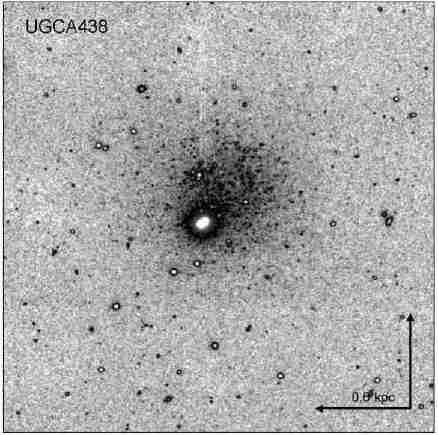}}}\\
\mbox{\scalebox{0.75}{\includegraphics[width=0.38\linewidth]{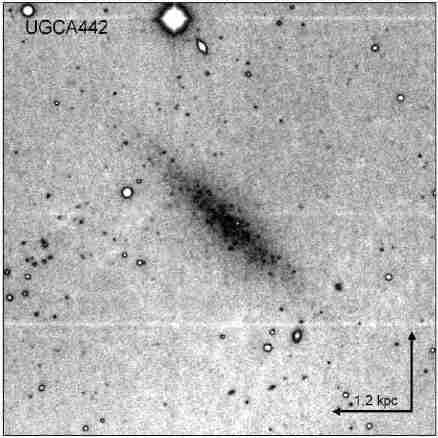}}}\\
  \end{tabular}
\caption{{LSI deep $H$-band images from the 3.9m Anglo-Australian Telescope. Here the scale represents 1 arcmin. The corresponding linear scale is also indicated. North is up and East is to the left. The intensity is represented by a greyscale which goes from white  (low intensity) to black (medium intensity) and then  back to white (high intensity). Higher resolution version available at  \url{http://www.mso.anu.edu.au/~emma/KirbyHband.pdf}}} 
\label{fig:galimages5}
\end{figure*}

Of the 68 galaxies observed, 11 remained undetected or could not be usefully analysed. This was either because our $H$-band surface brightness limit of 
$\mu_{lim}\approx 25$\,mag\,arcsec$^{-2}$ or 20$\,L_{\odot}$\,arcsec$^{-2}$ at a distance of 1\,Mpc (adopting $M_{H,\odot}=3.35$\,mag, \citealt{colina96}) was not low enough or the galaxy light was heavily contaminated by  Galactic foreground stars (see Table~\ref{badgals}). 
In  both situations the data was not processed further. For instance, the companion galaxies NGC2784 DW1 and KK98-73 were observed 
parallel to NGC2784. While NGC2784 and KK98-73 can be seen, NGC2784 DW1,  located between NGC2784 and KK98-73, is barely visible (see Figure~\ref{fig:galimages4}, second row, middle panel where KK98-73 is visible to the bottom left of the image). The images of 
the other 57 galaxies are shown in Figures~\ref{fig:galimages1} to~\ref{fig:galimages5}.

\begin{figure}[!htb]
\centering
\begin{tabular}{cc}
  \mbox{\includegraphics[width=0.45\linewidth]{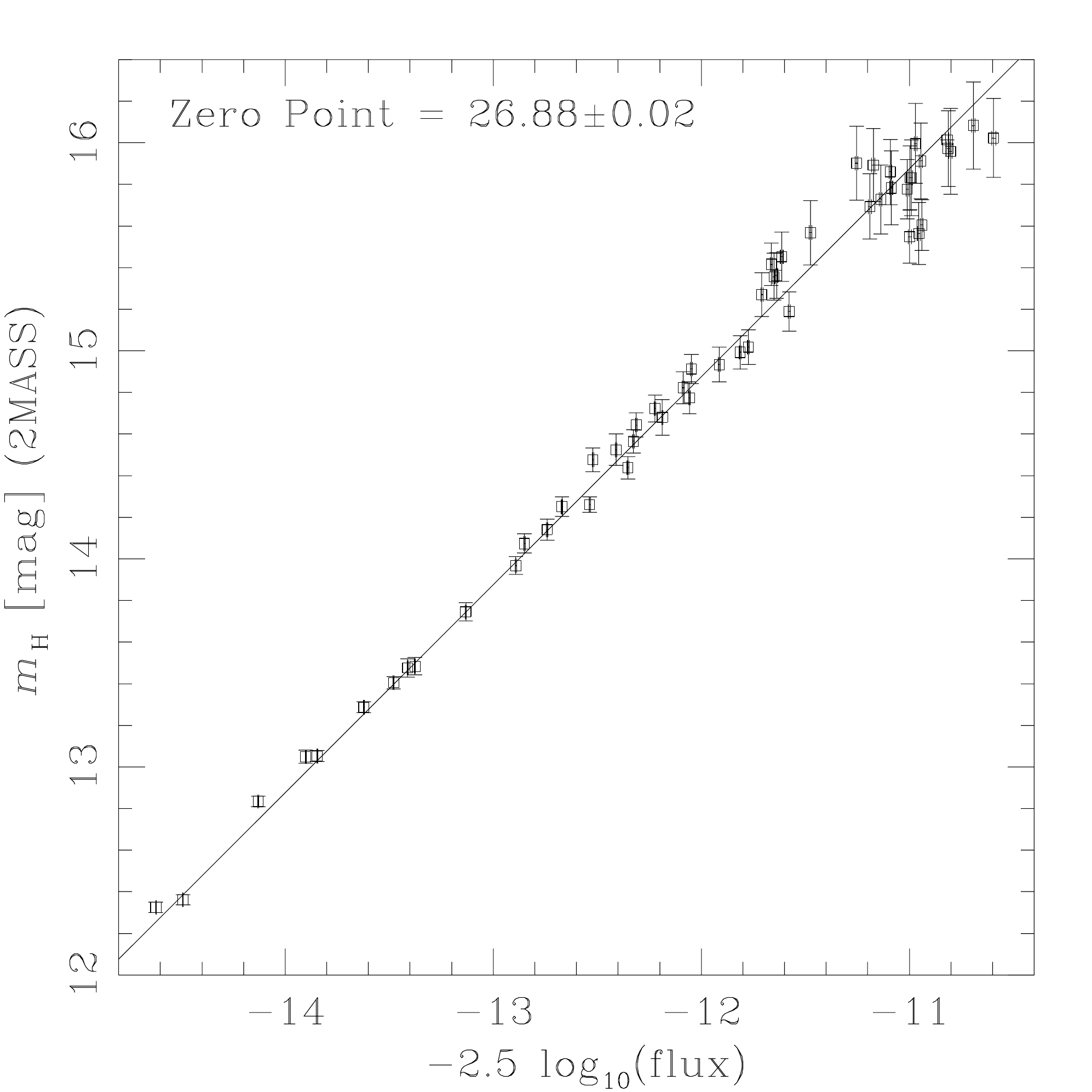}}&
  \mbox{\includegraphics[width=0.45\linewidth]{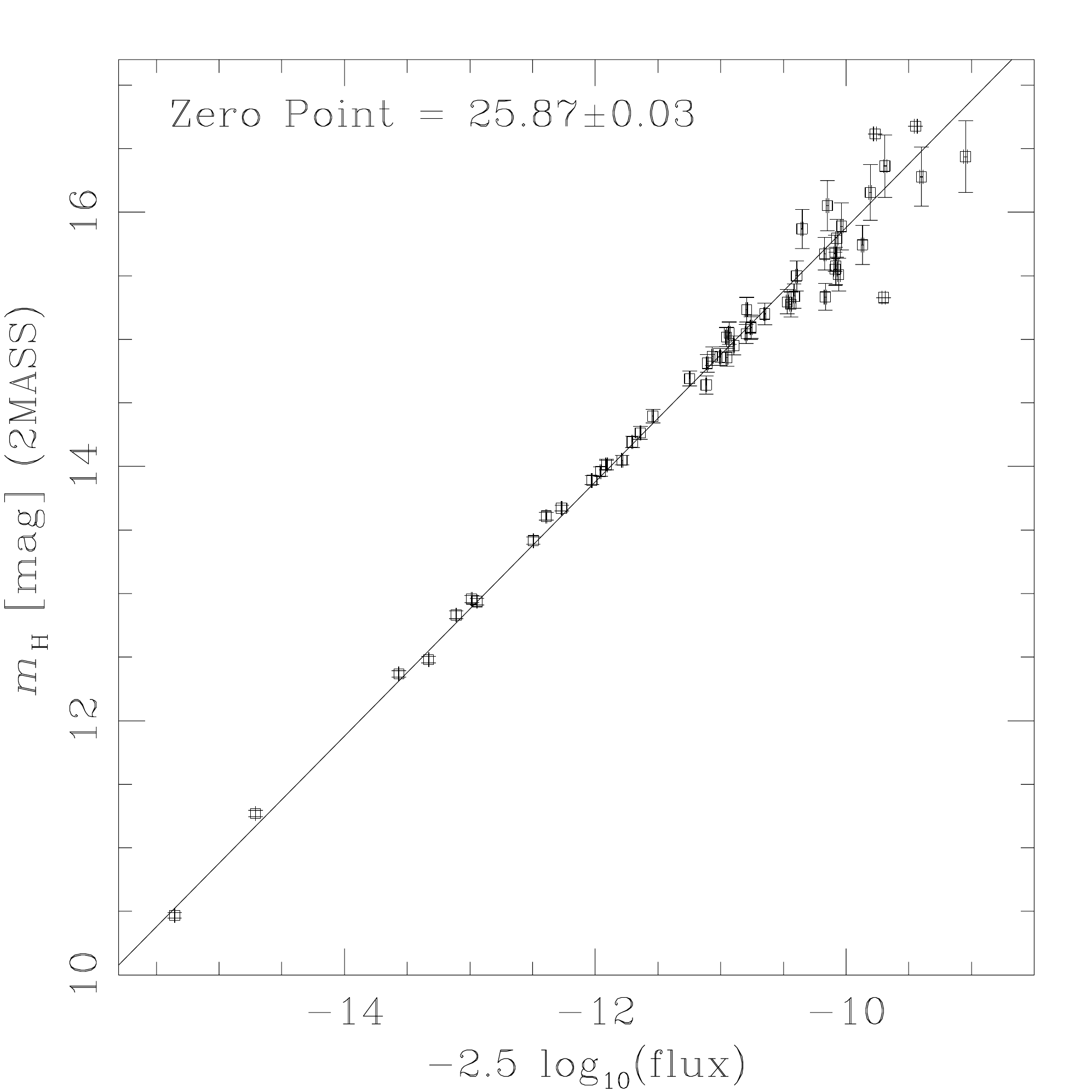}}\\
\end{tabular}
\caption{{\small Instrumental versus 2MASS $H$-band  magnitudes for foreground stars around the galaxies ESO121-G020 (left) and UGCA153 (right). Both galaxies were observed for the same amount of time and the difference of  one mag in zero-point  is due to the presence of thin clouds when UGCA153 was observed.}} 
\label{fig:zeropoint}
\end{figure}

On each image, instrumental  magnitudes for 50--100 field stars were measured employing standard IRAF PSF fitting routines. 
Cross-correlating the stellar positions with the 2MASS Point Source Catalog 
provided $H$-band magnitudes and allowed the photometric 
calibration of each field  (see Figure~\ref{fig:zeropoint}). The stars which deviate from the 45 degree line 
were usually either extremely red or blue where the transformation between 
2MASS and IRIS2 $H$-bands  \citep{ryder07} breaks down.  The 1$\sigma$ uncertainty in the zero point 
was calculated to be between 0.01 and 0.04\,mag depending on the number 
of stars used for the calibration.

\begin{figure}[!hbt]
\centering
\begin{tabular}{cc}
 \mbox{\includegraphics[width=0.45\linewidth]{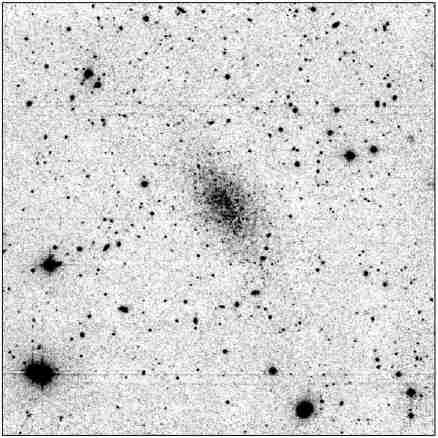}}&
 \mbox{\includegraphics[width=0.45\linewidth]{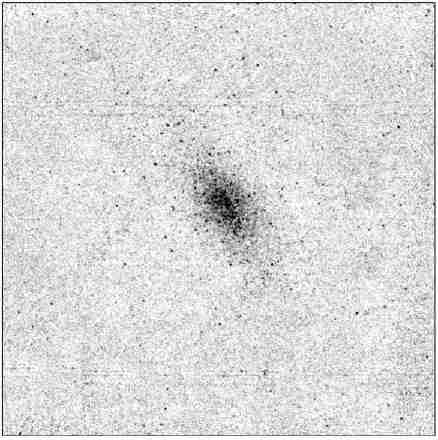}}\\
 \mbox{\includegraphics[width=0.45\linewidth]{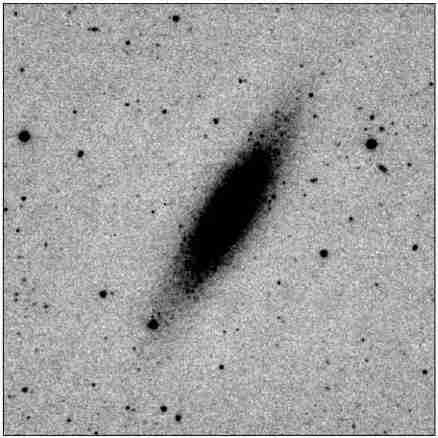}}&
 \mbox{\includegraphics[width=0.45\linewidth]{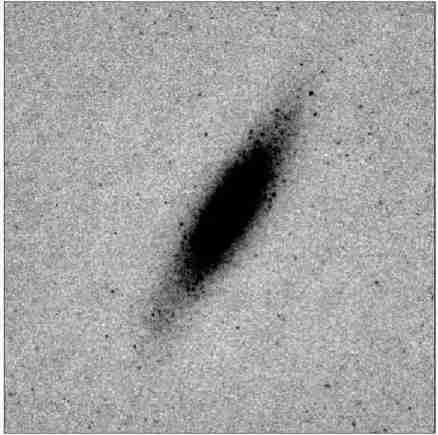}}\\
\end{tabular}
\caption{{\small Two examples of the effectiveness of the foreground cleaning process. Shown are the before and after cleaning images of ESO468-G020 [upper panel] and IC1959 [lower panel]. Higher resolution version available at  \url{http://www.mso.anu.edu.au/~emma/KirbyHband.pdf}}}
\label{fig:cleaning} 
\end{figure}

To ensure accurate galaxy surface photometry  down to the faintest possible isophotes, the images were cleaned of  foreground stars using procedures written within the IRAF package. Thereby, 
 stars in the field around a galaxy were carefully replaced with nearby patches of plain sky. If 
 superposed on the galaxy, the galaxy light under the star was restored by replacing the
contaminated area with its mirror image with respect to the galaxy
center. The galaxy center was defined as the center of the
luminosity-weighted light distribution.  The star removal process was monitored visually to identify small-scale structures and asymmetries,  and to ensure accurate removal of foreground stars whilst not removing sources associated with the galaxy itself. The effectiveness of the
cleaning procedure is illustrated in Figure~\ref{fig:cleaning} where
pre- and post- cleaning images are shown for the 
two galaxies ESO468-G020 and  IC1959.

The fields of IC5152 (Figure~\ref{fig:galimages3}) and UGCA438 (Figure~\ref{fig:galimages5}) both have a bright foreground star which
obscures a large portion of the galaxy. To obtain rough photometric
parameters for these  objects, the contribution of the bright stars to the total flux needed to be removed. Because such a large portion of both the galaxies is obscured, it was necessary to remove the affected
quarter of the image and replace it with the opposite quarter, which was
rotated by 180 degrees (i.e., in the case of IC5152, the top-right quarter was 
replaced with the rotated bottom-left quarter of the image).  This method failed in the case of the dIrr galaxy AM0521-343, where an even brighter star (CD-34 2225, $m_H=7.1$) near the faint dwarf galaxy prevented proper cleaning (see Figure~\ref{fig:galimages1}).  

\section{Surface Photometry}\label{s:photometry}

Simulated circular aperture photometry of the star-subtracted $H$-band
images produced a growth curve as a function of the geometric mean radius
$\sqrt{ab}$ (where $a$ and $b$ are the galaxy's major and minor axes). 
The asymptotic intensity corresponds to the total apparent magnitude, $m_H$, that 
can be recovered down to the background noise level of the image. The largest source 
of uncertainty is the sky level. By systematically varying the sky brightness we
determined which growth curve converges best to a plateau as far as
possible from the center of the galaxy.  We measure the half-light geometric mean radius,  $r_{eff}$, at  half the asymptotic intensity and calculated the mean surface
brightness within that radius: $\langle\mu_H\rangle_{eff}$.  The overall uncertainty for the total
magnitude, $m_H$, is between 0.05 and 0.30\,mag; for the mean effective
surface brightness, $\langle\mu_H\rangle_{eff}$, less than 0.2\,mag
arcsec$^{-2}$, and for the half-light radius, $r_{eff}$, is of the order of five percent. The image of AM0521-343 (Figure~\ref{fig:galimages1}), contains a bright foreground star and thus, the sky brightness plus the contribution from the stellar halo was estimated at the galaxy's position and simulated aperture photometry performed out to the radius of asymptotic intensity (defined on side of the galaxy opposite to the contaminating star).

\begin{figure*}
\centering
\begin{tabular}{ccccc}
  \mbox{\scalebox{0.24}{\includegraphics{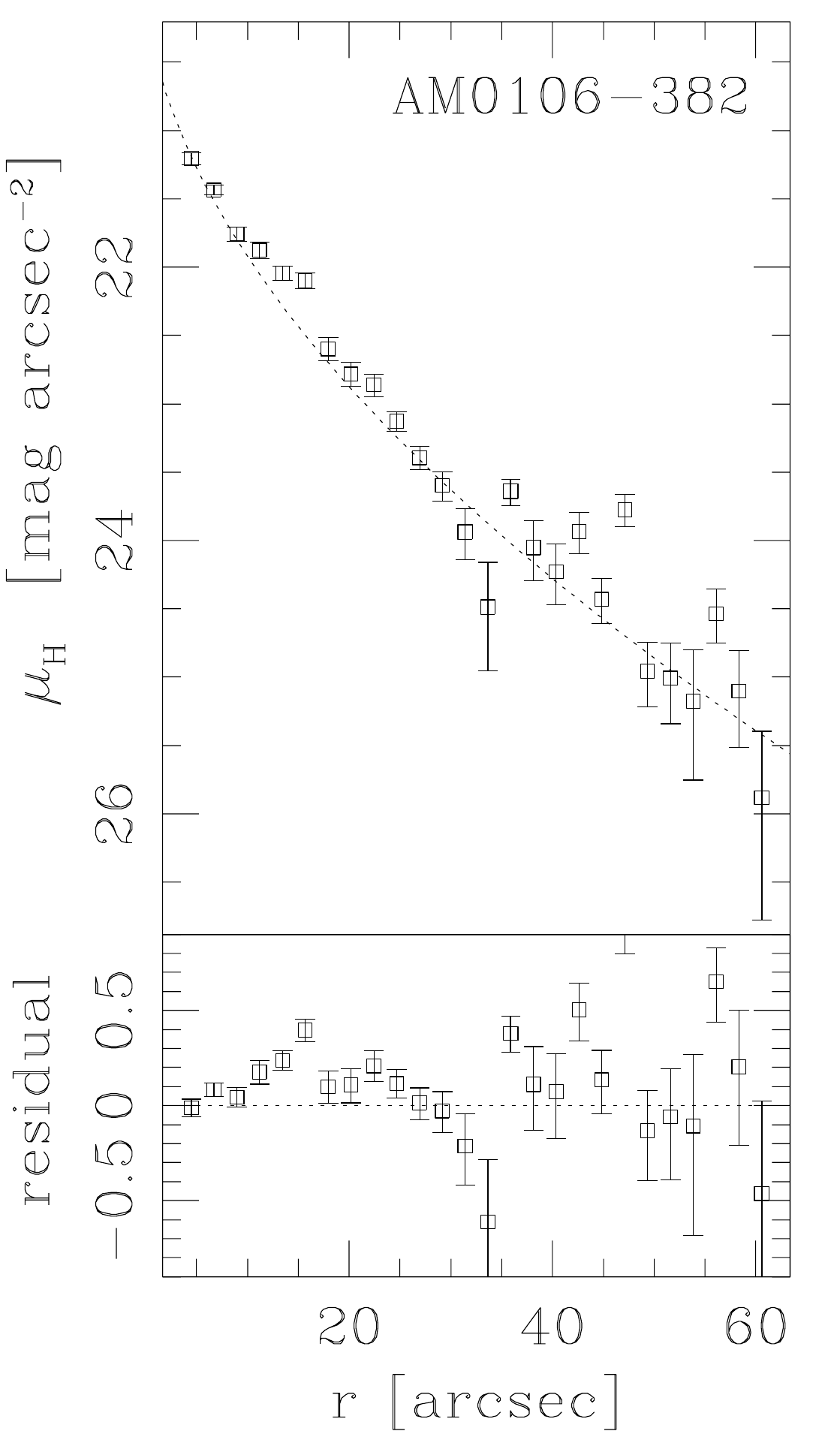}}} &
    \mbox{\scalebox{0.24}{\includegraphics{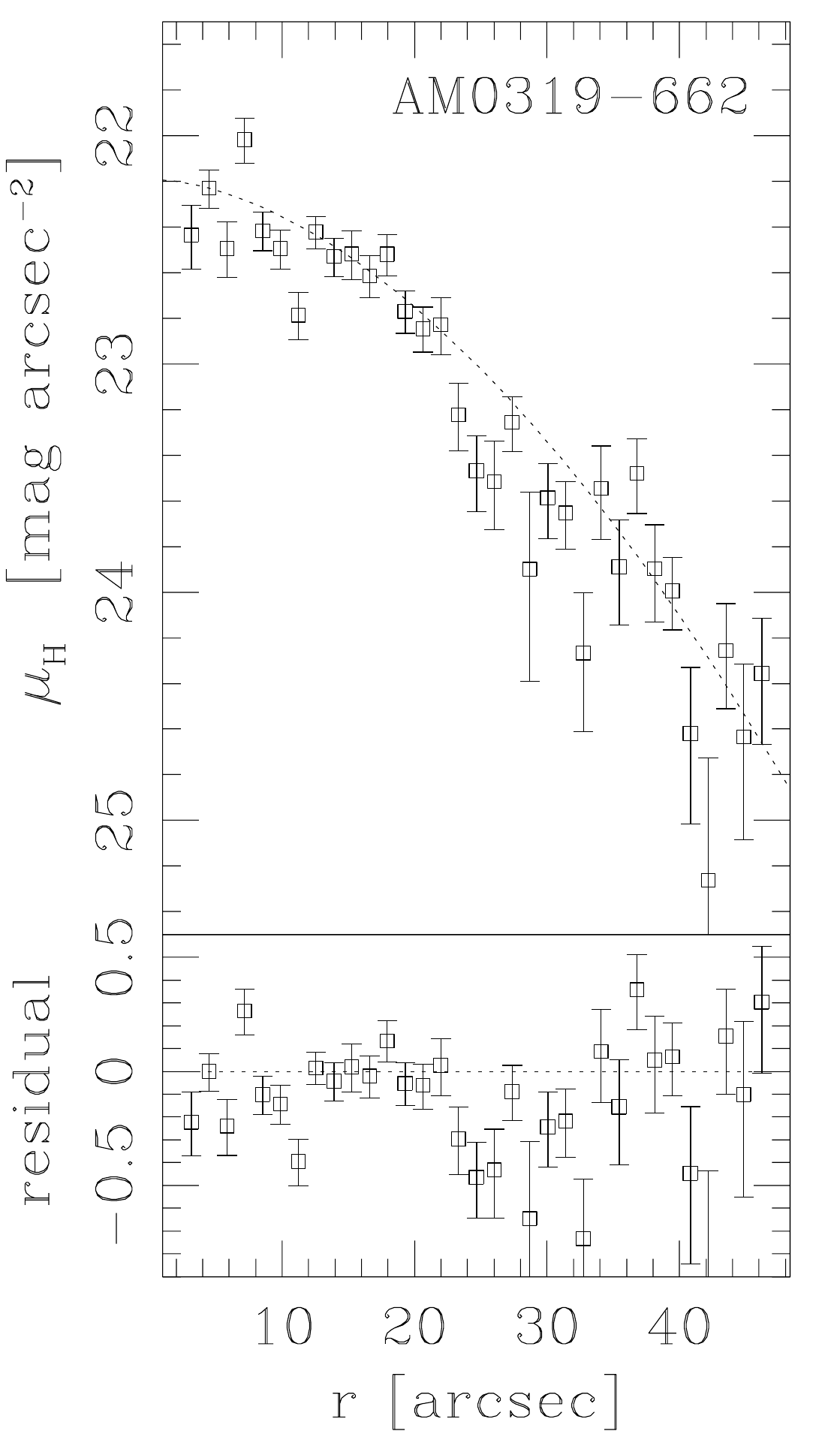}}} &
  \mbox{\scalebox{0.24}{\includegraphics{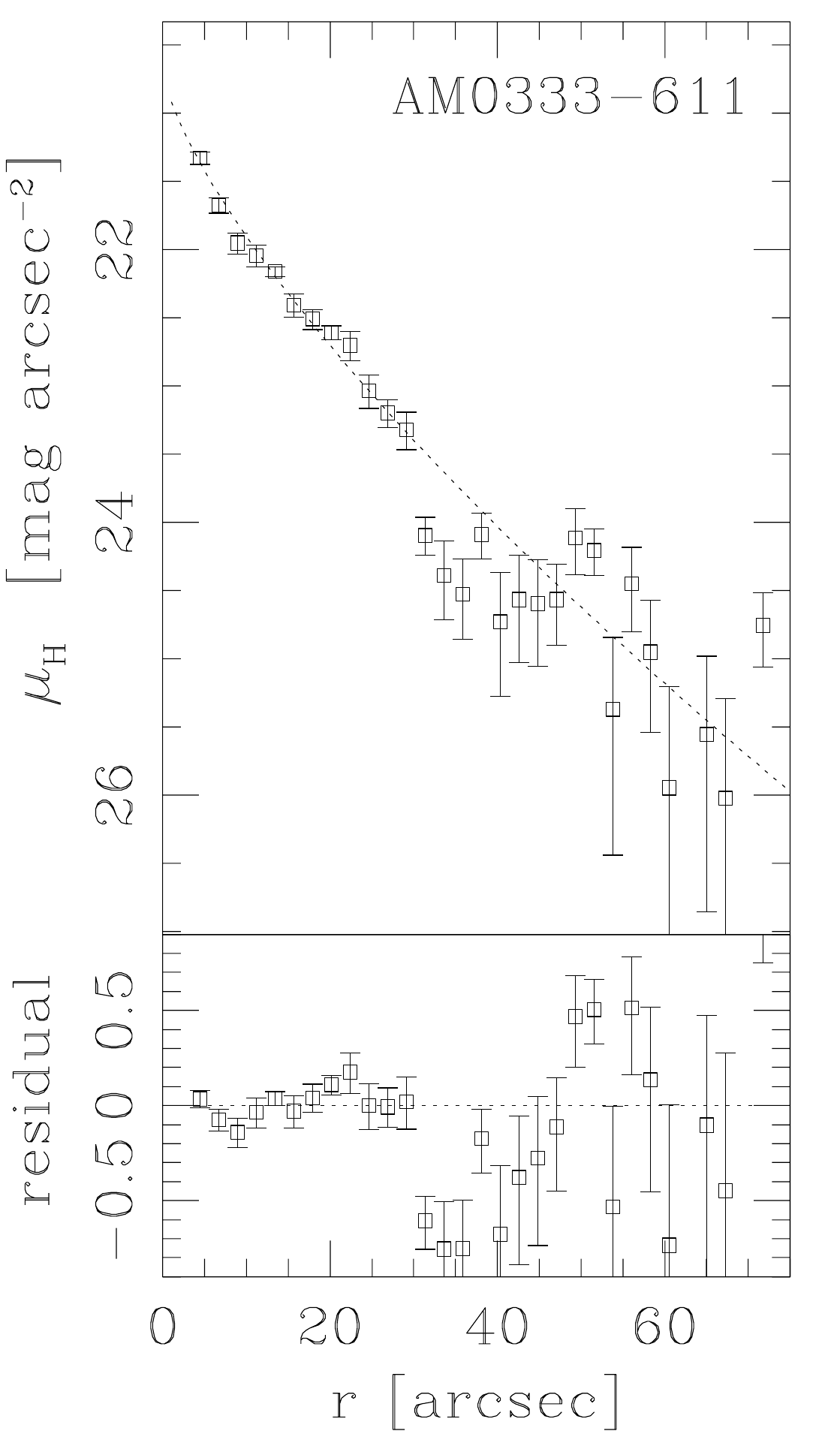}}}  &
  \mbox{\scalebox{0.24}{\includegraphics{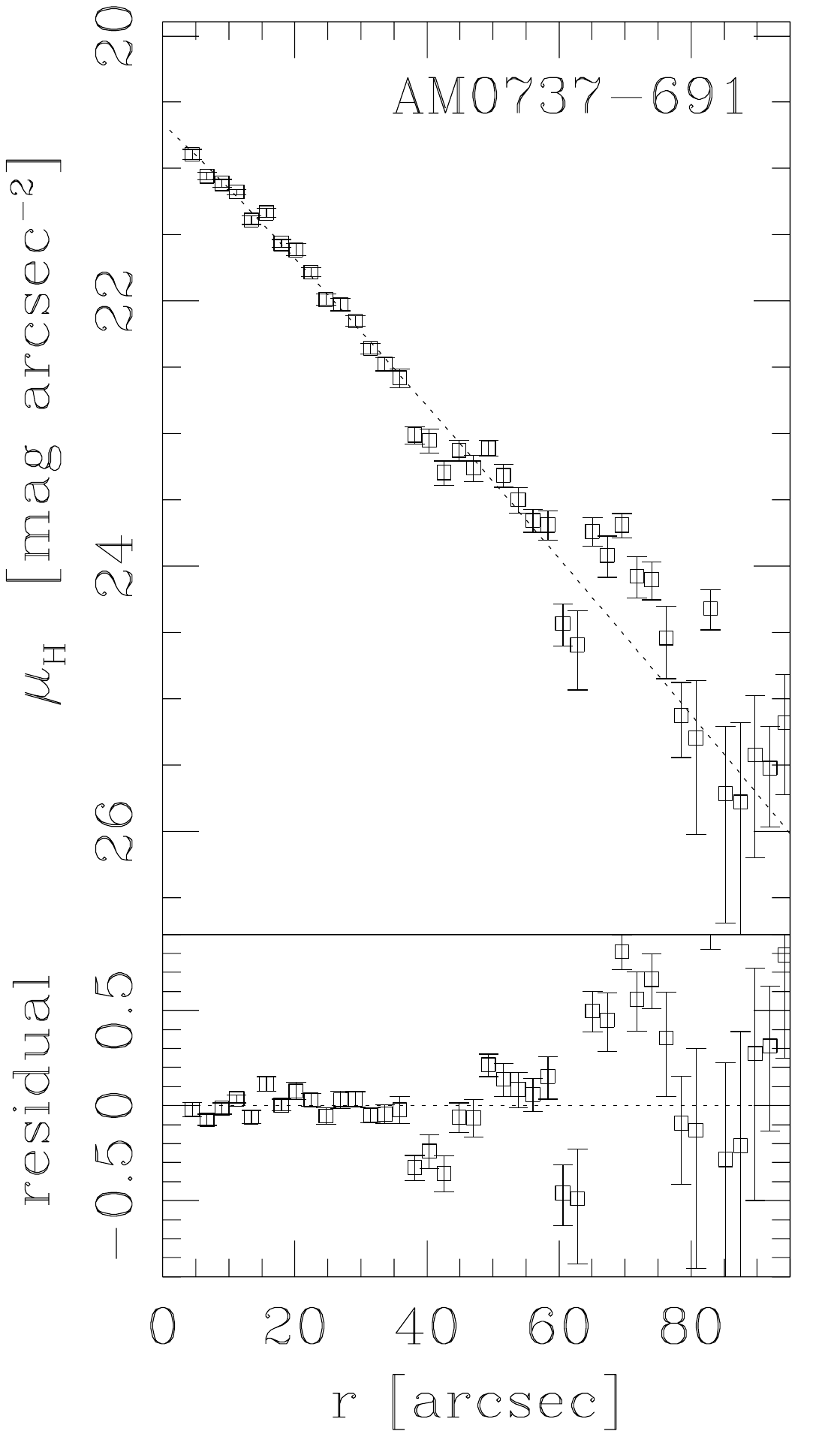}}}  &
  \mbox{\scalebox{0.24}{\includegraphics{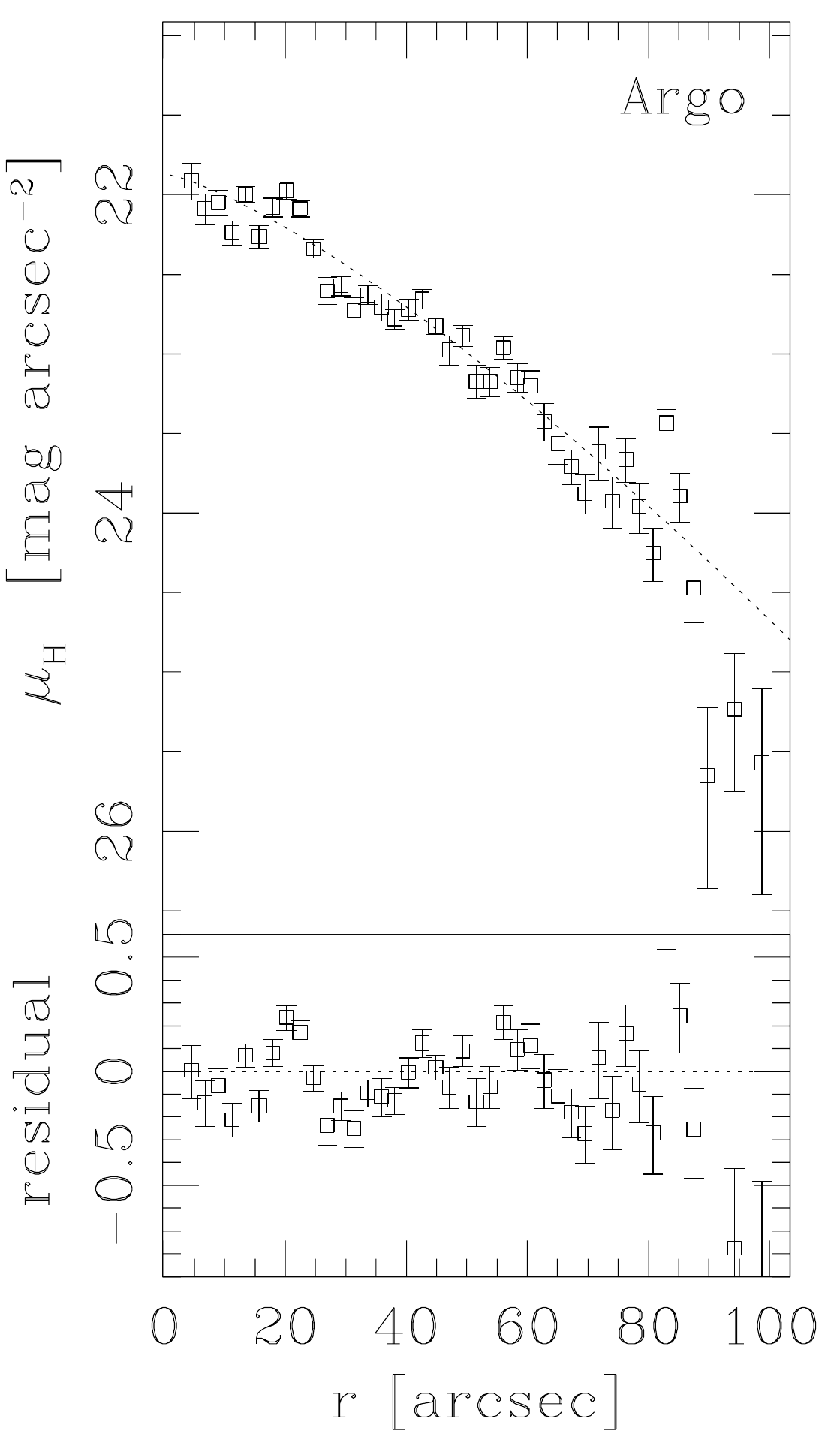}}}  \\
  \mbox{\scalebox{0.24}{\includegraphics{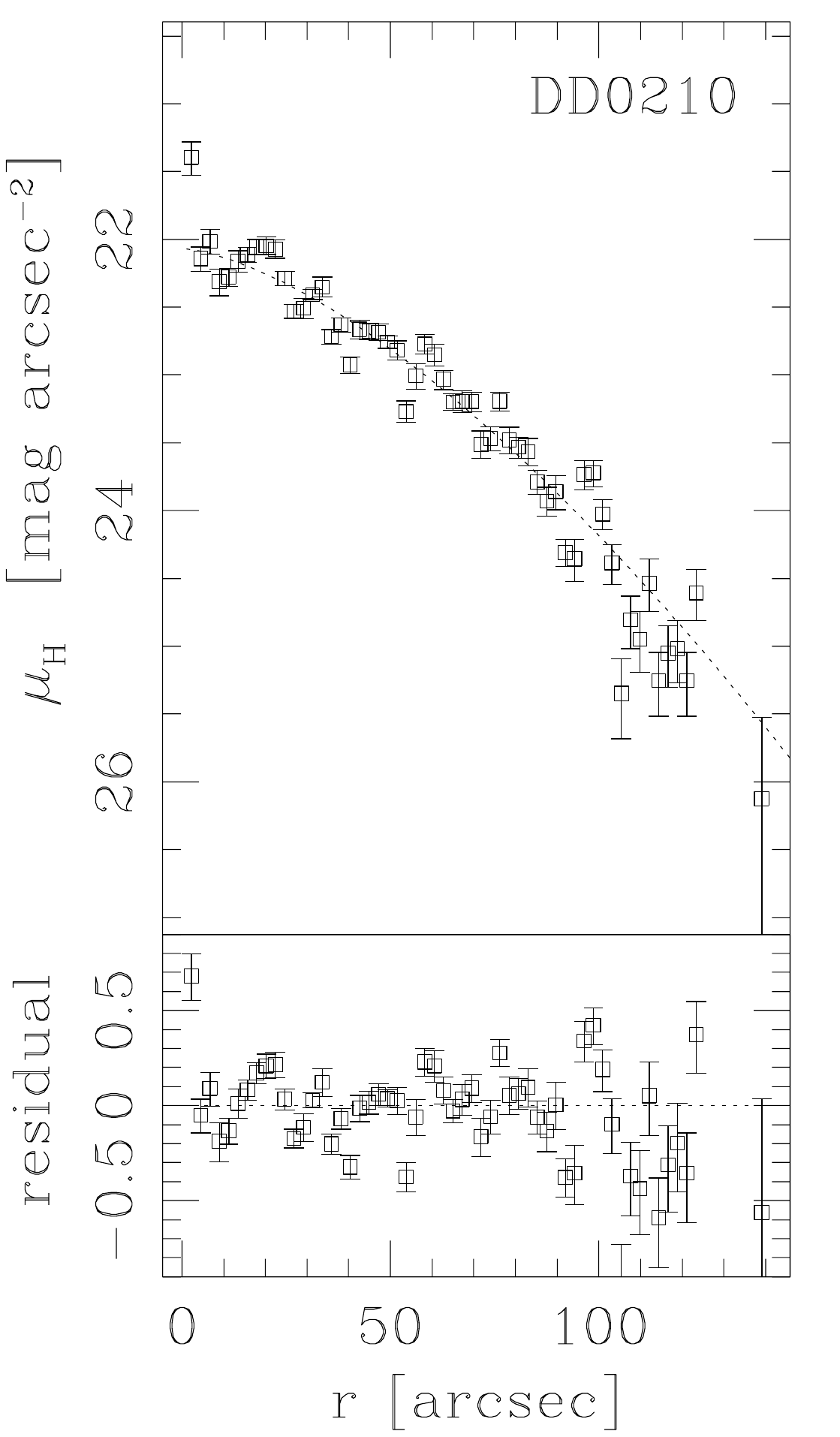}}} &
  \mbox{\scalebox{0.24}{\includegraphics{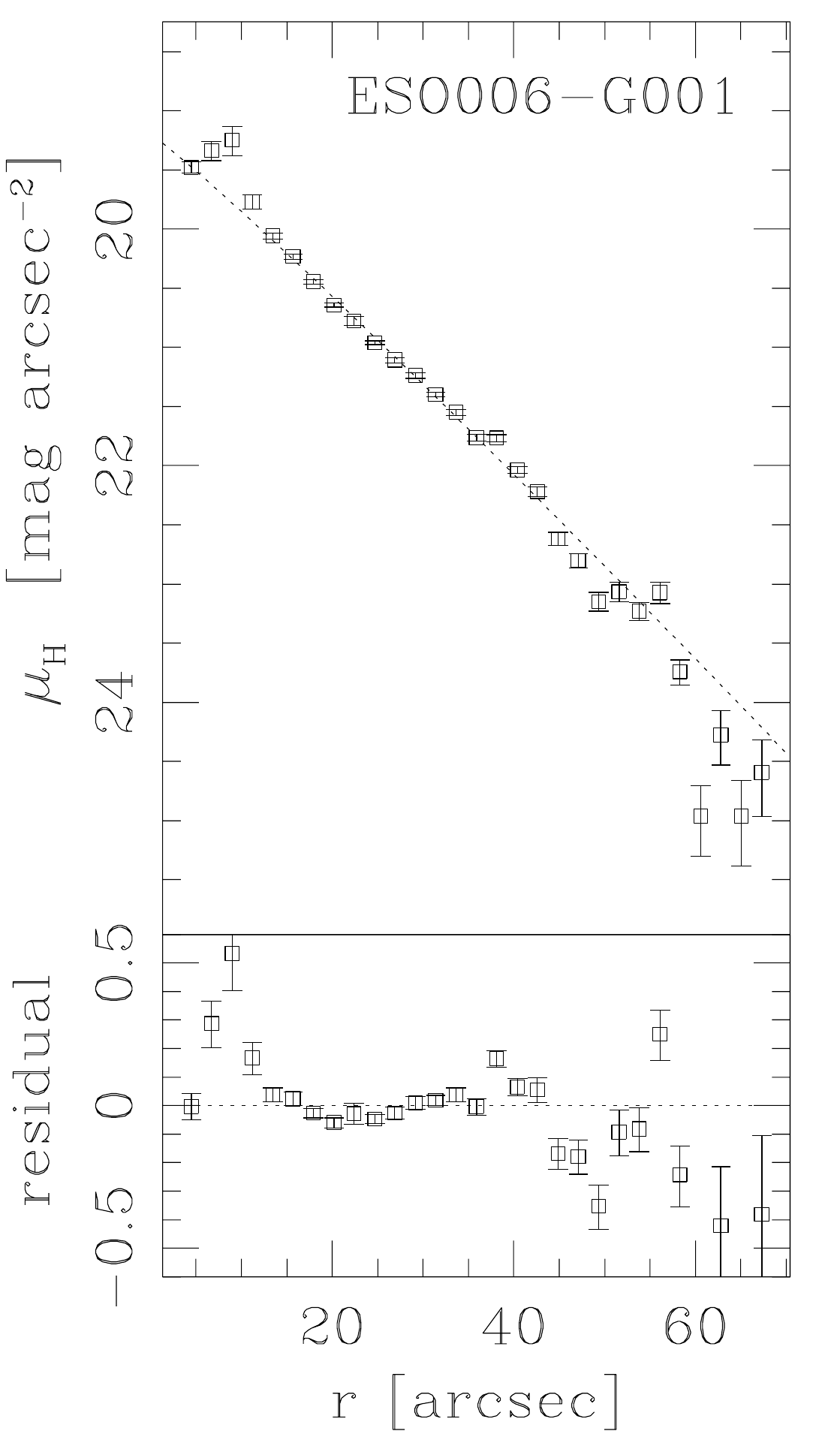}}} &
  \mbox{\scalebox{0.24}{\includegraphics{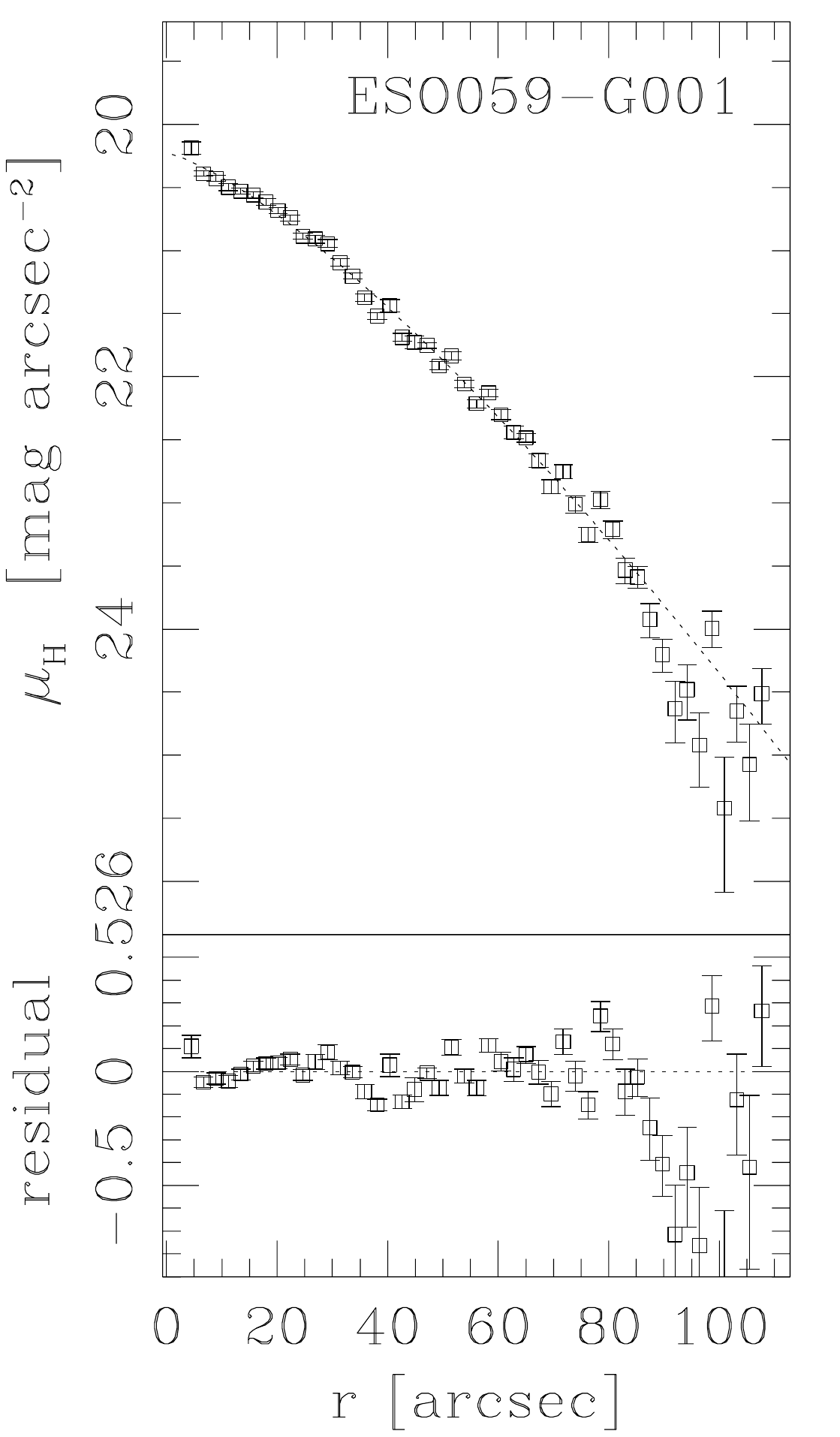}}} &
  \mbox{\scalebox{0.24}{\includegraphics{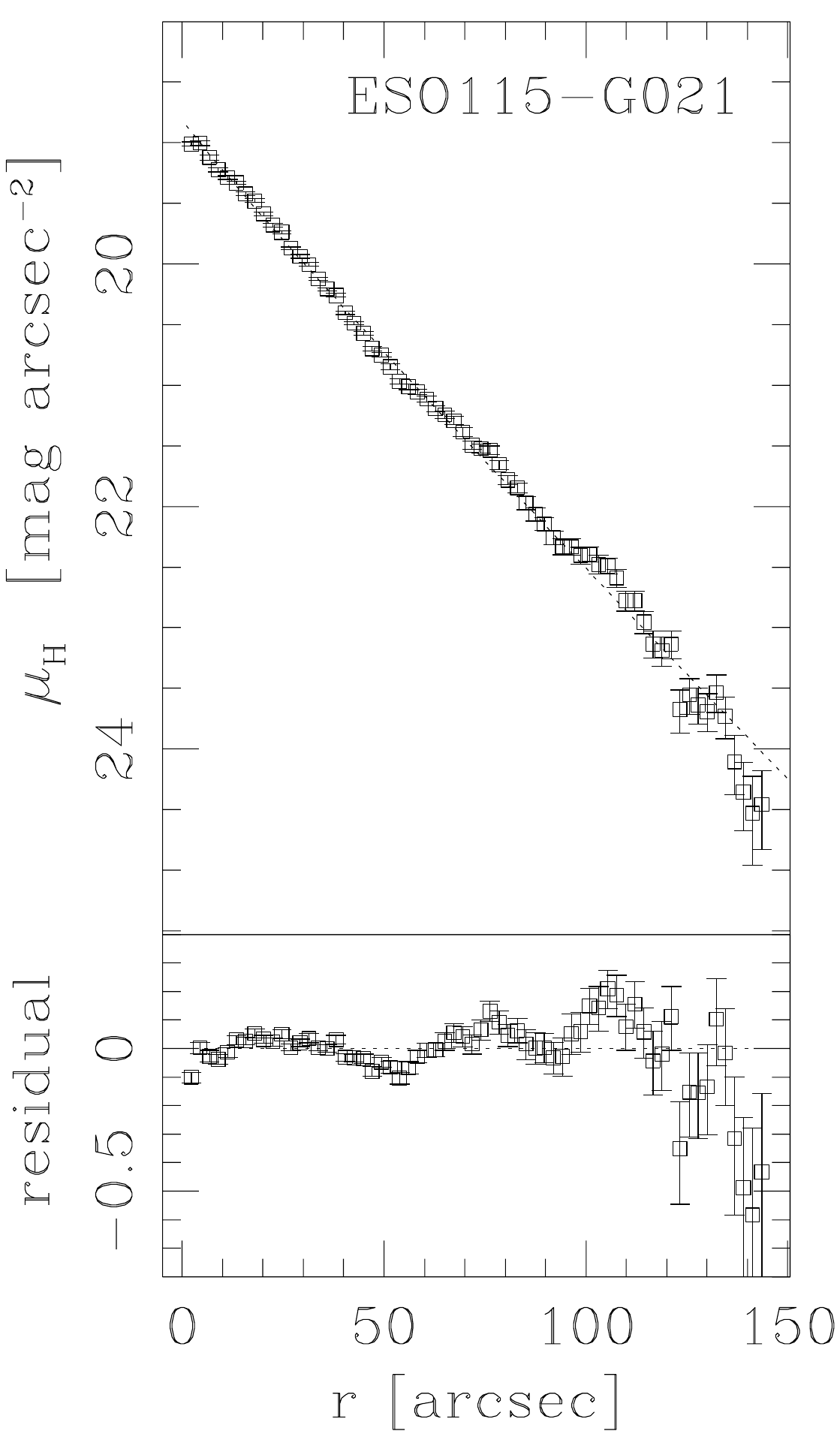}}} &
  \mbox{\scalebox{0.24}{\includegraphics{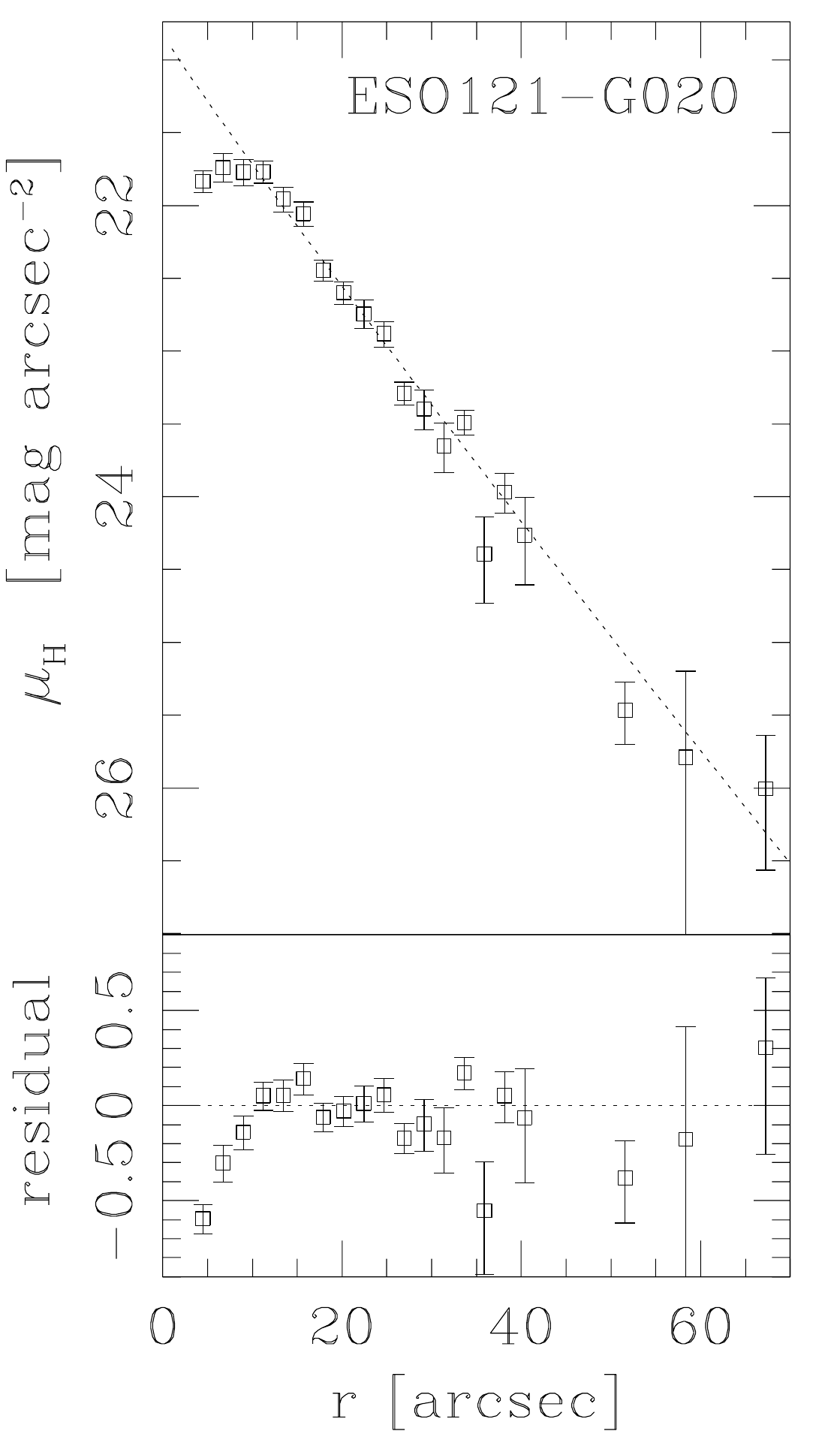}}}  \\
  \mbox{\scalebox{0.24}{\includegraphics{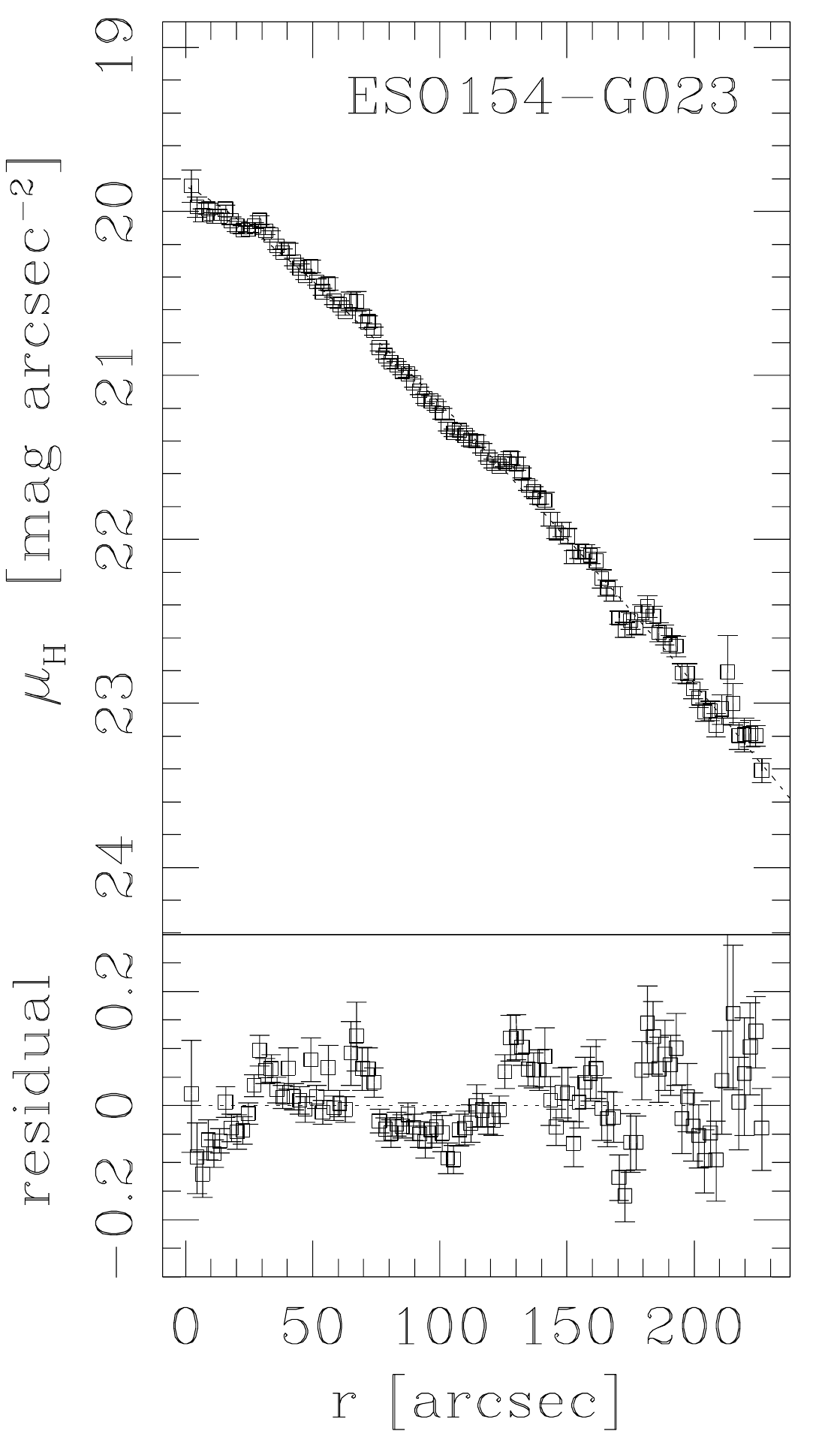}}} &
  \mbox{\scalebox{0.24}{\includegraphics{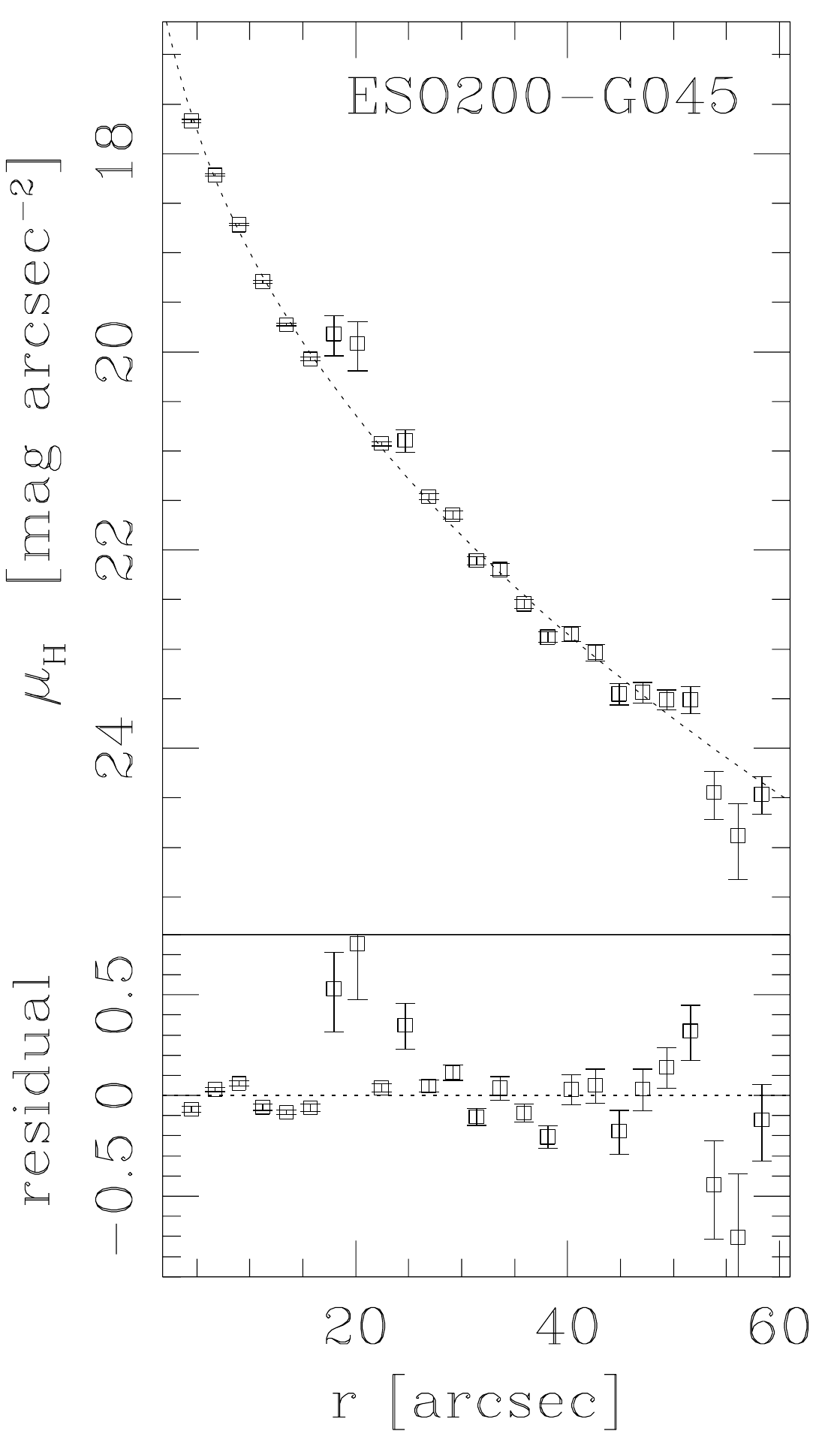}}} &  
  \mbox{\scalebox{0.24}{\includegraphics{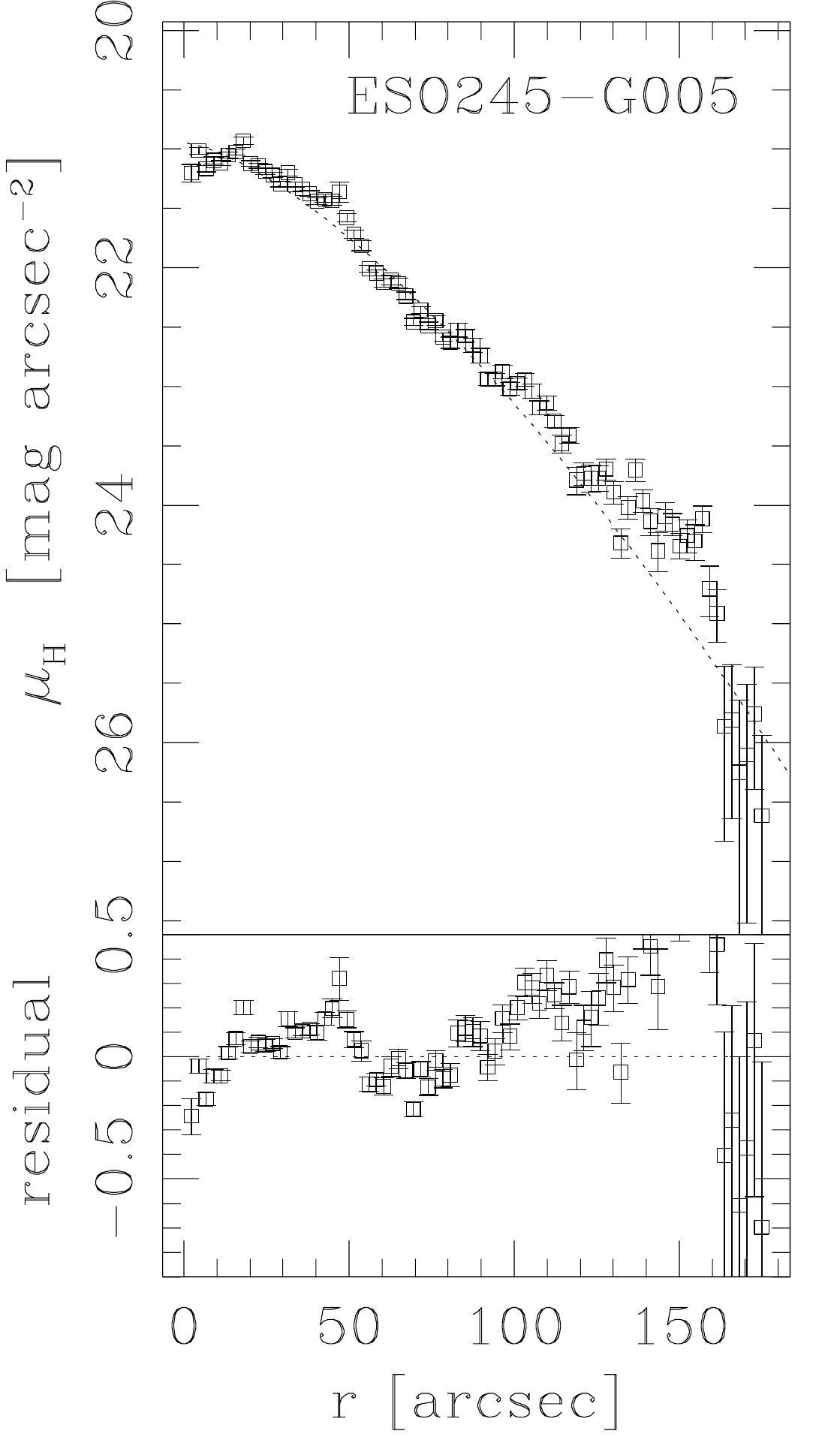}}} &
  \mbox{\scalebox{0.24}{\includegraphics{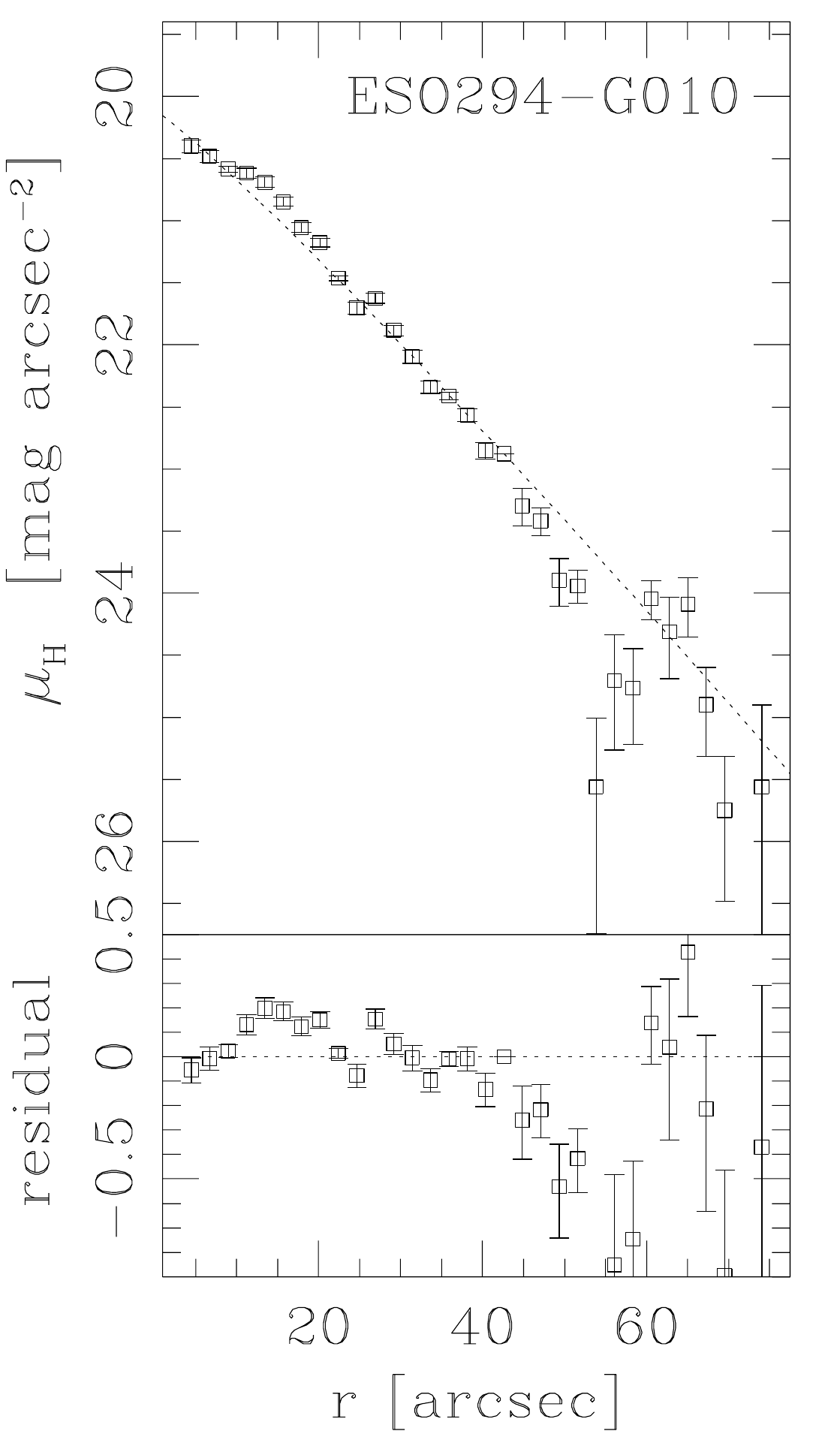}}} &
  \mbox{\scalebox{0.24}{\includegraphics{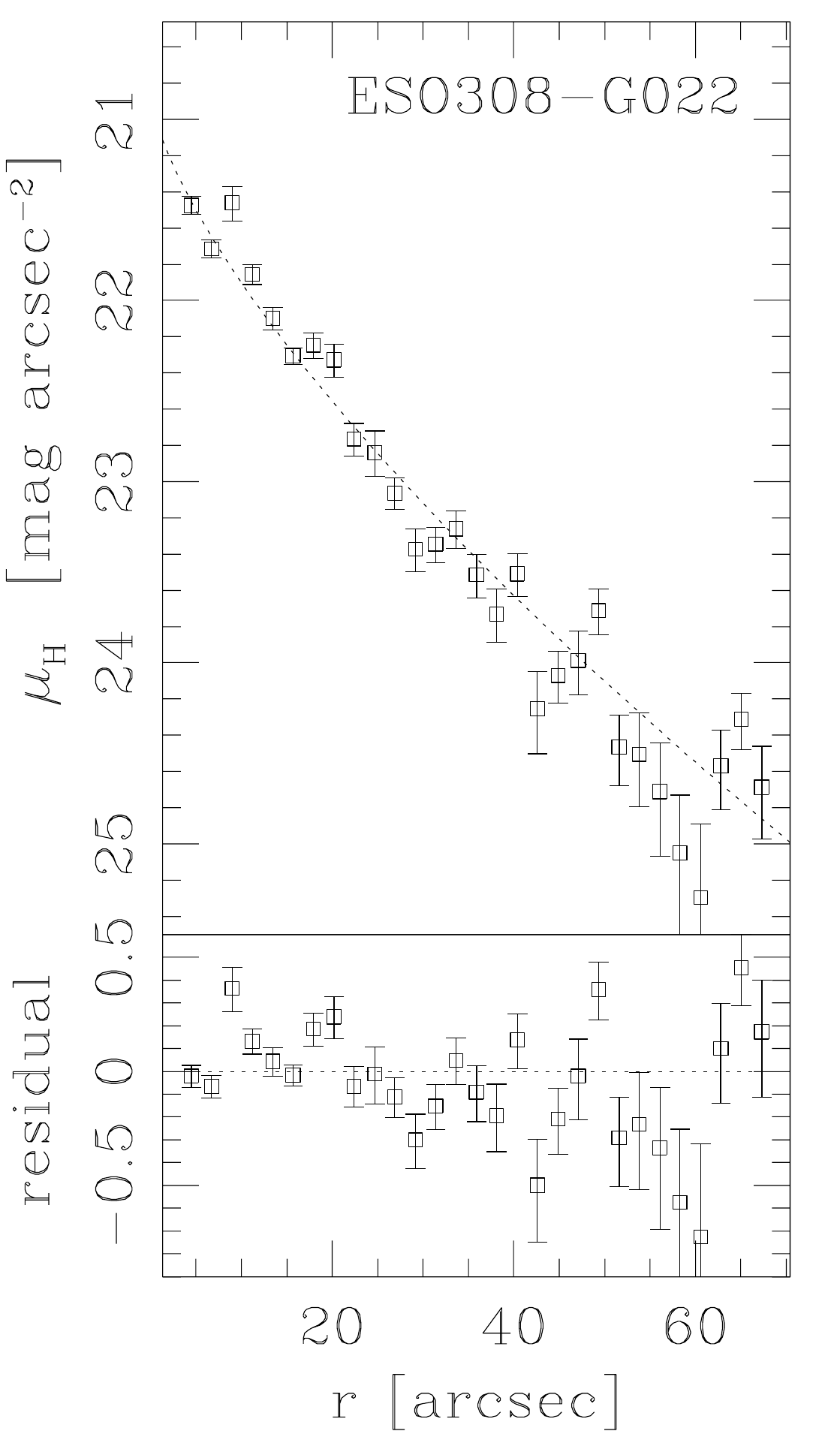}}} \\
  \mbox{\scalebox{0.24}{\includegraphics{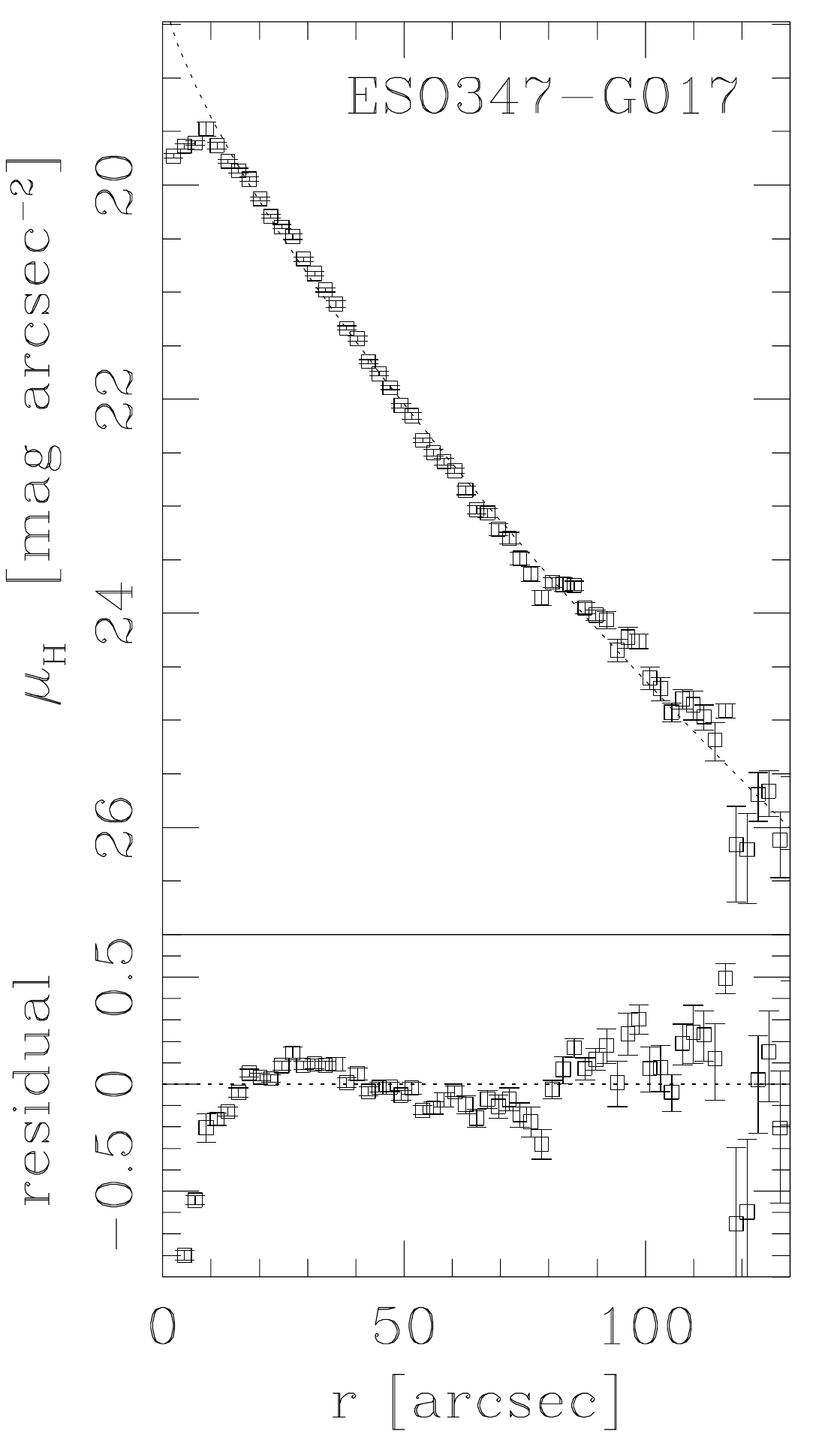}}} &
  \mbox{\scalebox{0.24}{\includegraphics{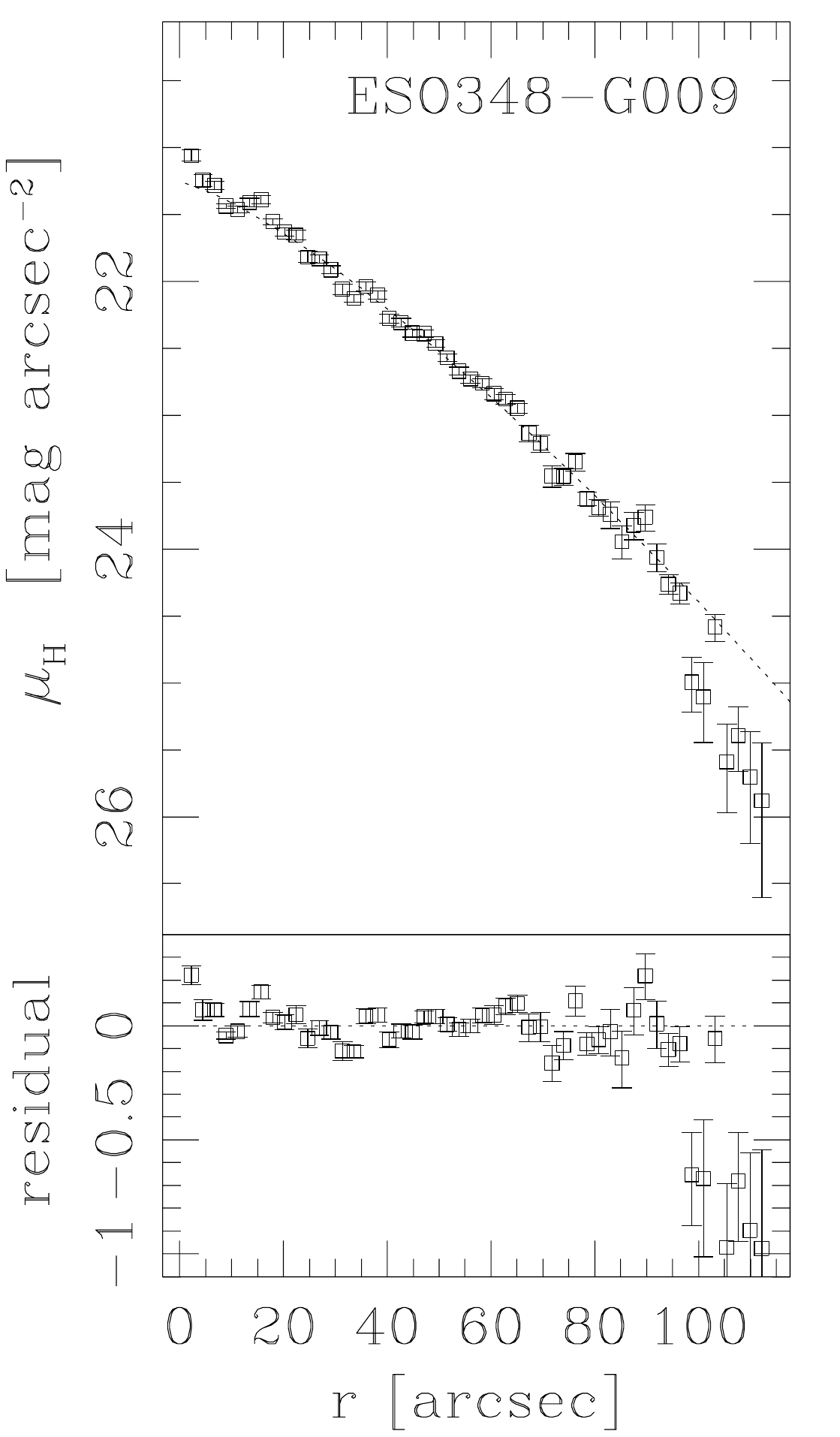}}} &
  \mbox{\scalebox{0.24}{\includegraphics{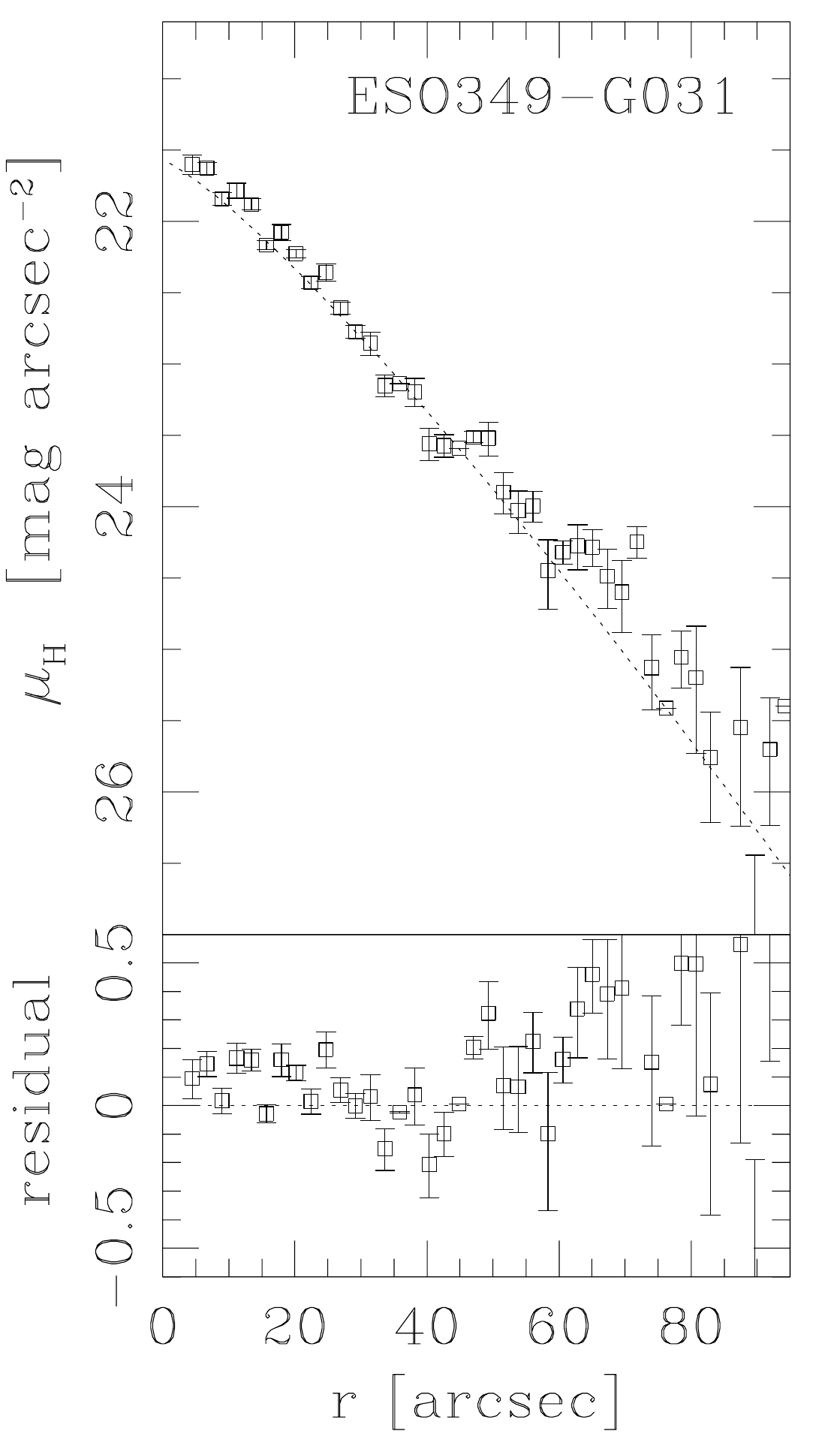}}} &
  \mbox{\scalebox{0.24}{\includegraphics{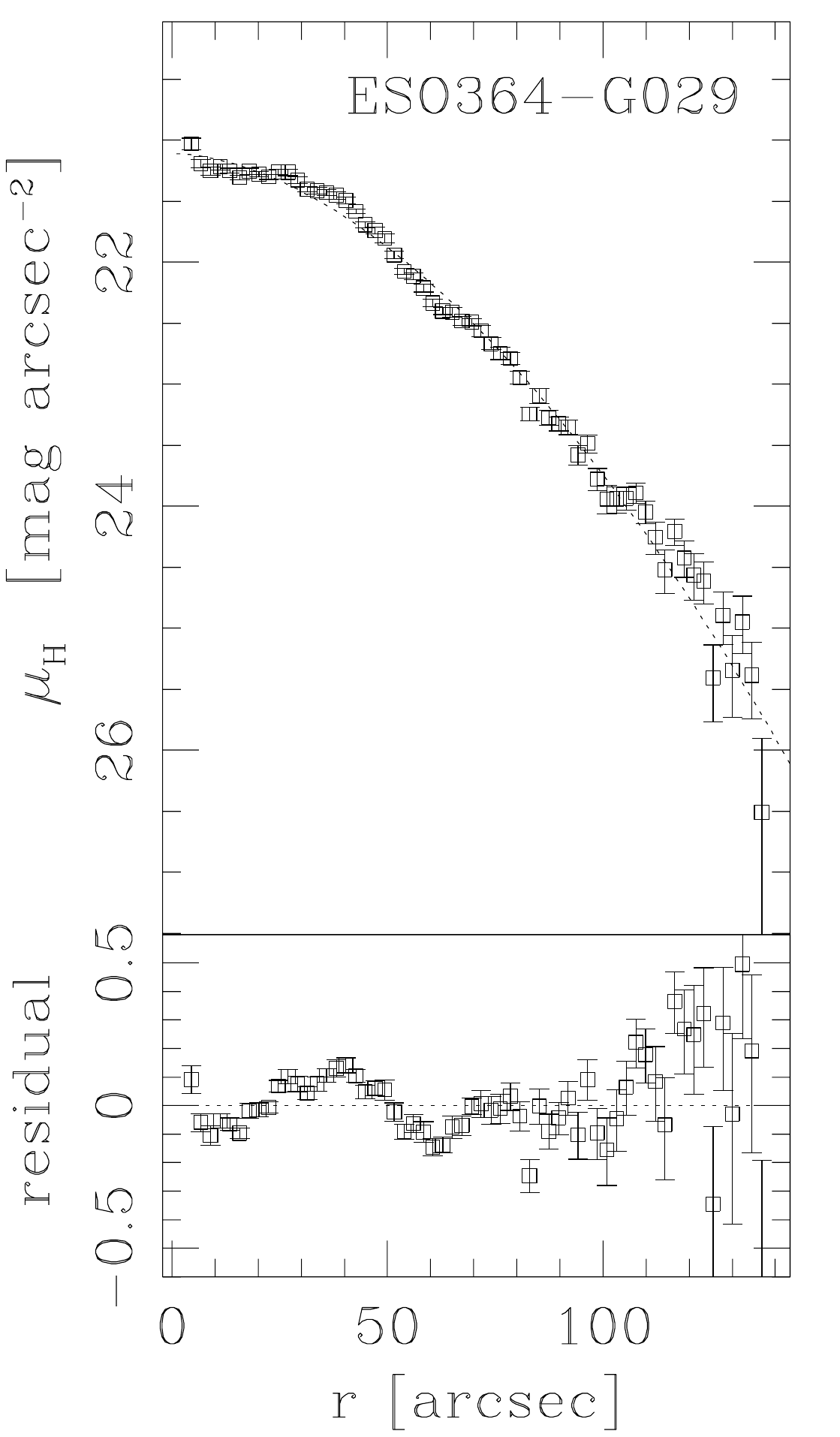}}} &
  \mbox{\scalebox{0.24}{\includegraphics{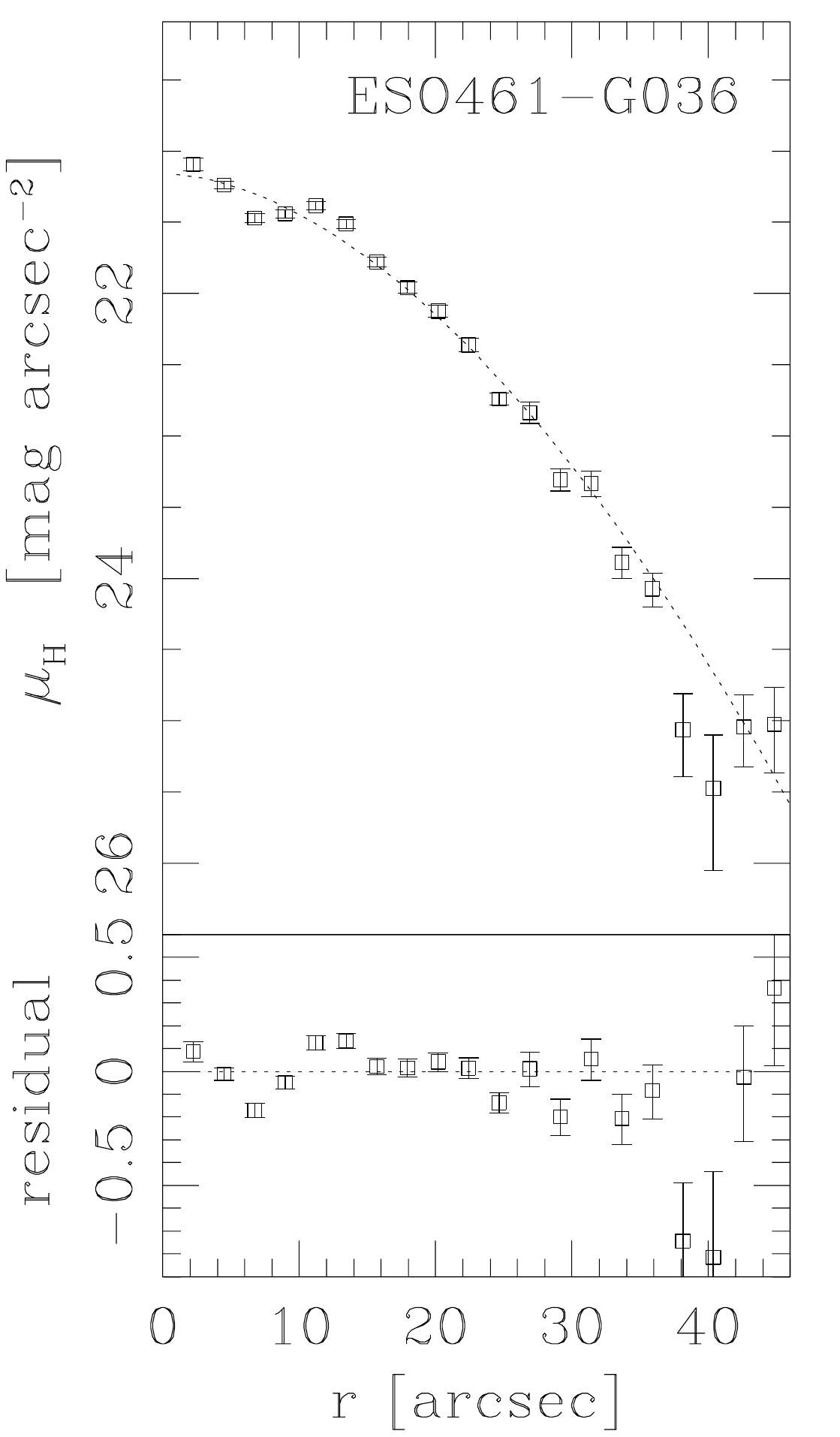}}}  \\
\end{tabular}
\caption{$H$-band surface brightness profiles for all program galaxies except AM0521-343 (see text). The best fitting  S\'ersic profile is shown as solid line together with the residuals.}
 \label{figsbprof1}
\end{figure*}

\begin{figure*} 
\centering
\begin{tabular}{ccccc}
  \mbox{\scalebox{0.24}{\includegraphics{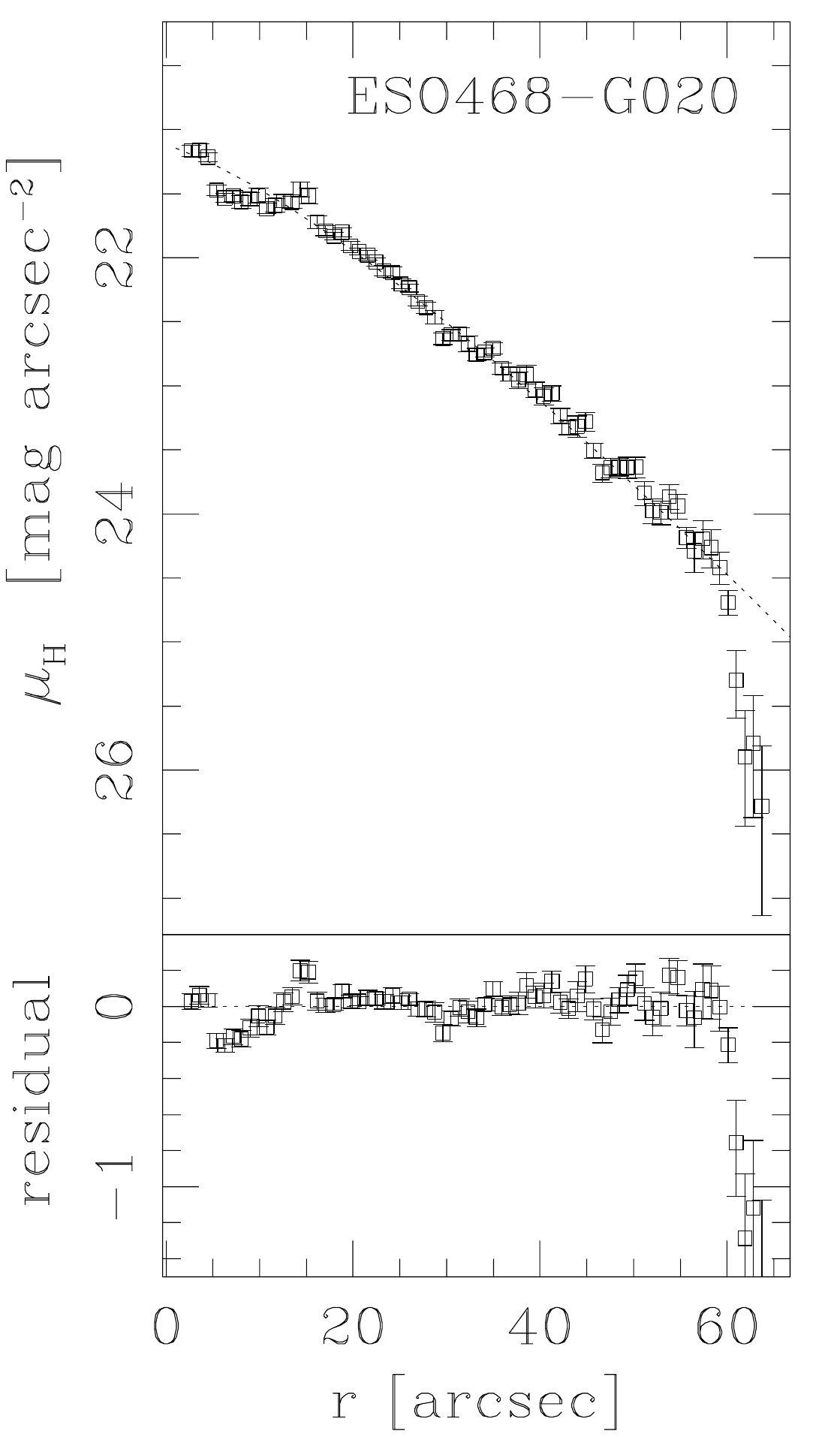}}} &
    \mbox{\scalebox{0.24}{\includegraphics{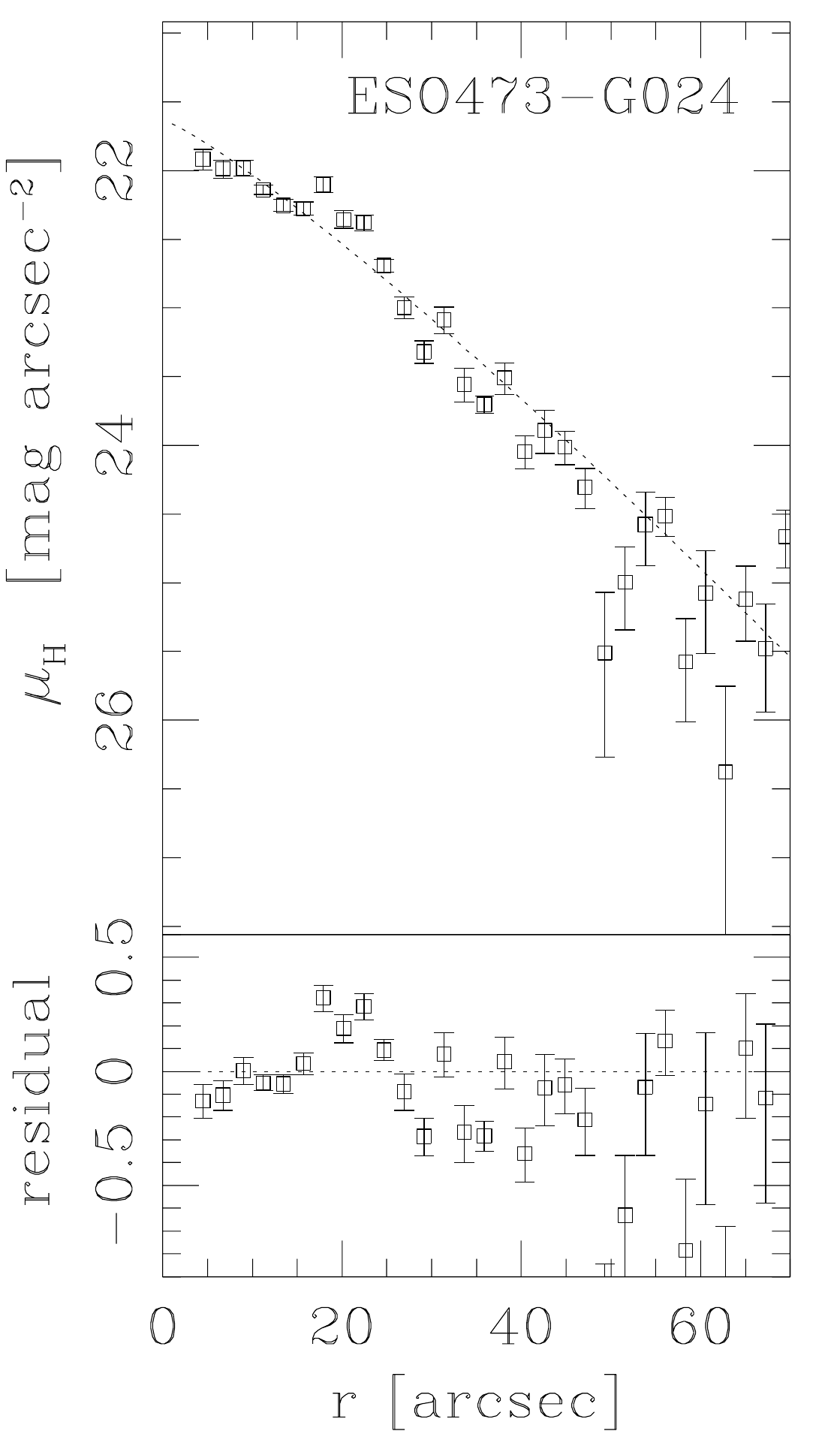}}} &
  \mbox{\scalebox{0.24}{\includegraphics{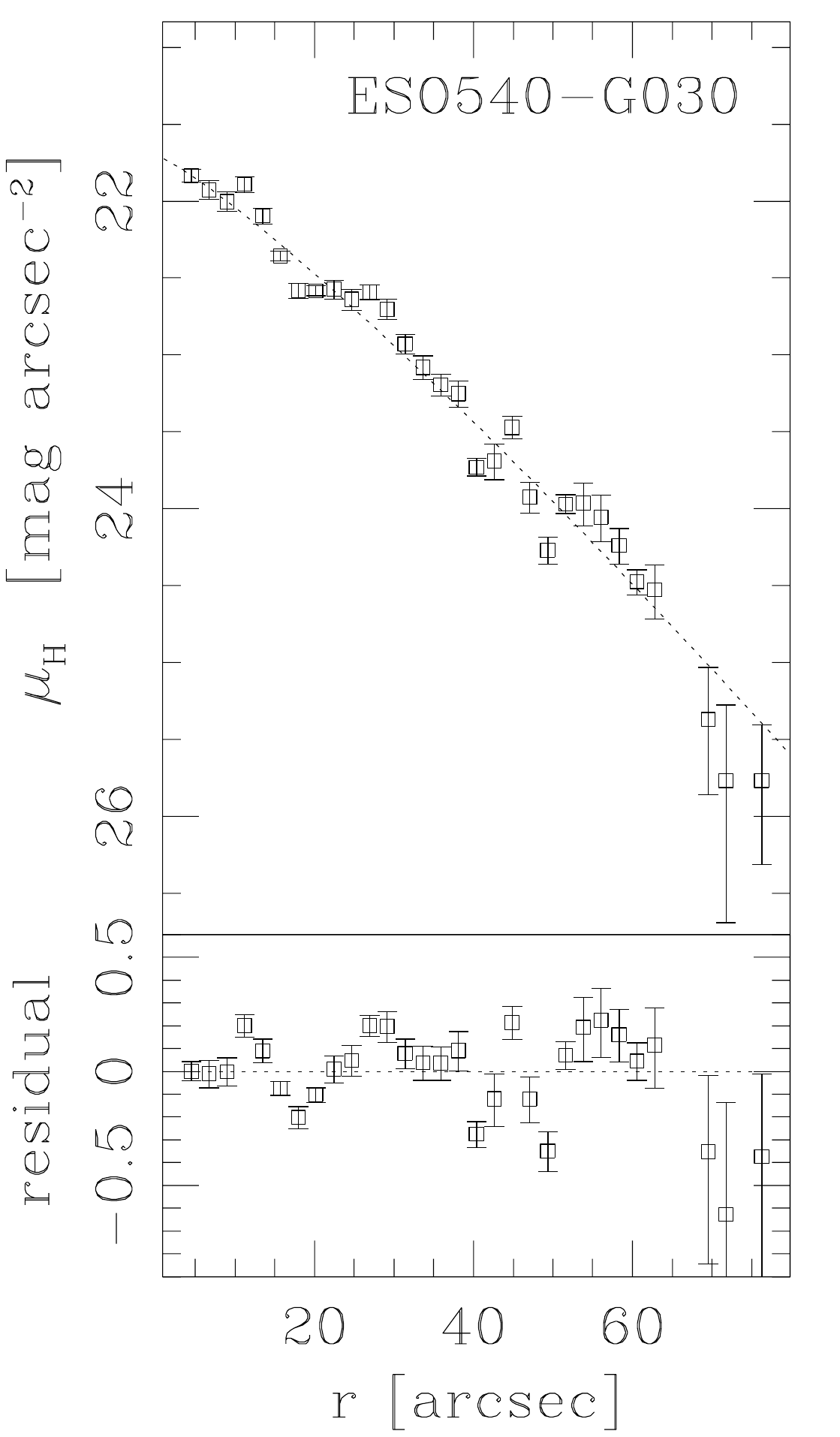}}}  &
  \mbox{\scalebox{0.24}{\includegraphics{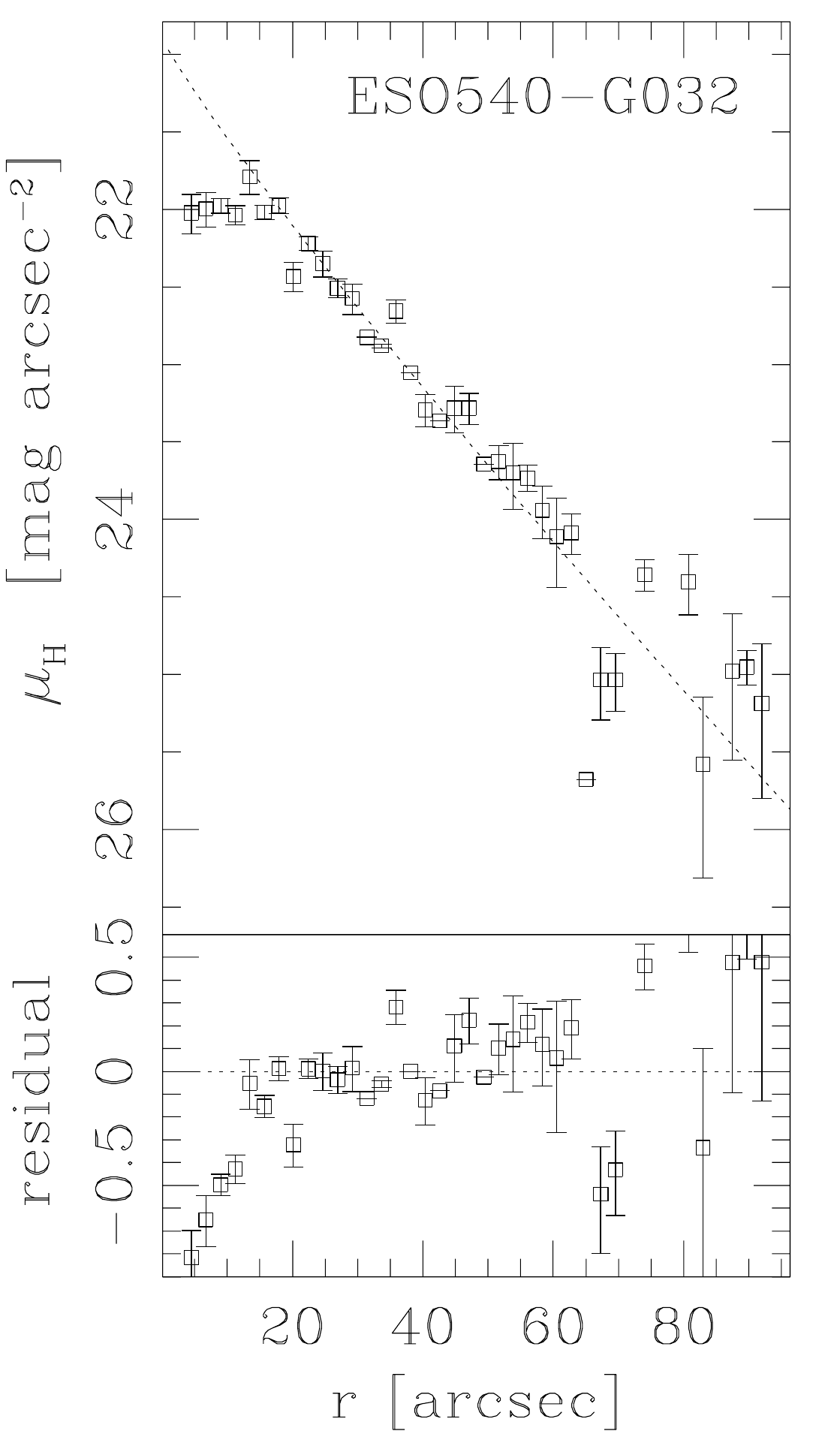}}}  &
  \mbox{\scalebox{0.24}{\includegraphics{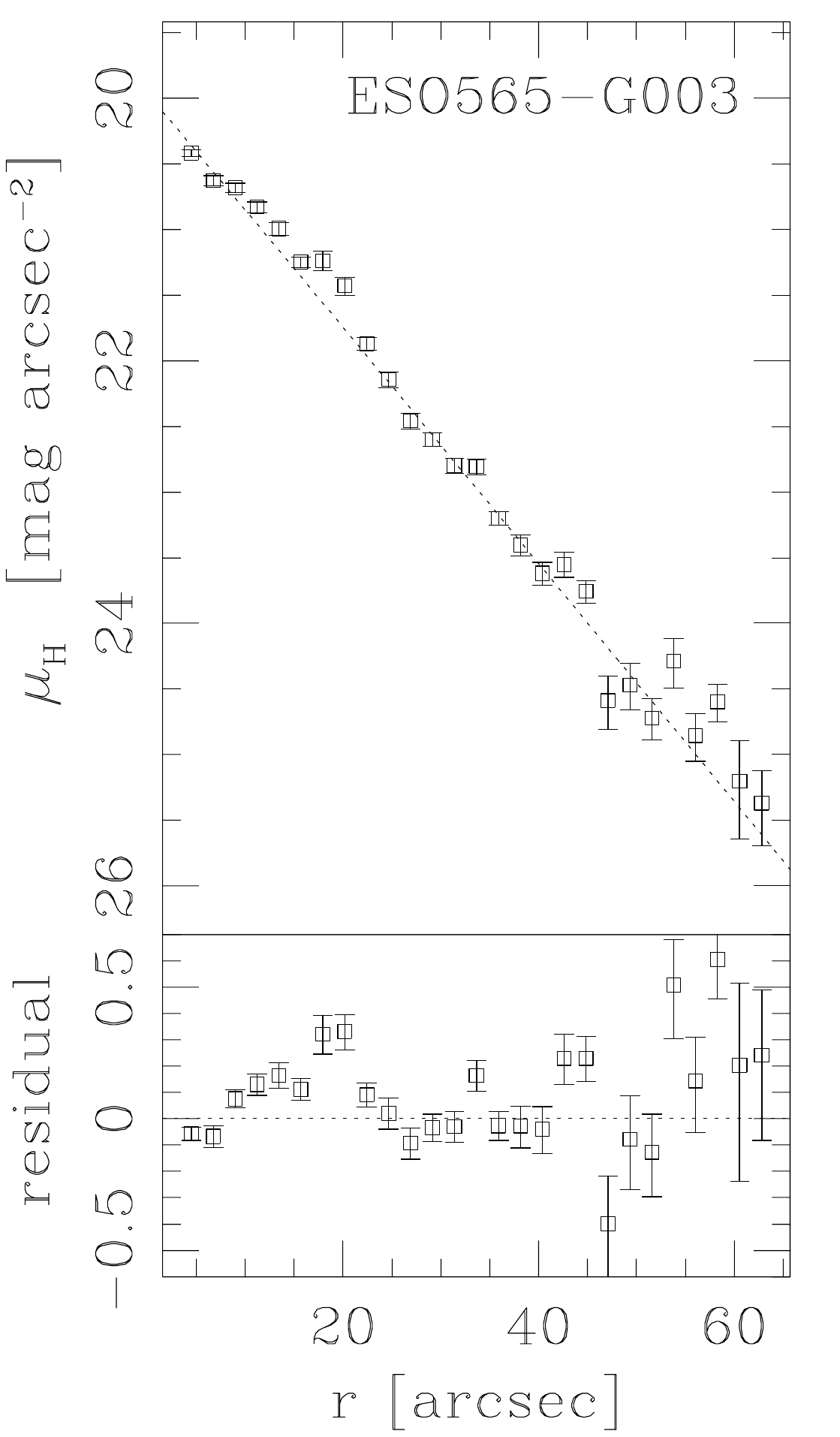}}}  \\
  \mbox{\scalebox{0.24}{\includegraphics{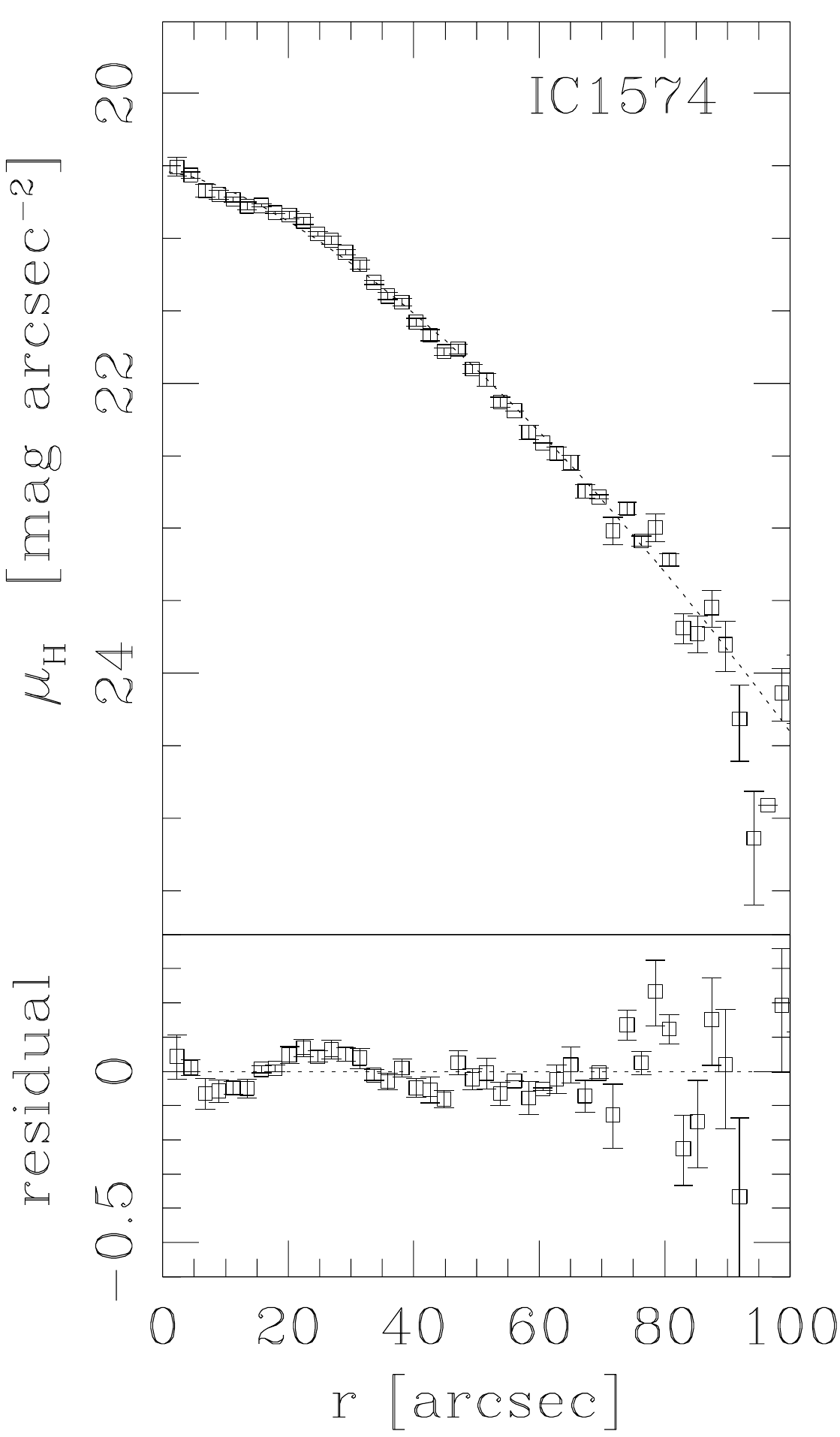}}} &
  \mbox{\scalebox{0.24}{\includegraphics{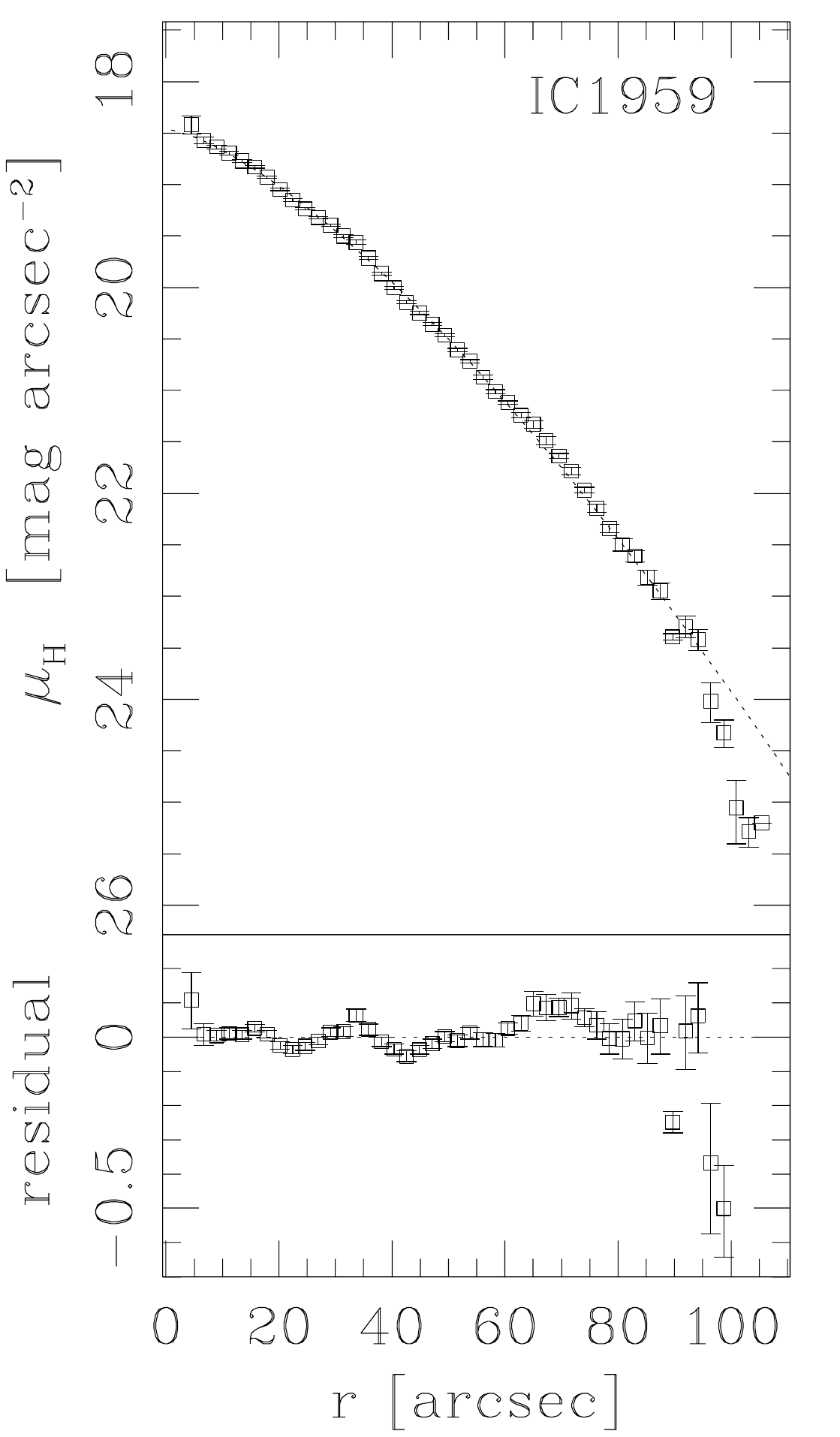}}} &
  \mbox{\scalebox{0.24}{\includegraphics{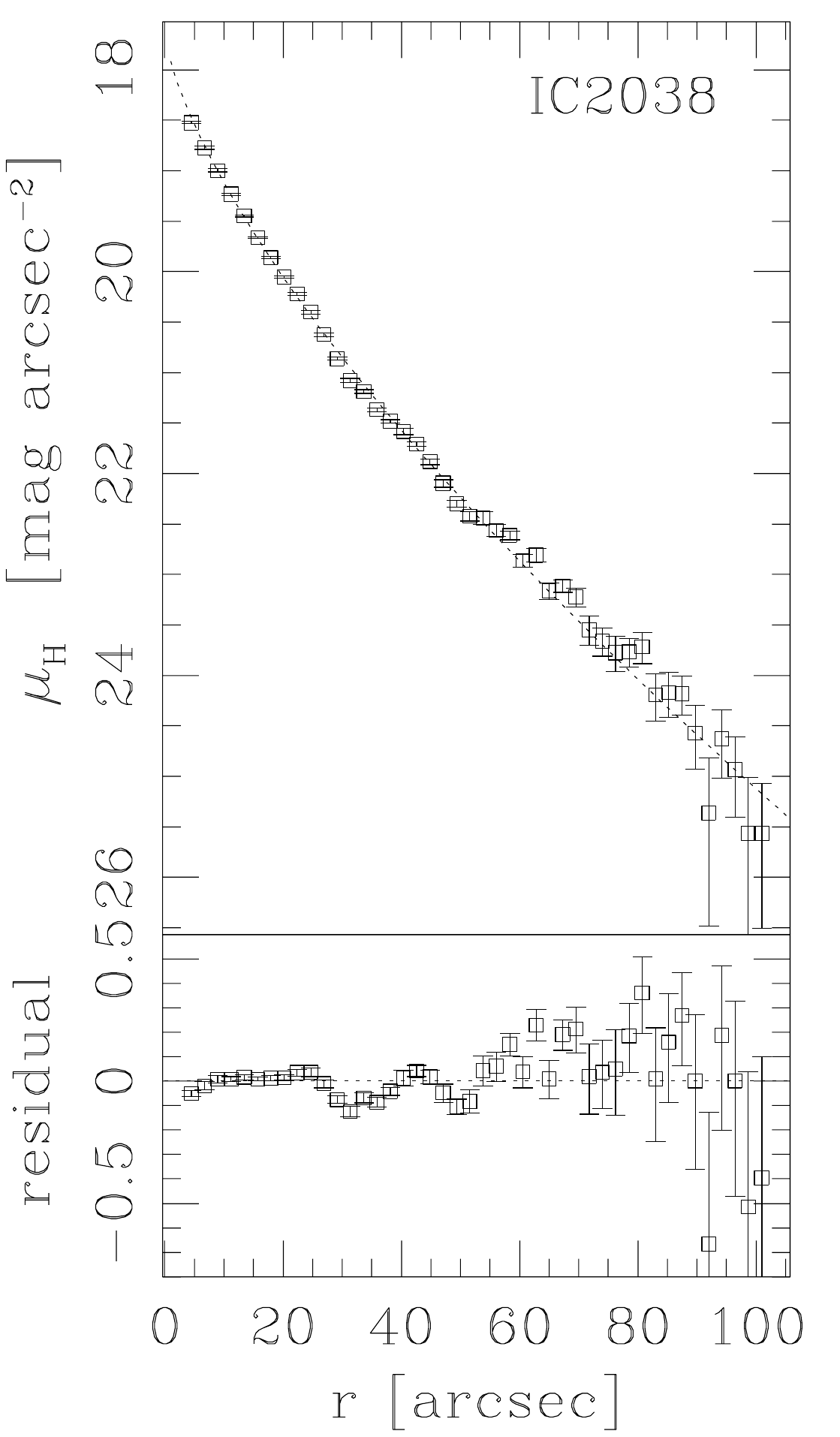}}} &
  \mbox{\scalebox{0.24}{\includegraphics{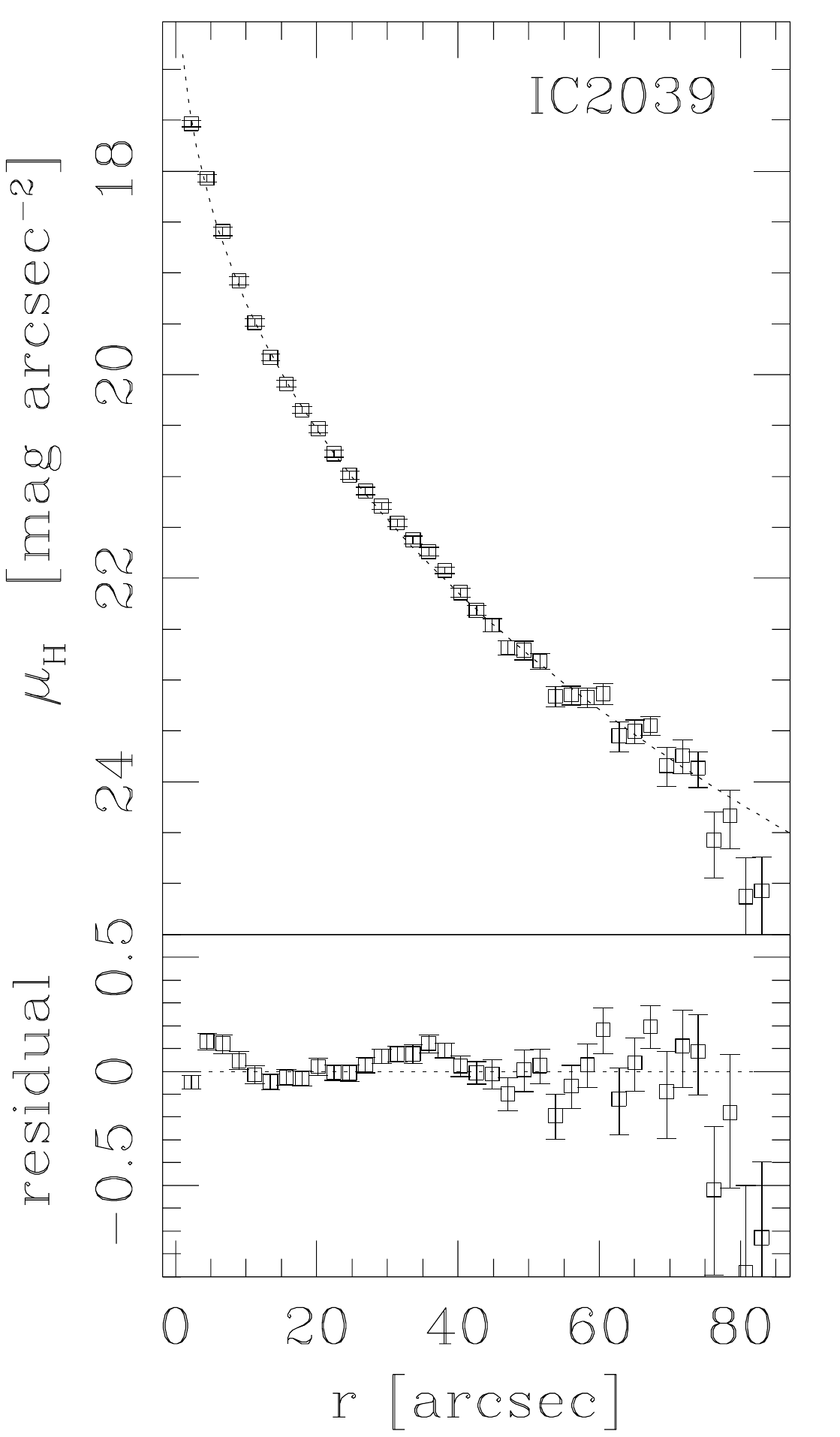}}} &
  \mbox{\scalebox{0.24}{\includegraphics{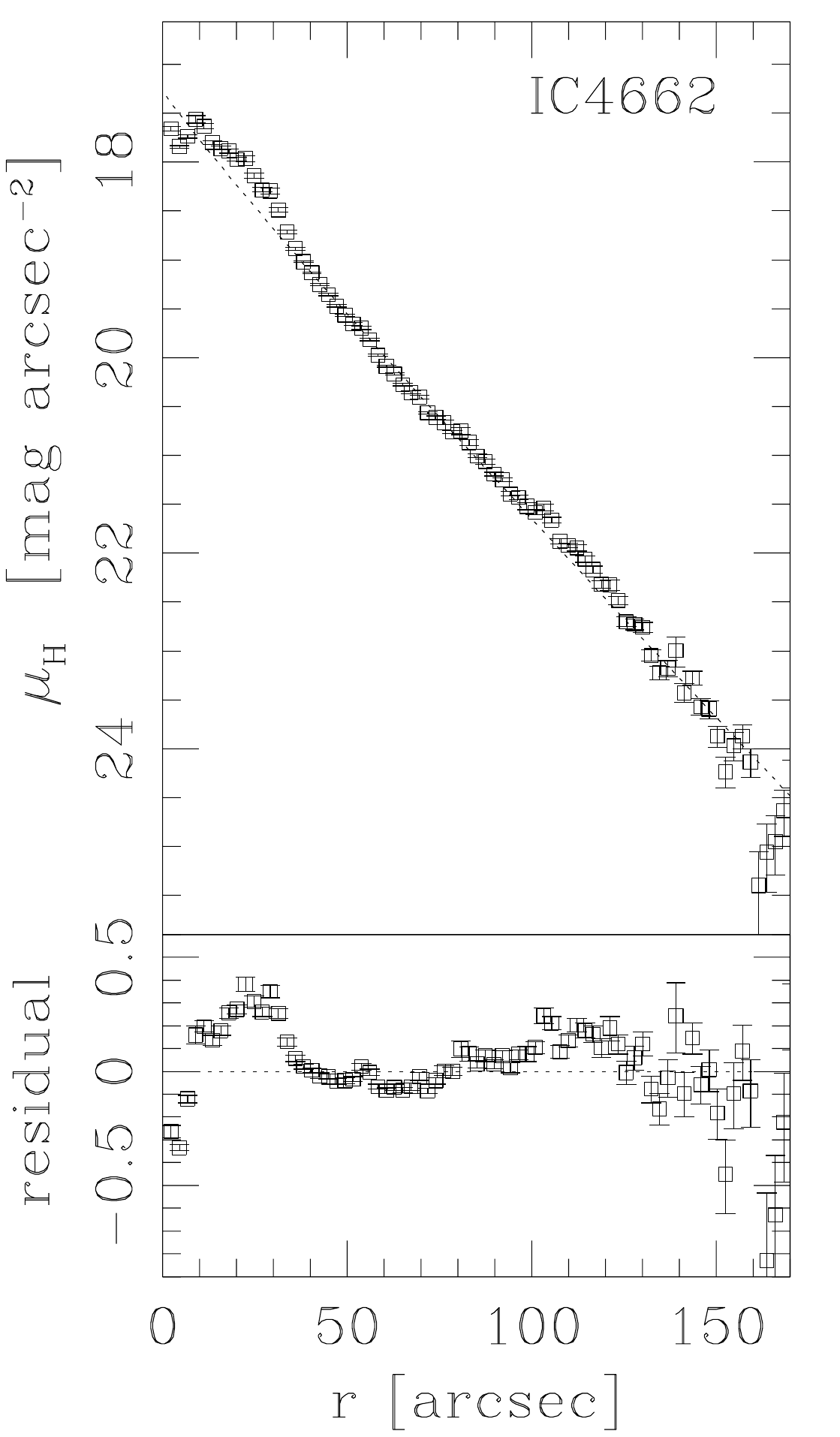}}}  \\
  \mbox{\scalebox{0.24}{\includegraphics{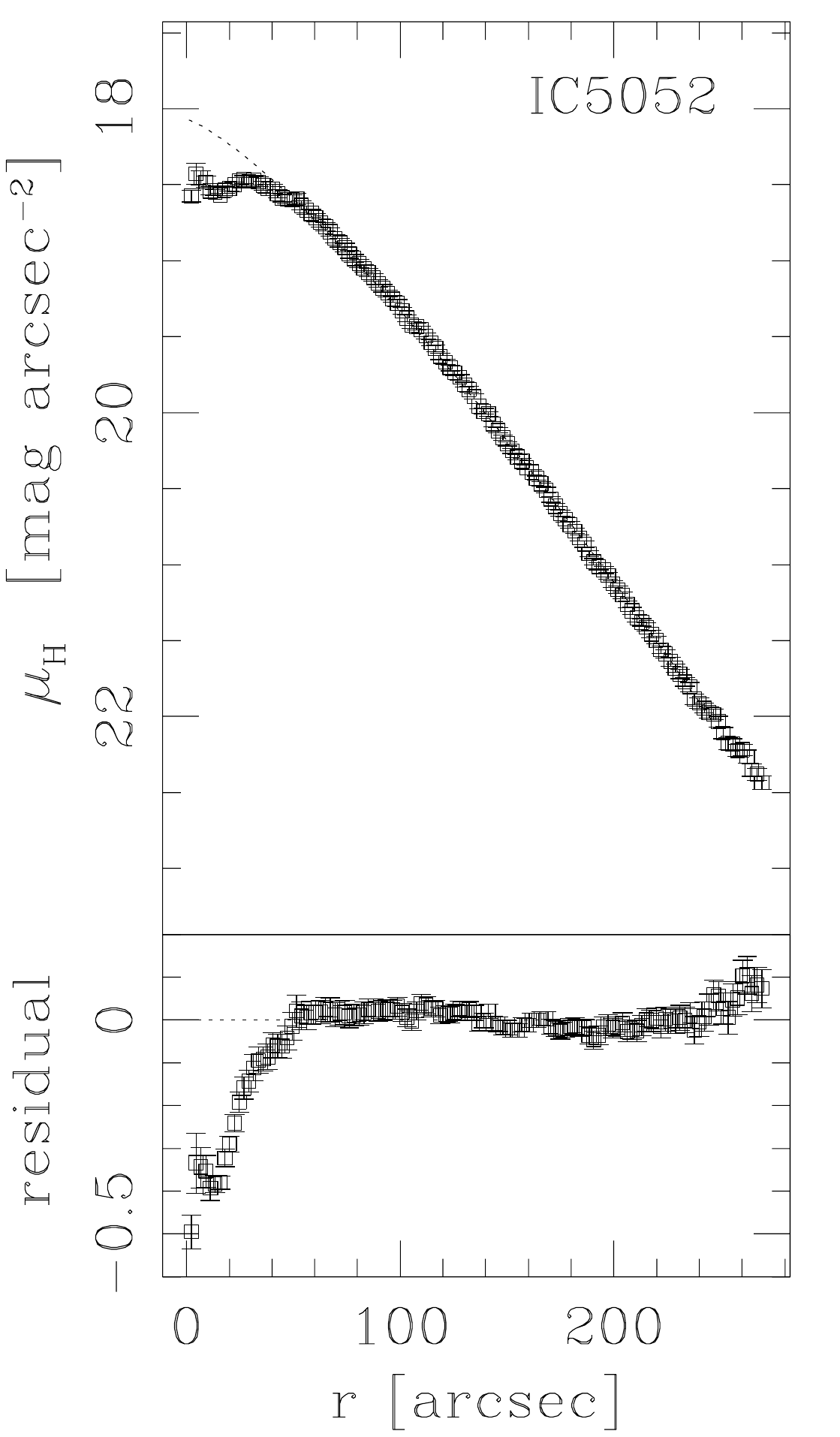}}} &
  \mbox{\scalebox{0.24}{\includegraphics{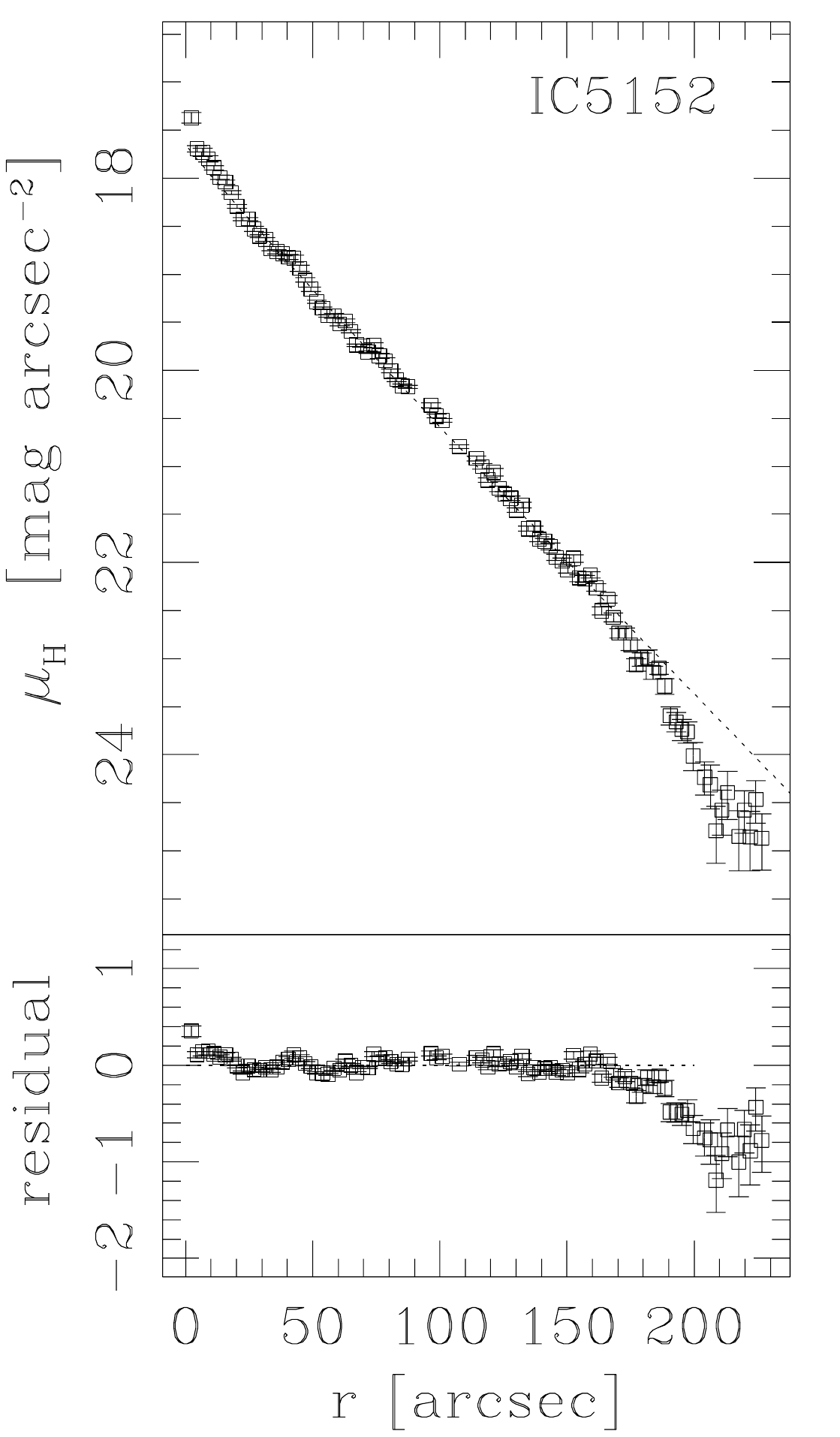}}} &  
  \mbox{\scalebox{0.24}{\includegraphics{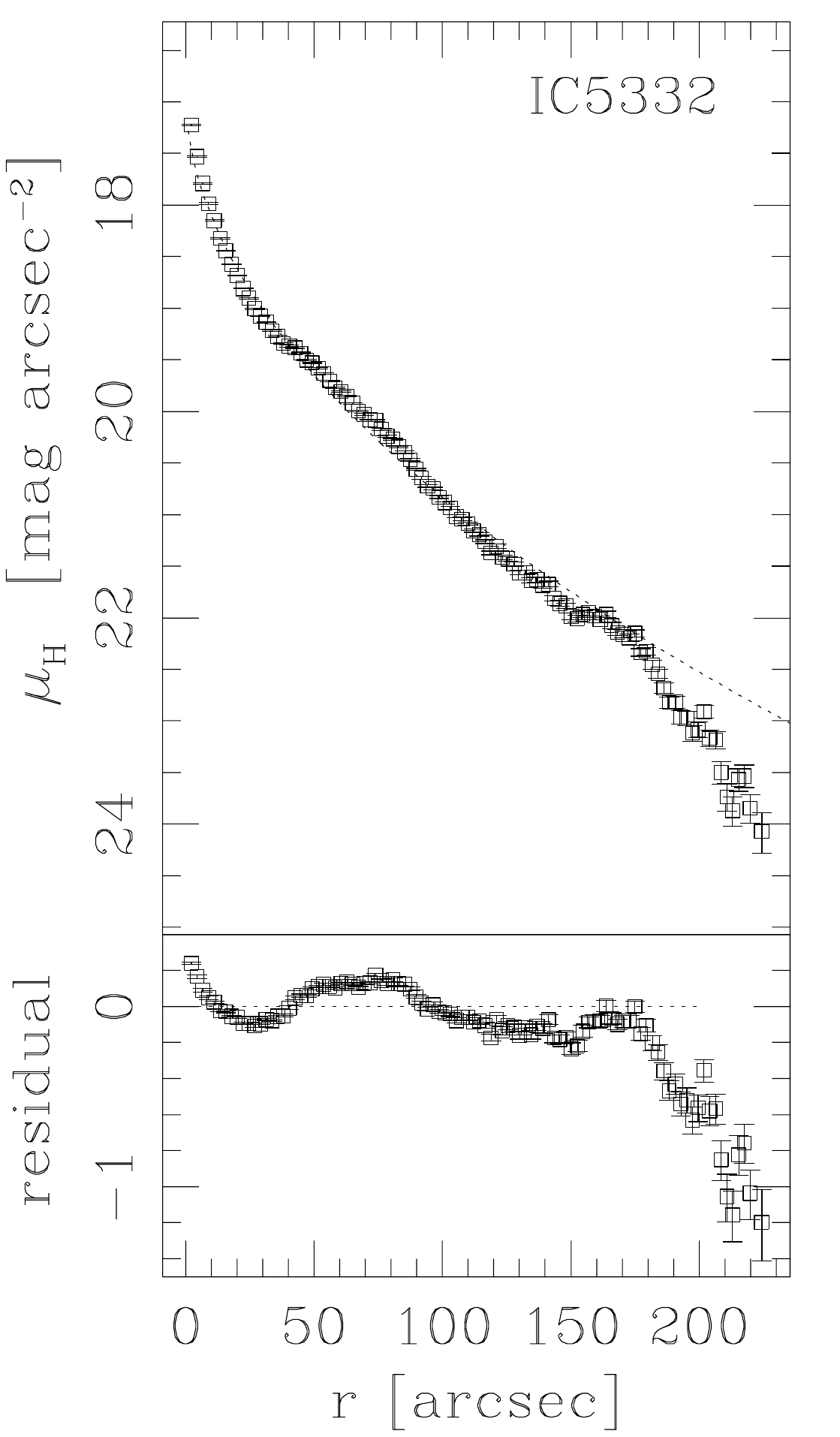}}} &
  \mbox{\scalebox{0.24}{\includegraphics{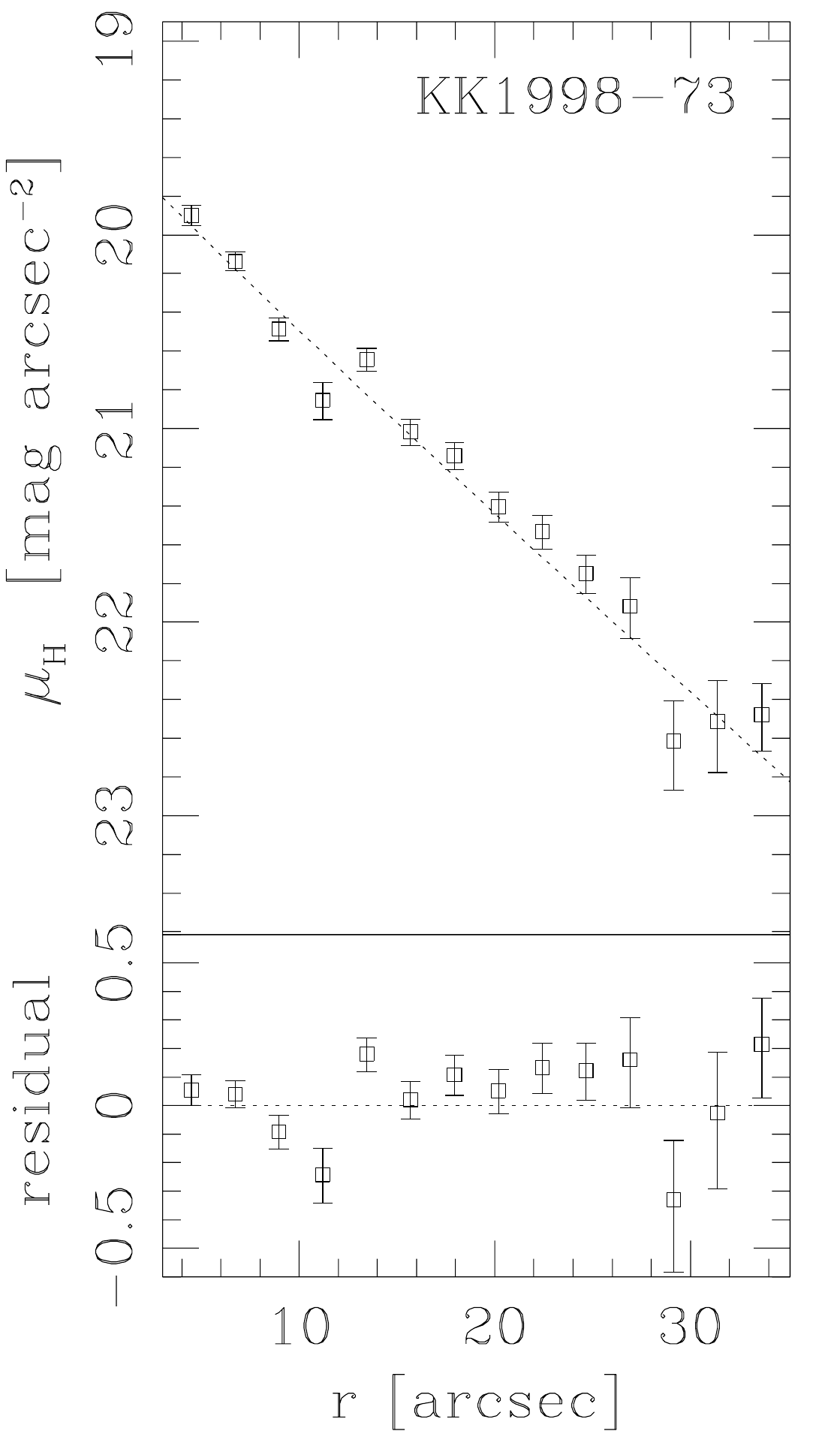}}} &
  \mbox{\scalebox{0.24}{\includegraphics{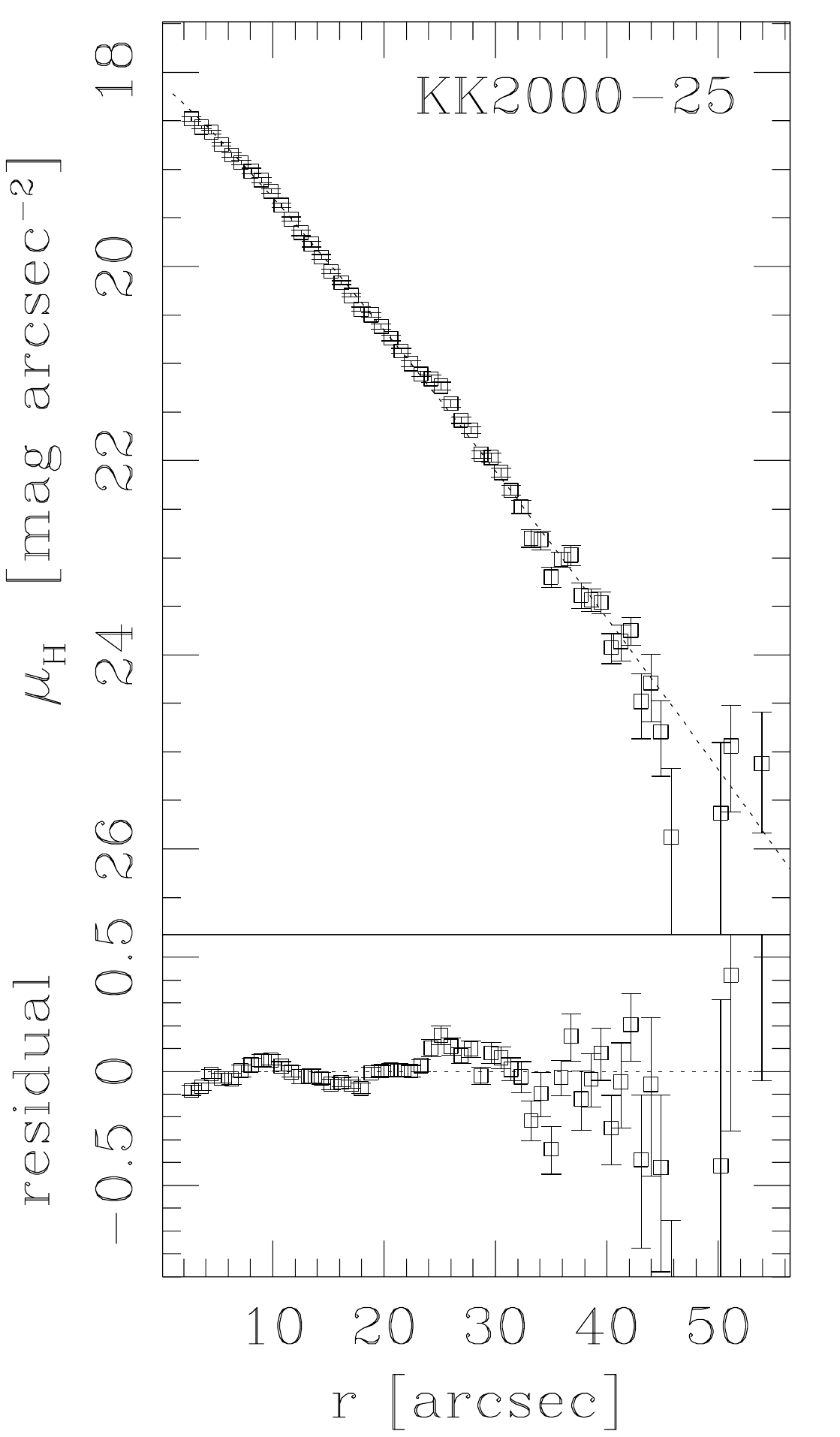}}} \\
  \mbox{\scalebox{0.24}{\includegraphics{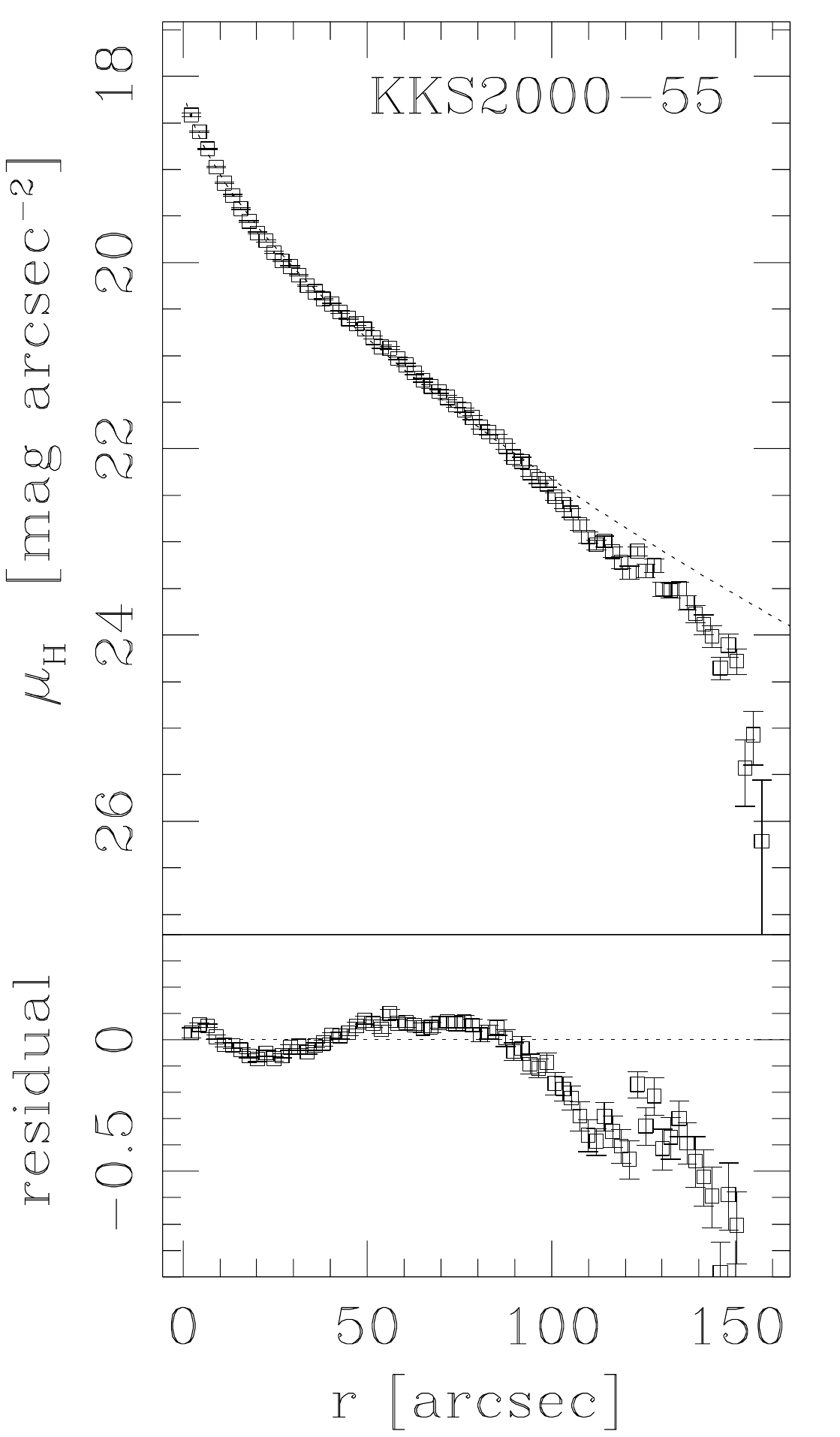}}} &
  \mbox{\scalebox{0.24}{\includegraphics{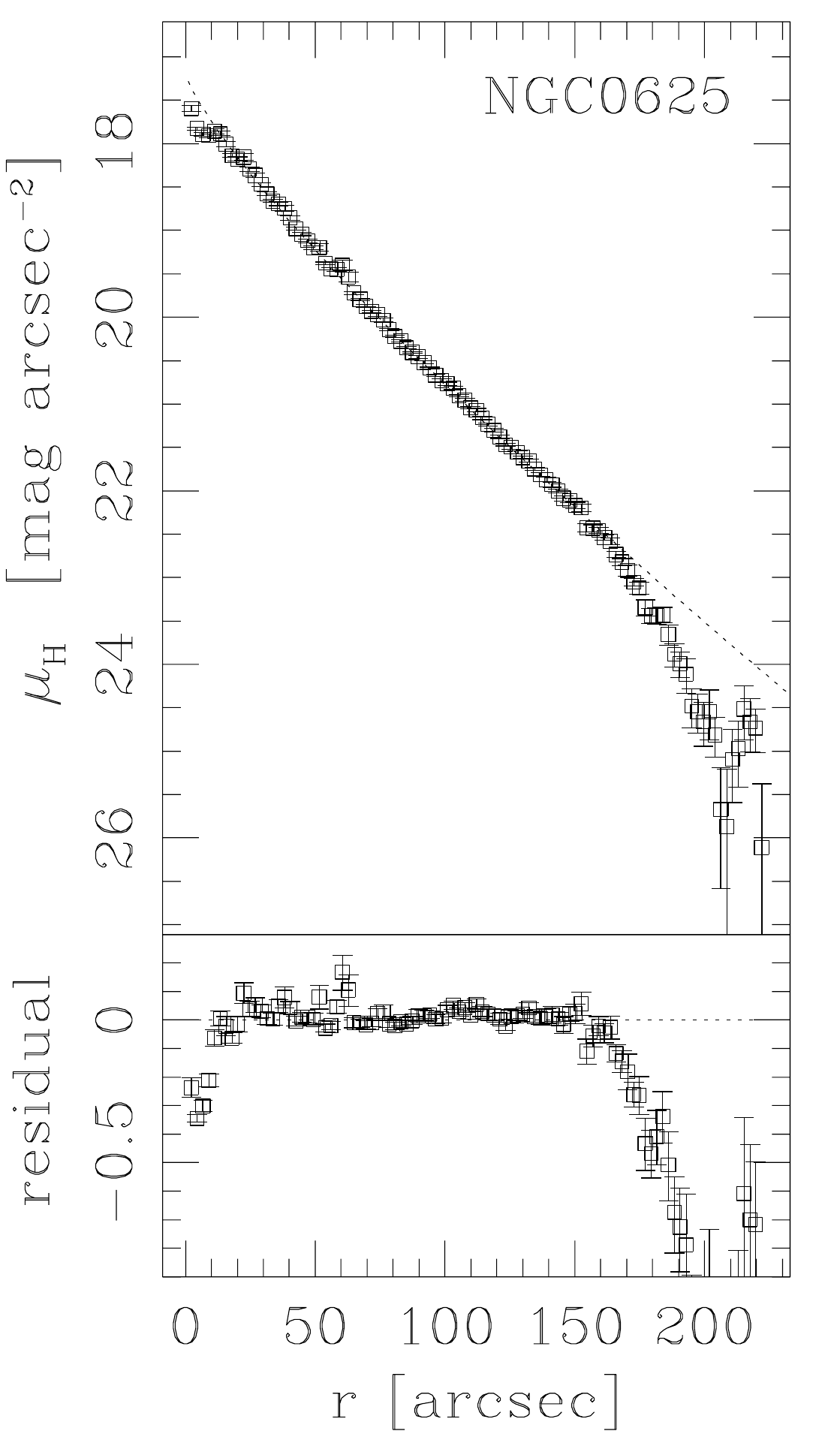}}} &
  \mbox{\scalebox{0.24}{\includegraphics{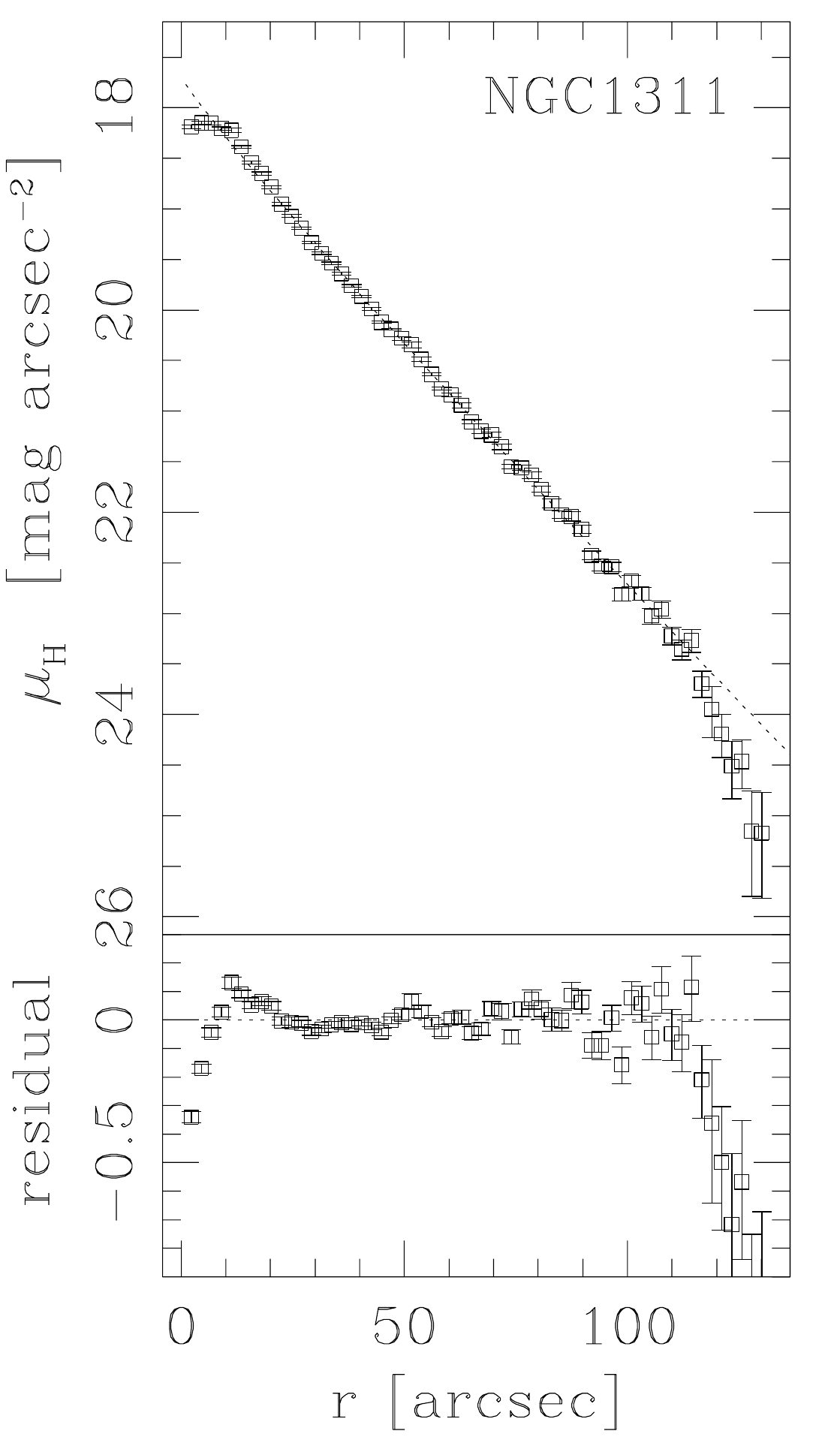}}} &
  \mbox{\scalebox{0.24}{\includegraphics{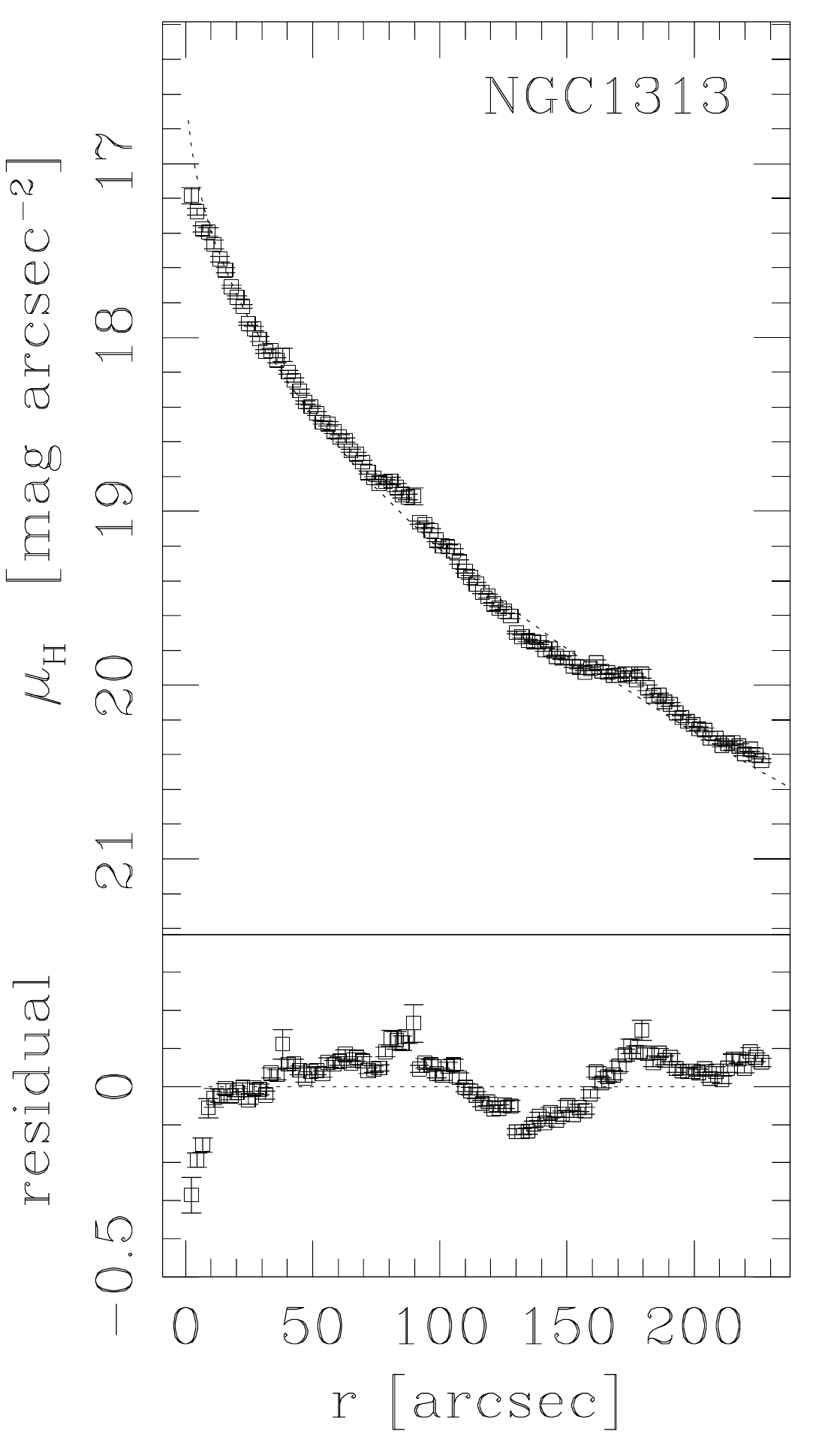}}} &
  \mbox{\scalebox{0.24}{\includegraphics{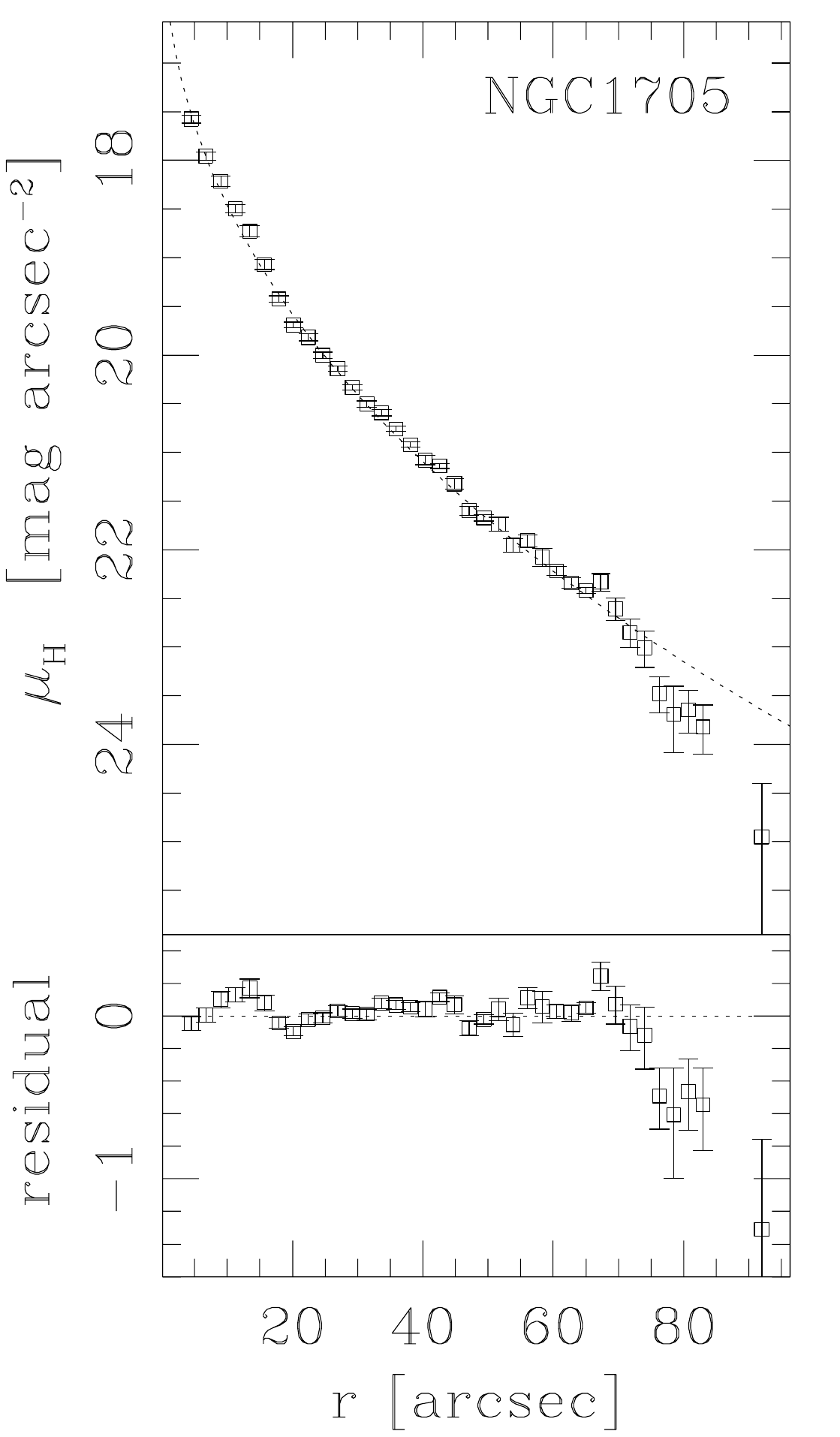}}}  \\
\end{tabular}
\caption{$H$-band surface brightness profiles for all program galaxies except AM0521-343 (see text). The best fitting  S\'ersic profile is shown as solid line together with the residuals.}
\label{figsbprof2}
\end{figure*}

\begin{figure*} 
\centering
\begin{tabular}{ccccc}
  \mbox{\scalebox{0.24}{\includegraphics{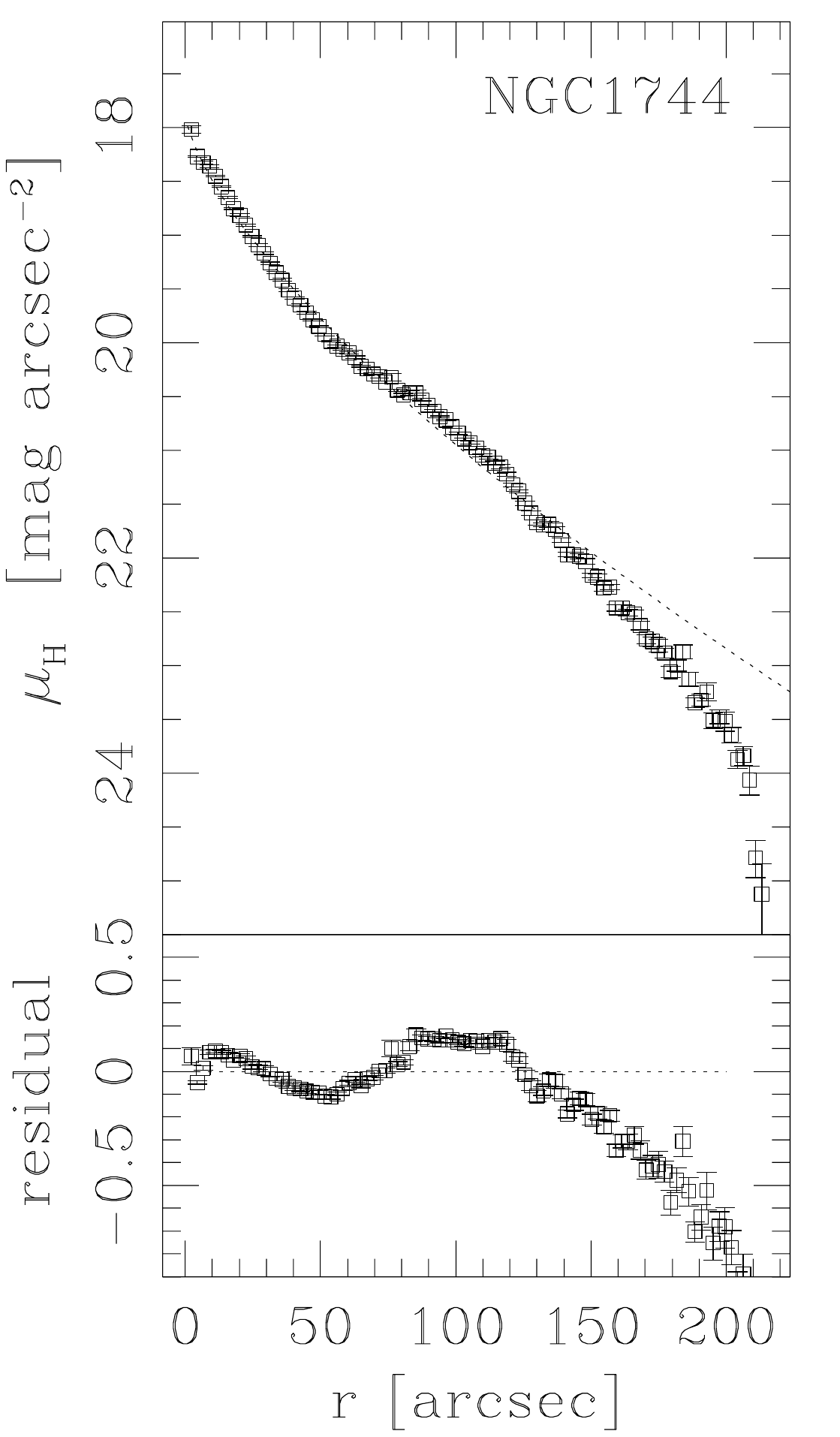}}} &
    \mbox{\scalebox{0.24}{\includegraphics{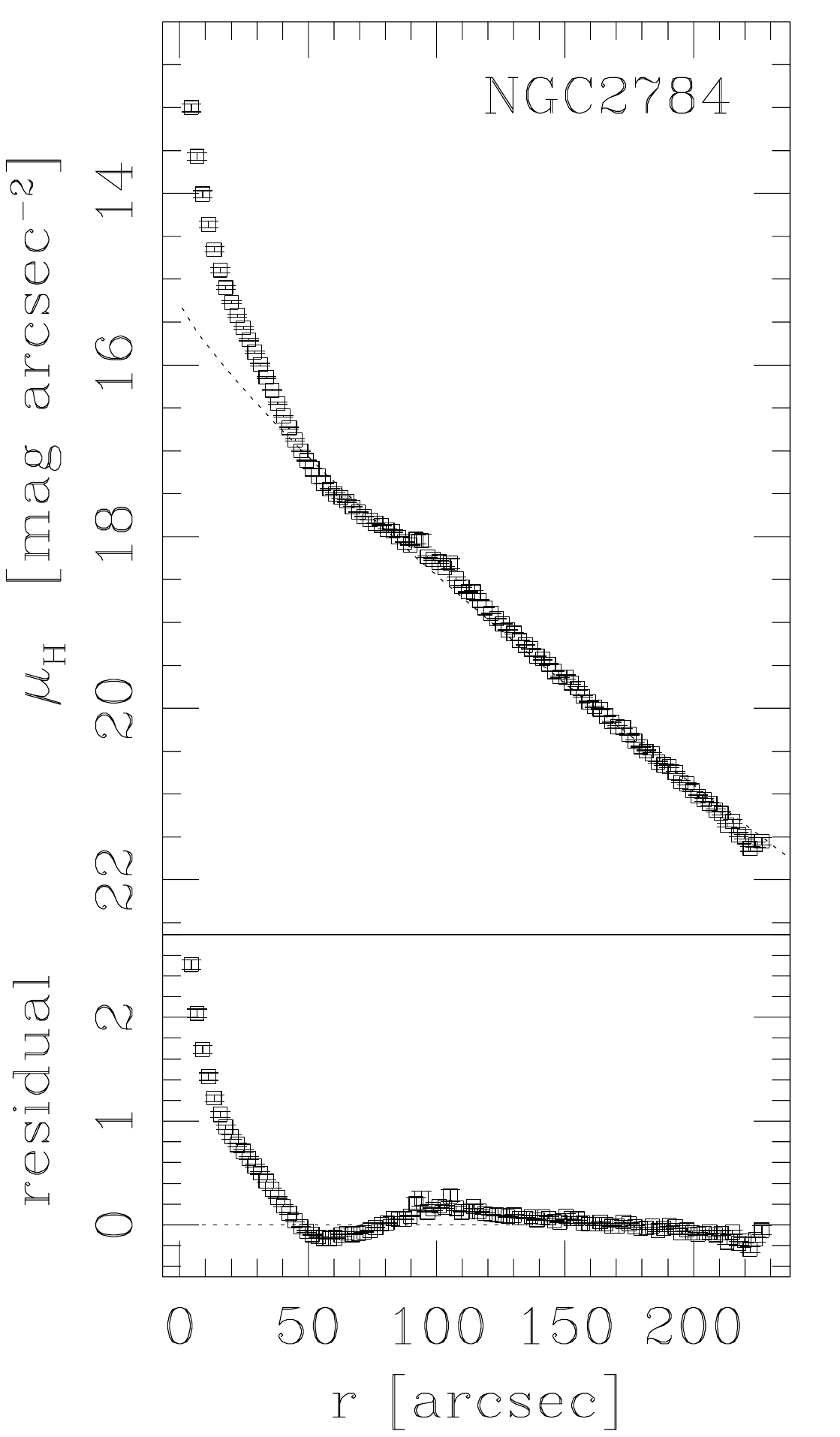}}} &
  \mbox{\scalebox{0.24}{\includegraphics{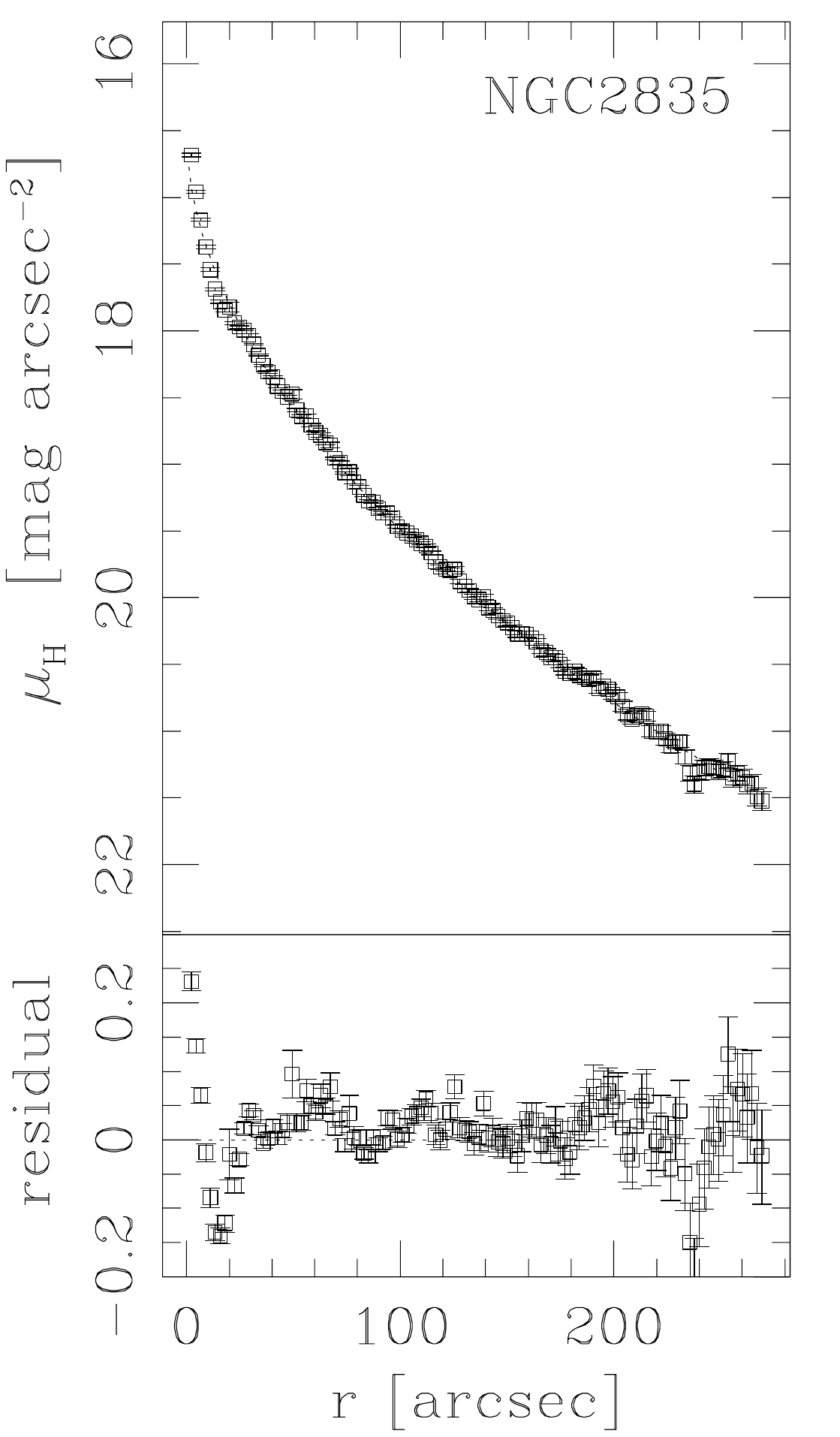}}}  &
  \mbox{\scalebox{0.24}{\includegraphics{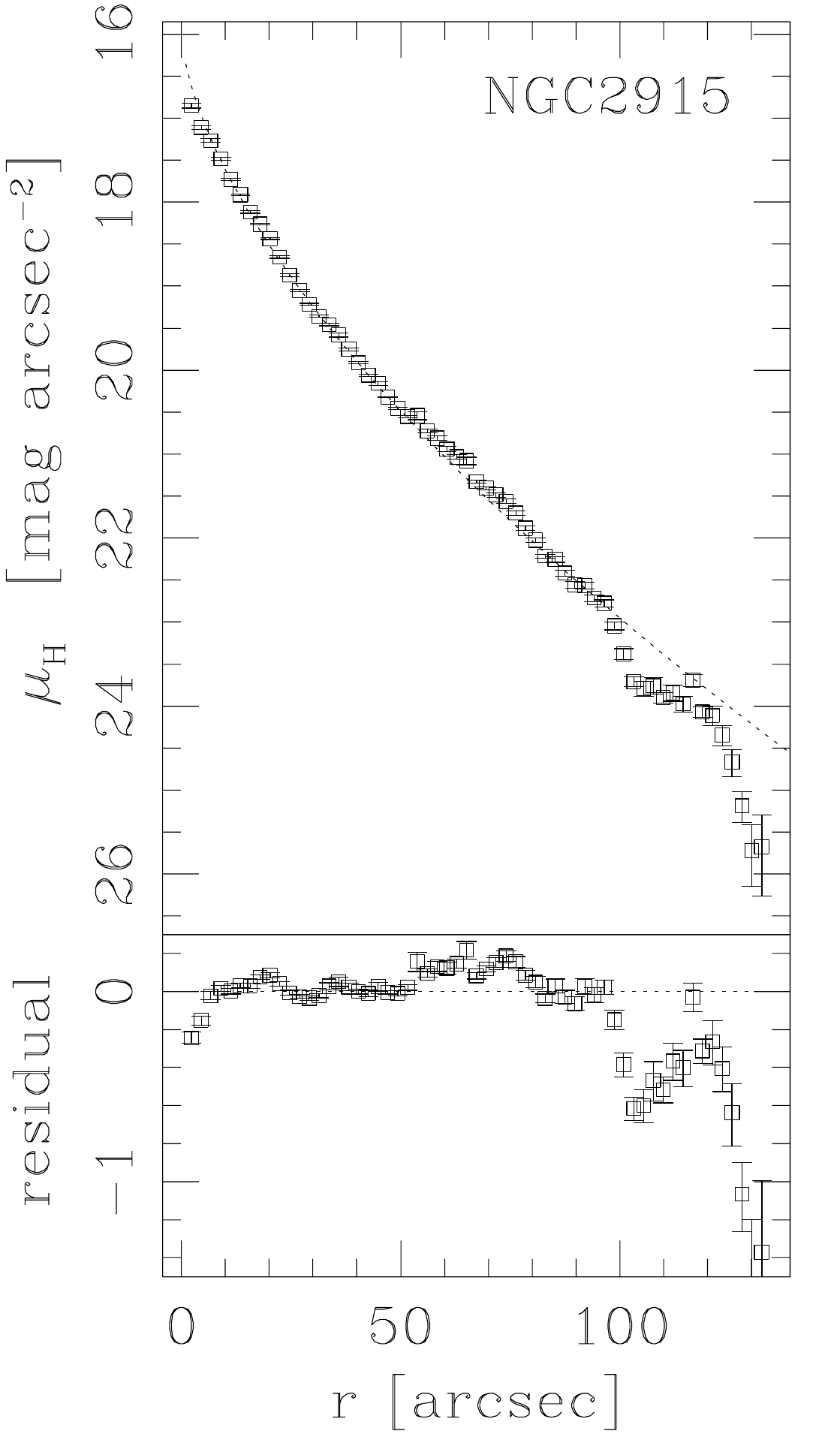}}}  &
  \mbox{\scalebox{0.24}{\includegraphics{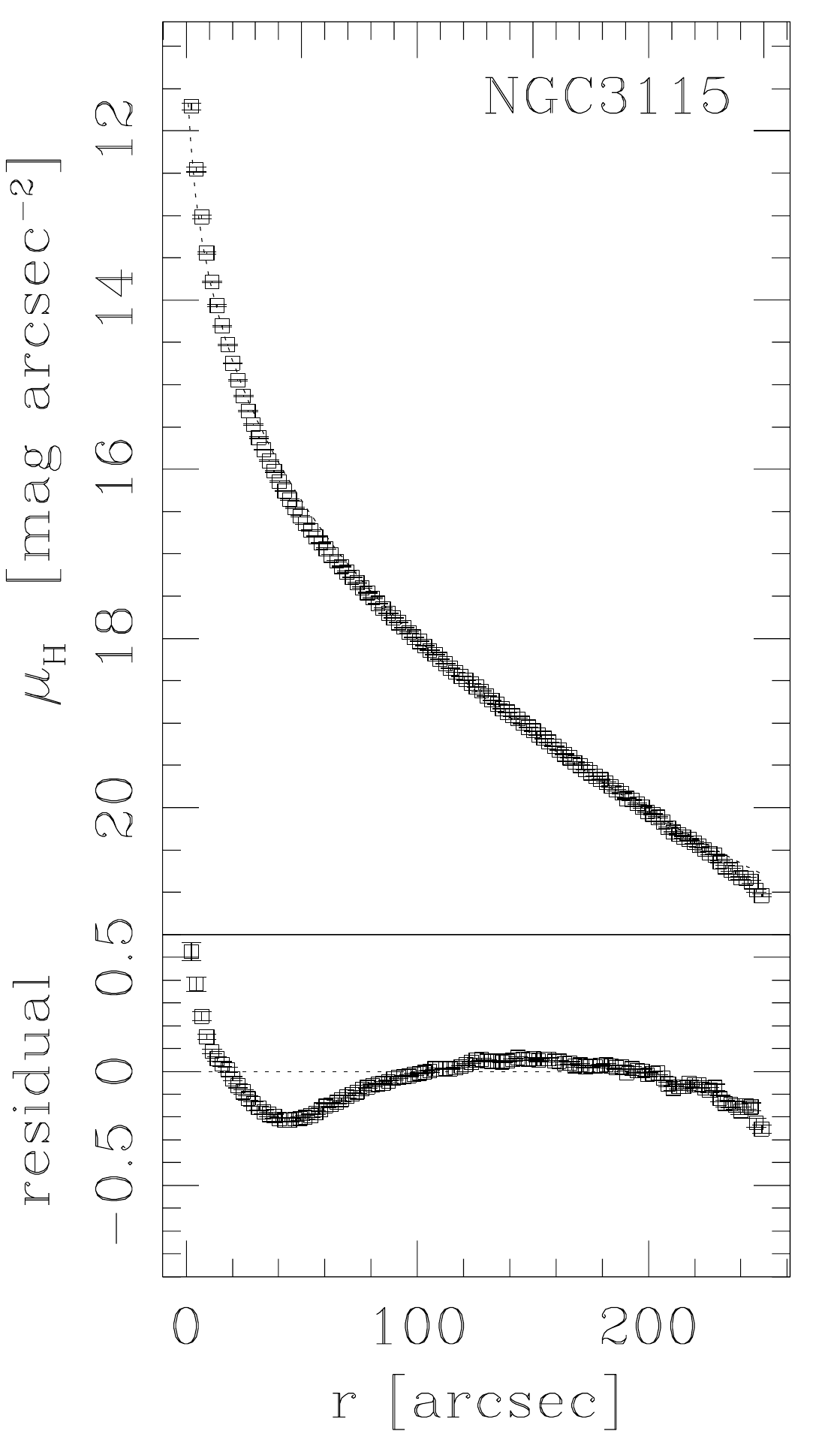}}}  \\
  \mbox{\scalebox{0.24}{\includegraphics{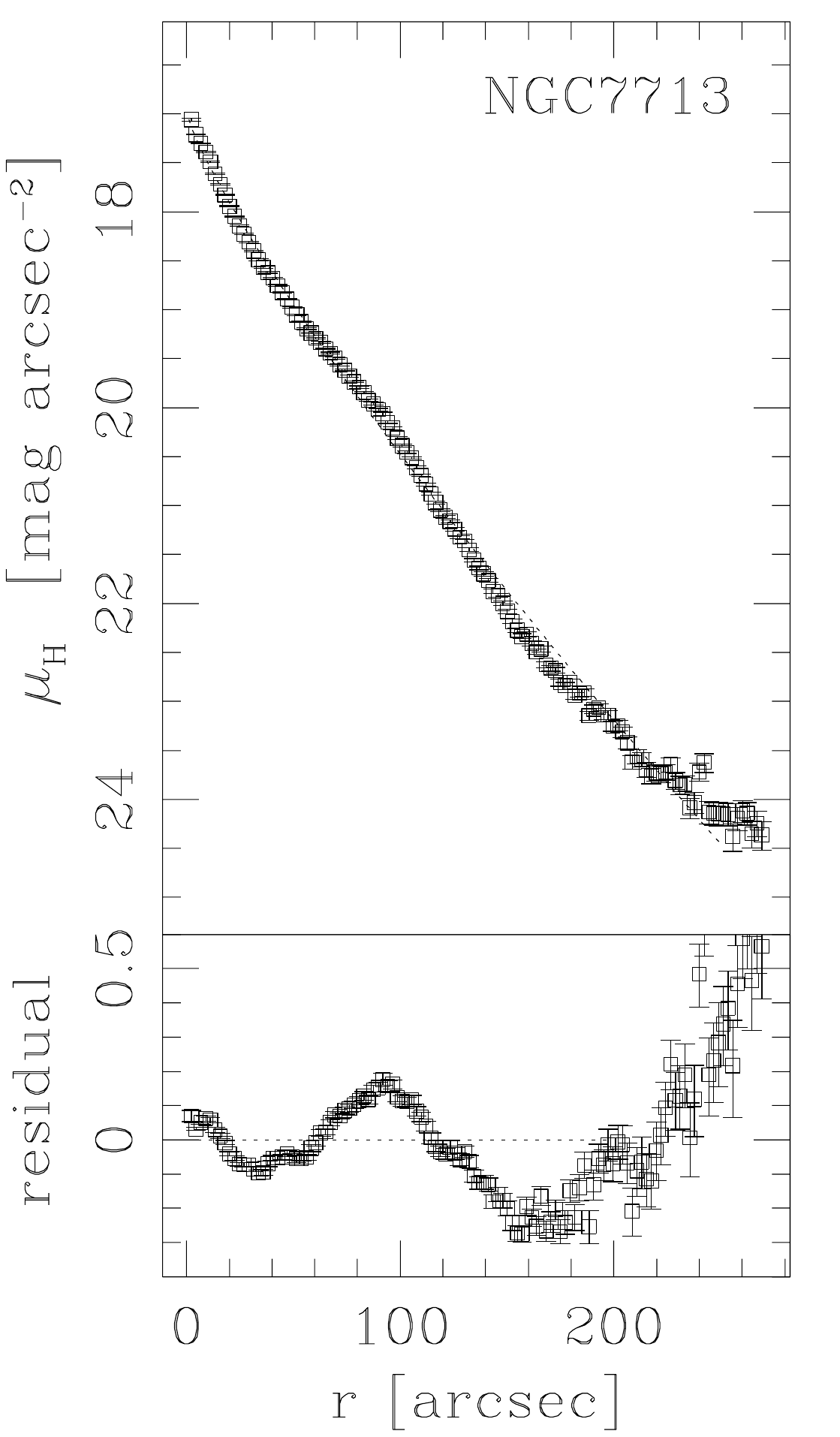}}} &
  \mbox{\scalebox{0.24}{\includegraphics{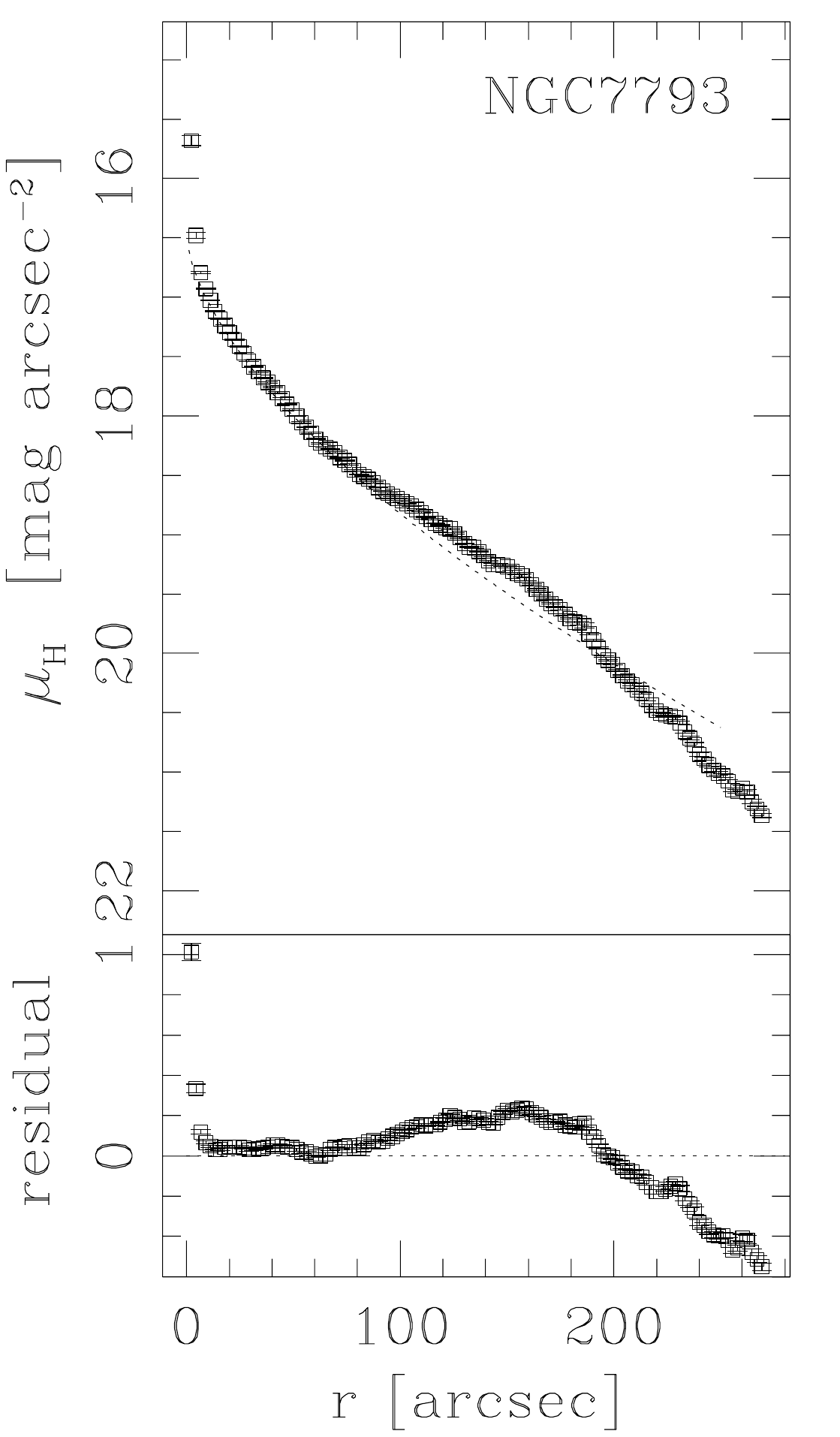}}} &
  \mbox{\scalebox{0.24}{\includegraphics{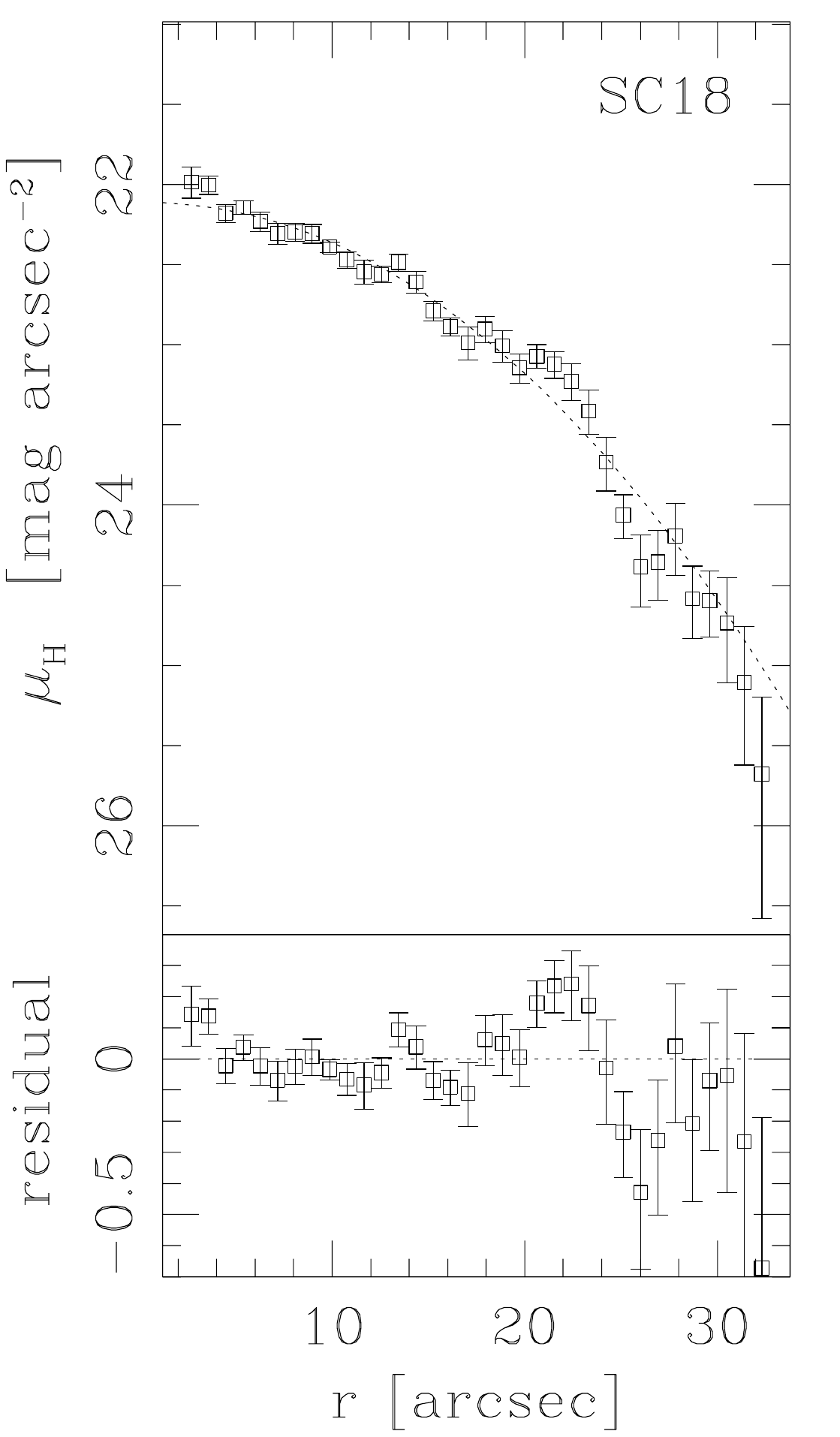}}} &
  \mbox{\scalebox{0.24}{\includegraphics{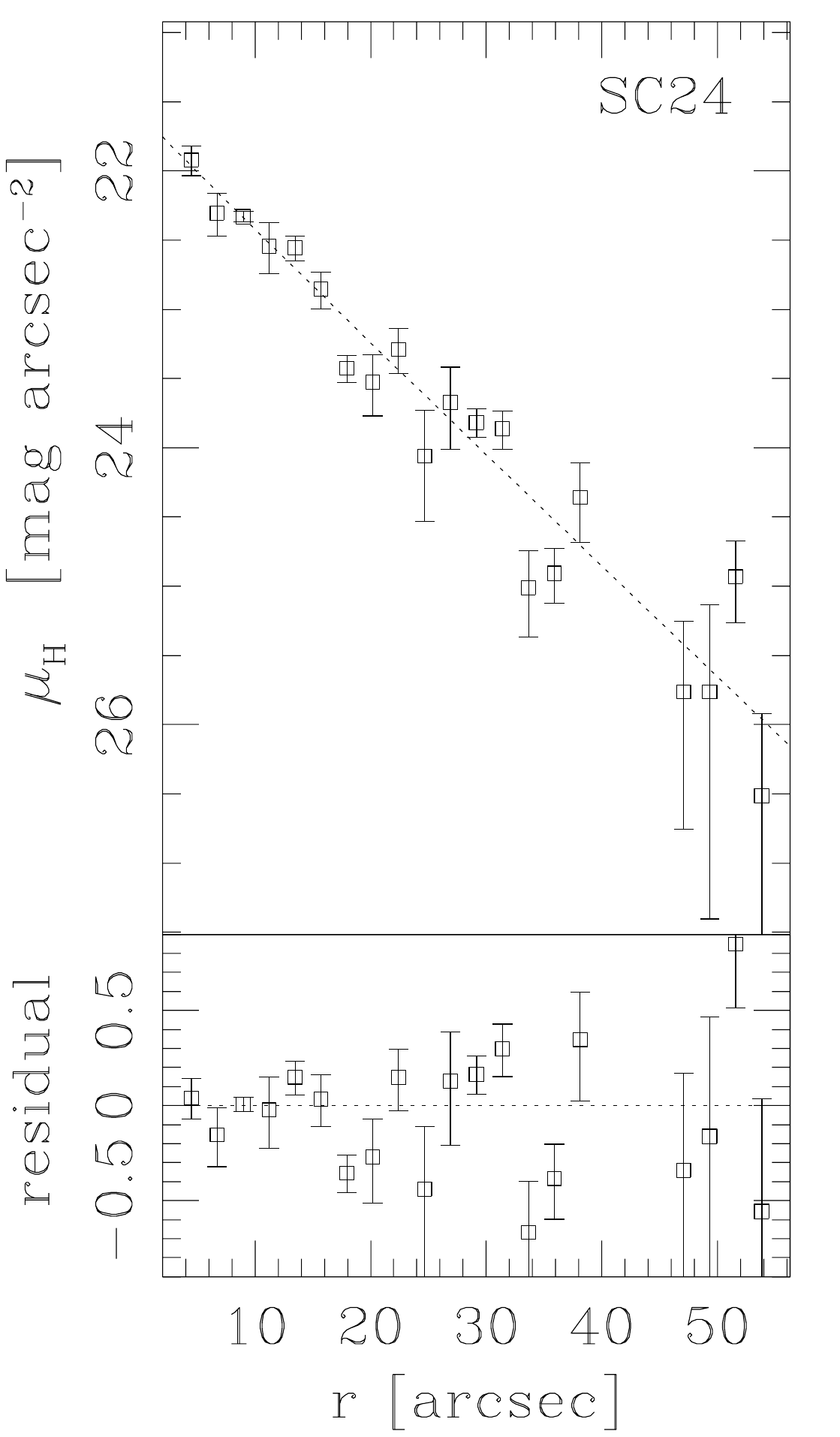}}}  &
  \mbox{\scalebox{0.24}{\includegraphics{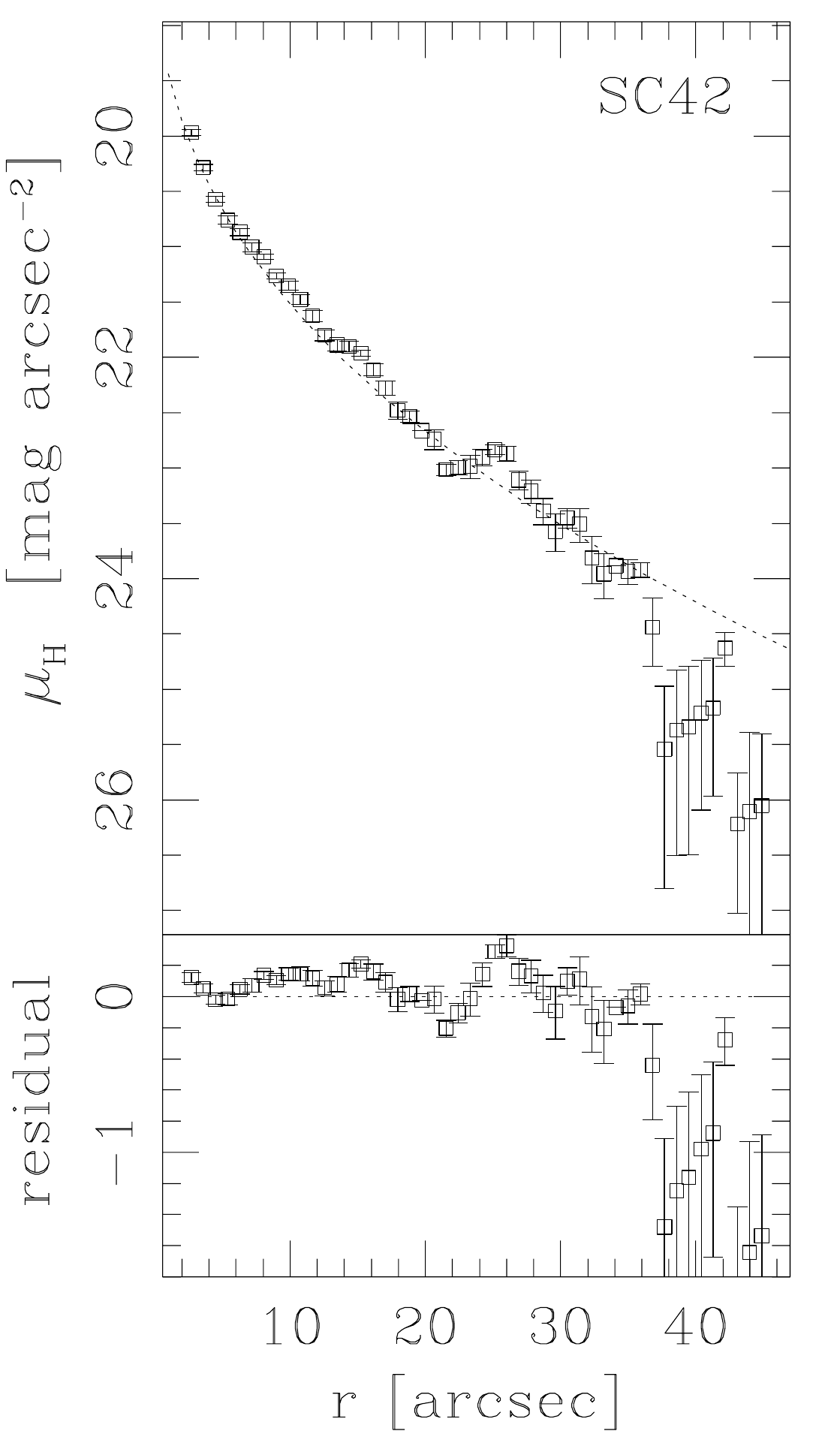}}} \\
  \mbox{\scalebox{0.24}{\includegraphics{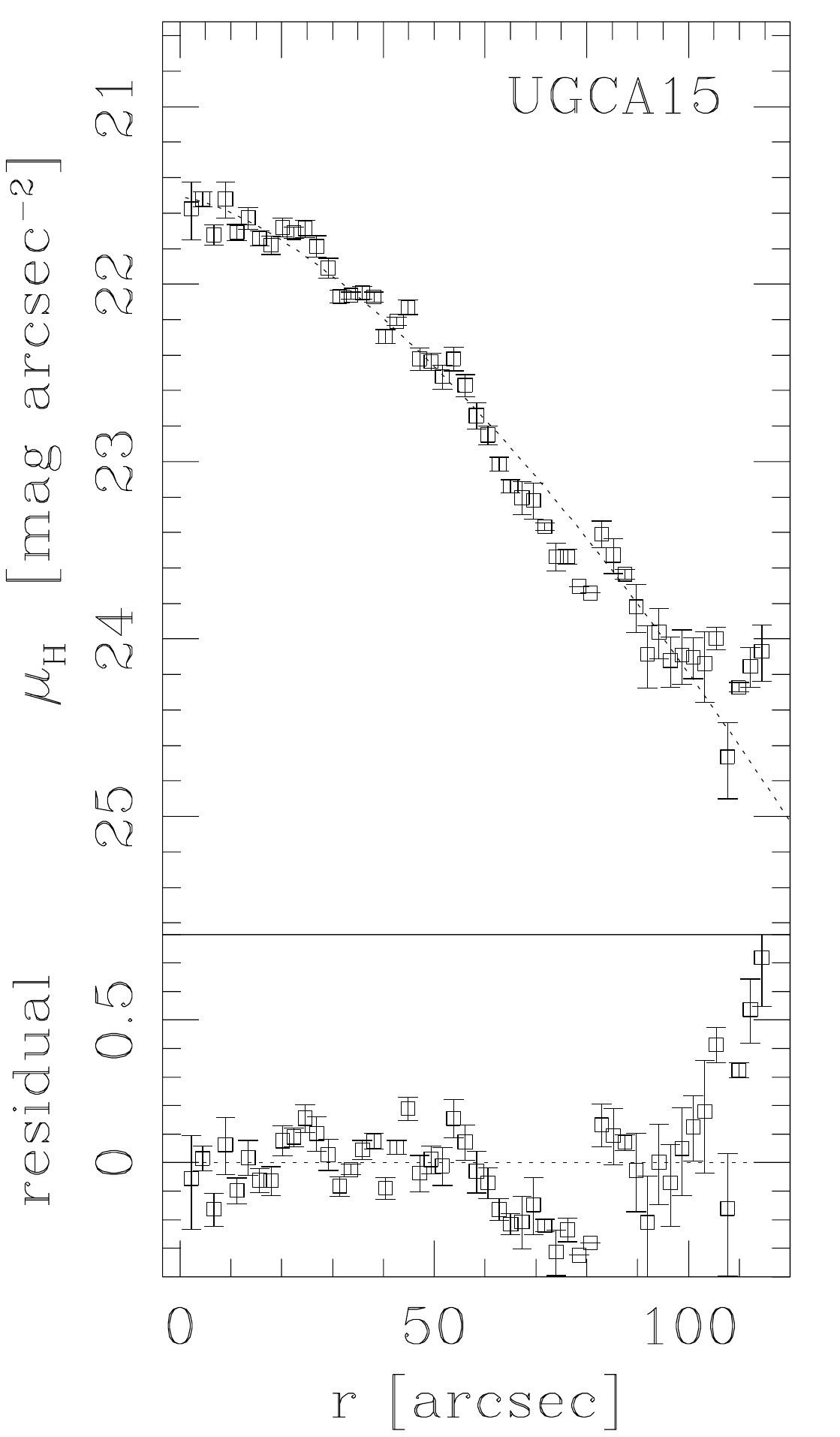}}} &  
  \mbox{\scalebox{0.24}{\includegraphics{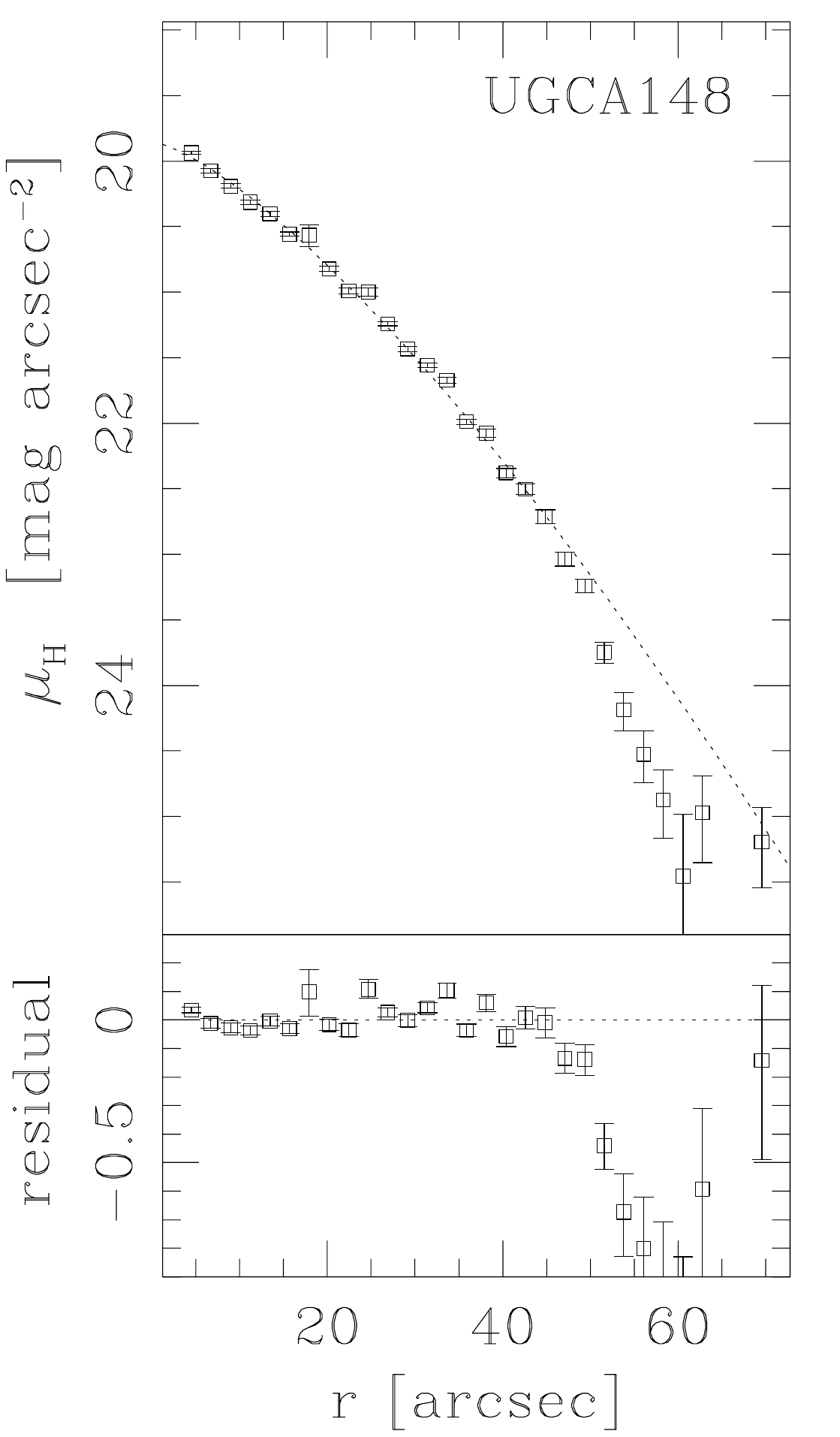}}} &
  \mbox{\scalebox{0.24}{\includegraphics{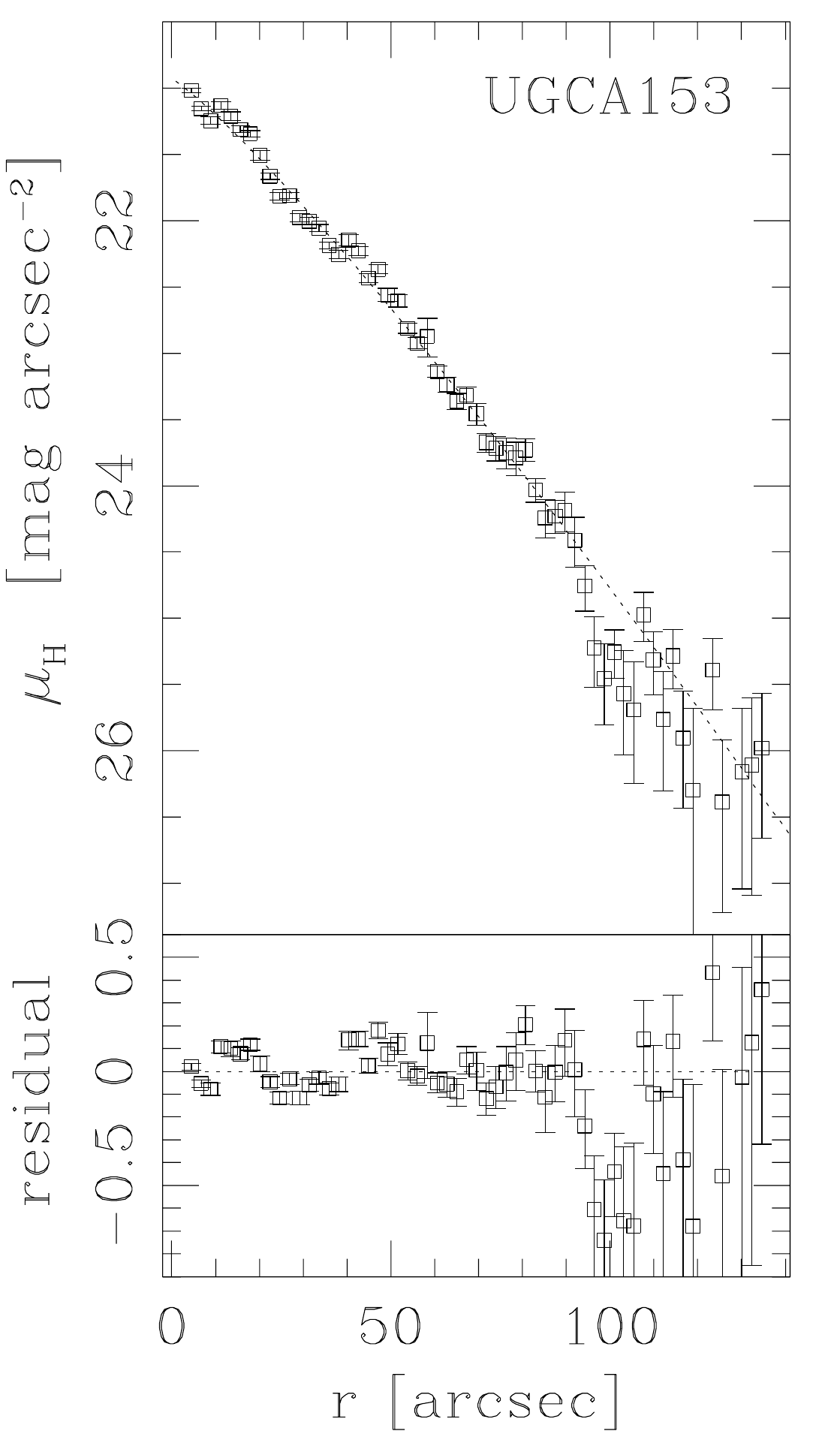}}} &
  \mbox{\scalebox{0.24}{\includegraphics{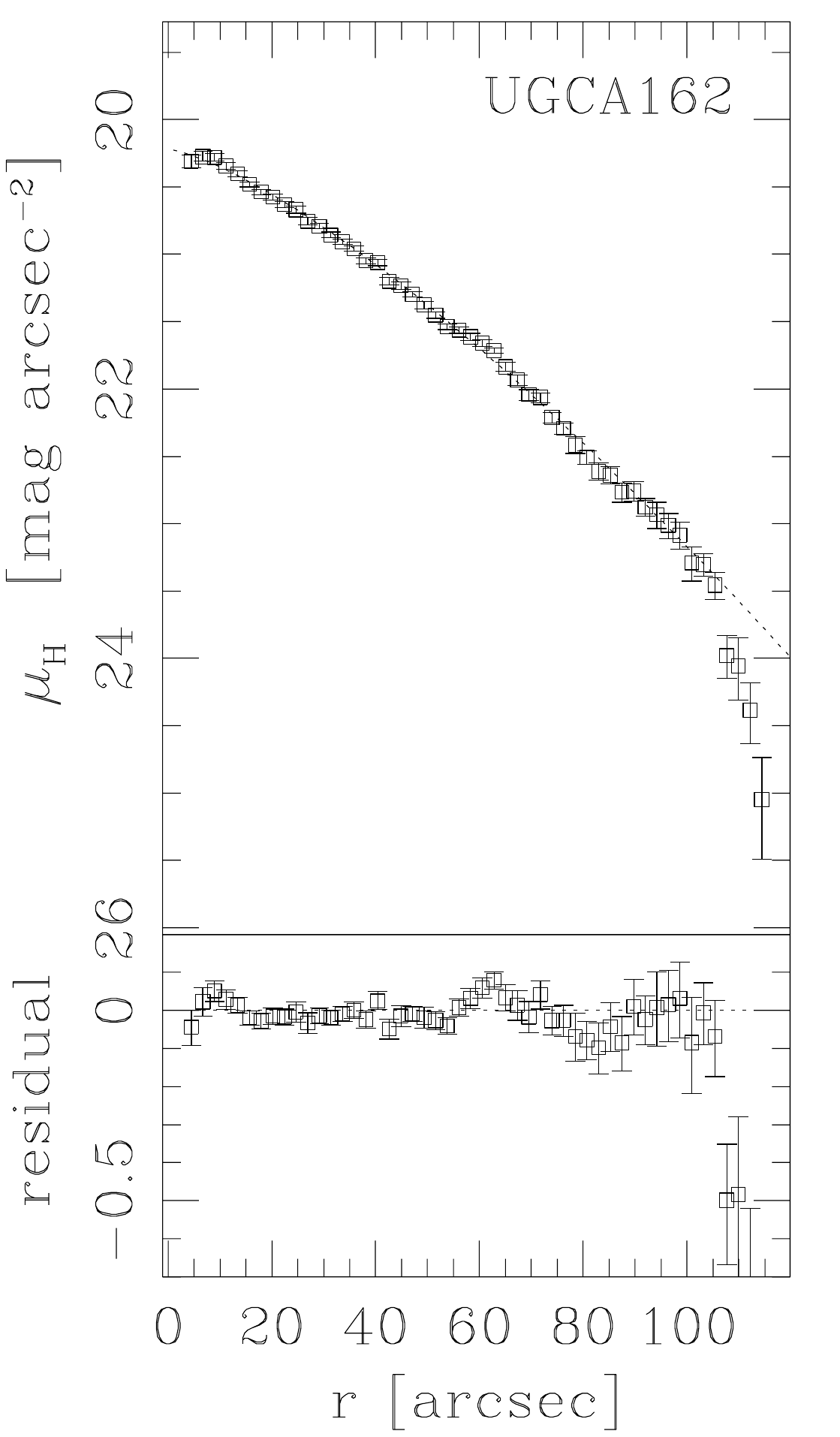}}} &
  \mbox{\scalebox{0.24}{\includegraphics{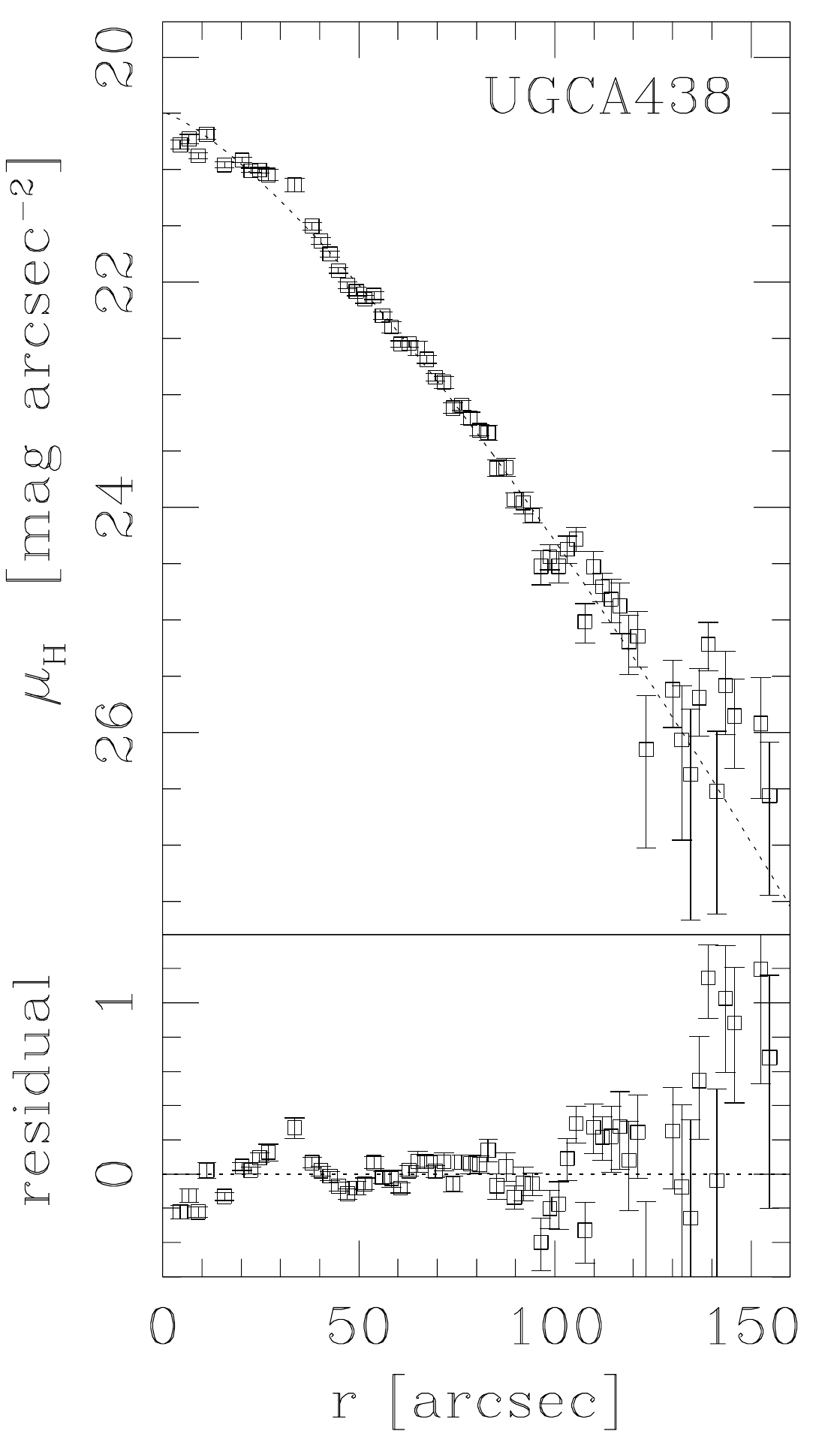}}} \\
  \mbox{\scalebox{0.24}{\includegraphics{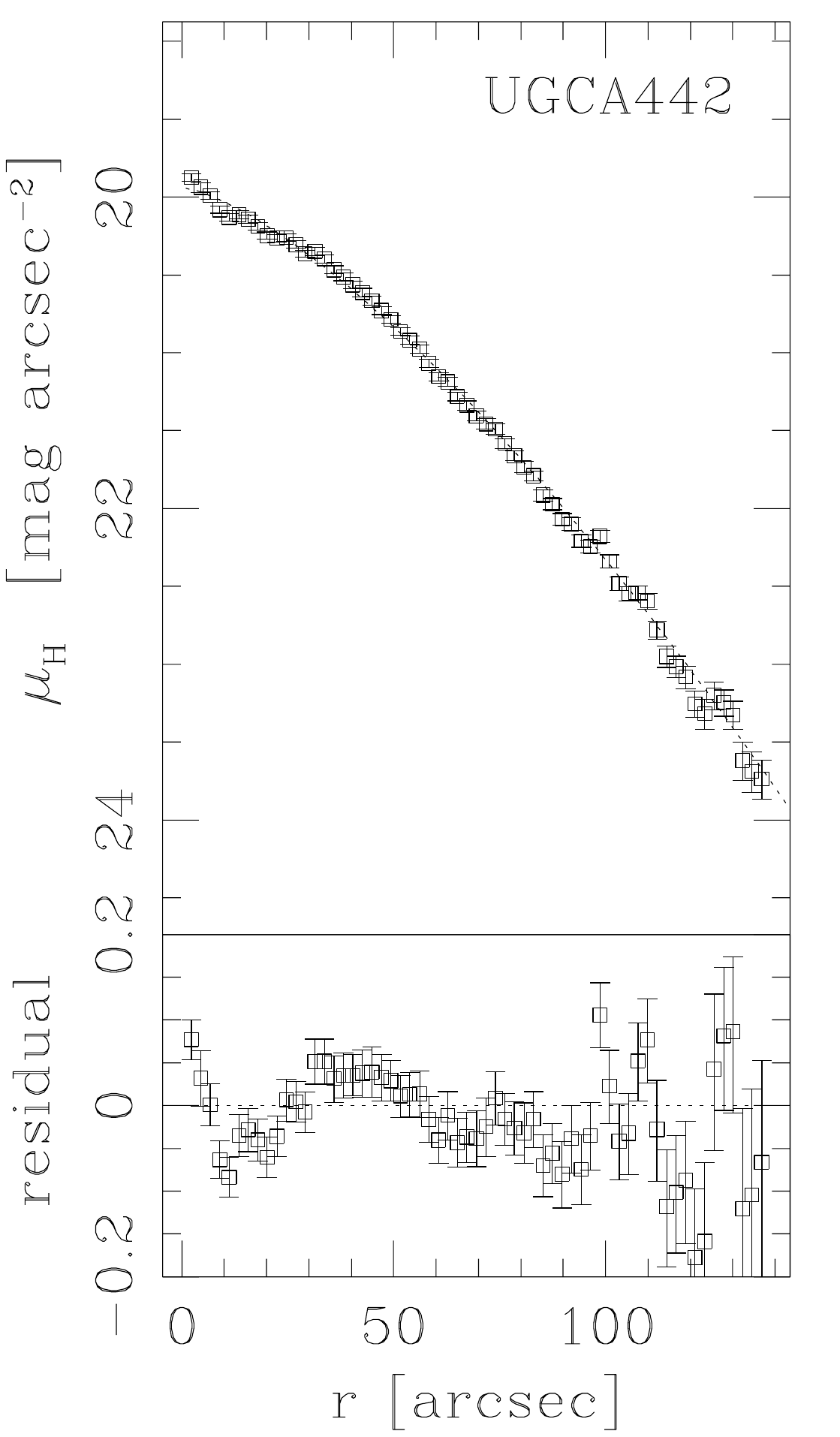}}}   \\
\end{tabular}
\caption{$H$-band surface brightness profiles for all program galaxies except AM0521-343 (see text). The best fitting  S\'ersic profile is shown as solid line together with the residuals.}
\label{figsbprof3}
\end{figure*}

The radial surface brightness profile of a sample galaxy was determined by
differentiating the growth curves with respect to radius. 
Depending on the total integration time of the image, the profiles 
could be reconstructed down to a surface brightness limit between 
$24.5$\,mag arcsec${}^{-2}< \mu_{lim}<26$\,mag arcsec${}^{-2}$. They 
are shown, with a linear radius scale in Figures~\ref{figsbprof1},~\ref{figsbprof2} and~\ref{figsbprof3}. The error bars are calculated as the rms scatter of the intensity along
each isophote. The position angle and ellipticity for each isophote, with the galaxy center fixed, were calculated with IRAF's ELLIPSE package as a function of radius. The ellipticity and position angle were, in general, settled in the outer regions of the galaxies, however, in some cases they varied significantly in the inner regions. 

Table~\ref{tab:phot1}  lists the measured properties of the 57 sample galaxies: column (1) galaxy name, (2) observed integrated apparent magnitude, $m_{H,obs}$, (3) effective radius, $r_{eff}$,  (4) mean effective surface brightness, $\langle \mu_H \rangle_{eff}$, (5) ellipticity and (6), position angle of the major axis 
measured in degrees from North through East (PA$=90^\circ$). The listed ellipticity and position angle is that of the outermost isophote fitted.

The surface brightness profile for each galaxy was fitted with the
S\'ersic function, $\mu(r)=\mu_0+1.086 (r/\alpha)^n$
(or $I(r)=I_0\exp(-(r/\alpha)^n)$), using IRAF's NFIT1D procedure. Extrapolation of the surface brightness profile to infinity with the help of the S\'ersic function allows us to
make an accurate estimate of the amount of flux that remained undetected in the
sky background noise. Using the S\'ersic parameter, $n$, the scale length, $\alpha$, and the central surface brightness, 
$\mu_0$,  we calculated the magnitude of the galaxy between the 
maximum radius and the limit $r=\infty$:
\begin{eqnarray*}
\Delta m &=& -2.5\log_{10}(I_{\rm missing}/I_{\rm tot})\nonumber\\
~&=&-2.5\log_{10}(\Gamma [2/n,(r_{max}/\alpha )^n]/\Gamma [2/n])
\end{eqnarray*}
where:
\begin{eqnarray}
I_{\rm missing} &=& \int_{r_{max}}^{\infty} 2\pi I_0 \exp [-(r/\alpha)^n ] rdr \nonumber\\
~&=&\frac{2\pi I_0 \alpha^2}{n}\Gamma [2/n,(r_{max}/\alpha )^n]
\end{eqnarray}
and 
\begin{eqnarray}
I_{\rm tot} &=& \int_{0}^{\infty} 2\pi I_0 \exp [-(r/\alpha)^n ] rdr \nonumber\\
~&=&\frac{2\pi I_0 \alpha^2}{n}\Gamma [2/n].
\end{eqnarray}
Thereby $\Gamma(a,x)$ is the upper incomplete gamma function
and $r_{max}$ was taken to be the radius at which the growth curve reaches within 5\% of the asymptotic intensity. The missing flux 
introduces a systematic error  to the total magnitude $m_{H,obs}$. This correction 
$\Delta m$ was typically less than 0.2\,mag (see Table~\ref{tab:phot2}),
except in the few cases, e.g.~NGC3115, where the galaxy's angular size extends beyond the 
IRIS2 FoV. We note that 
those galaxies have been observed in the CSJ mode to estimate the sky level from 
dedicated blank sky observations. AM0521-343 was  not corrected for missing flux as the bright foreground star prevented the fitting of its surface brightness profile.

The corrected total apparent magnitude, $m_H=m_{H,obs}-\Delta m$, was converted into a luminosity using 
the standard equations
\begin{eqnarray*}
M_{H,0}= m_H-5\log_{10} D - 25 - A_H
\end{eqnarray*}
and 
\begin{eqnarray*}
 L=10^{0.4(M_{H,\odot}-M_{H,0})}
\end{eqnarray*}

where $M_{H,\odot}= 3.35$\,mag is the $H$-band luminosity of the sun \citep[][]{colina96} and $A_H=0.576\cdot E(B-V)$ is the Galactic extinction \citep{schlegel98}.  The accuracy of  the reddening-corrected absolute magnitude, $M_{H,0}$, is dominated by the accuracy of the distance. 

The effective radius, $r_{eff}$, is the aperture radius which encloses half the total light of a galaxy. This quantity is systematically underestimated unless corrected for the amount of undetected flux. The corrected effective radius, $R_{eff}$ is defined implicitly by the equation
\begin{equation}\label{eq:halflight}
\frac{I_{\rm tot}}{2} \,=\,\frac{2\pi I_0 \alpha^2}{n}\cdot\Gamma \left[\frac{2}{n}, \left(\frac{R_{eff}}{\alpha}\right)^{n}\right]
\end{equation}
where $I_0$, $\alpha$, and $n$ are the S\'ersic parameters for a particular galaxy. The solution for all $n \in [0.25, 2]$ can be approximated by the equation
\begin{equation}\label{eq:halflight2}
R_{eff} \approx \alpha \left(\frac{2}{n} -0.33211\right)^{1/n}.
\end{equation}
Within the quoted uncertainties, there is no deviation between the approximate analytic solution (equation~\ref{eq:halflight2}) and the numerical solution of equation~\ref{eq:halflight}.

The $H$-band luminosity of each galaxy was converted into a stellar mass using a mass-to-light ratio of  $\Upsilon_{\ast}^H=1.0\pm 0.4$. This conversion factor is discussed in detail in Section~\ref{ss:mtol}. The derived parameters are listed in Table~\ref{tab:phot2} for the 57 LSI galaxies: column (1) galaxy name, (2) effective radius, $r_{eff}$, in kpc, (3), (4) and (5) the S\'ersic parameters $\mu_0$, $n$ and $\alpha$, (6) the missing flux, $\Delta m$, (7) and (8) the corrected effective radius, $R_{eff}$, in arcsec and kpc, (9) absolute H magnitude, $M_{H,0}$, and (10) the total stellar mass, $\log_{10}{\mathcal{M}}_{\ast}$.

 \subsection{Galactic extinction correction}\label{ss:extinct}

The extinction correction used in our study is that of \cite{schlegel98}. We prefer these IR emission maps over the older \cite{burstein78, burstein82, burstein84} models as the reddening is derived directly from dust emission rather than H\,{\sc i} column densities and galaxy counts. The \cite{schlegel98} maps have a typical uncertainty of 16\%. However, for low latitudes, $| b|< 5^{\circ}  $, most contaminating sources were not removed from the maps leading to larger errors for that part of the sky. This difficulty can not be circumvented by using the Burstein \& Heiles maps as they do not include latitudes below $| b|<10^{\circ}  $.
Consequently, the reddening corrections applied to the two sample galaxies KK2000-25 and KKS2000-09  are less secure. This uncertainty will particularly affect the quoted $B$-band magnitudes. The $H$-band results will be affected to a lesser extent as the correction is of order one-tenth of that in the $B$-band.

Should one choose to use the  Burstein \& Heiles models the difference in our results is minimal. There are only two sample galaxies (KKS2000-09 and ESO461-G036) which have significantly different absolute $B$-band magnitudes. In these two cases we apply an average of the Burstein and Heiles and the Schlegel et al.\ corrections. The choice of the reddening estimate does not change the $H$-band results within the quoted errors.

\subsection{The $H$-band mass-to-light ratio}\label{ss:mtol}
For our analysis we will adopt a $H$-band mass-to-light ratio that is well supported by observations and theory. Assuming a typical 12 Gyr old, solar metallicity, stellar population with a constant star formation rate and a Salpeter initial mass function,  the  \cite{dejong96} model yields a mass-to-light ratio of  $\Upsilon_{\ast}^H=1.0$. This is consistent with the empirically derived value of $\Upsilon_{\ast}^H=0.9 \pm 0.6$ obtained for our sample by using each galaxy's $B-H$ colour (Tables~\ref{basicdata} and \ref{tab:phot1}) and adopting the colour-dependent stellar mass-to-light ratio relation from the \cite{bell01} galaxy evolution models.  Finally, these two mass-to-light ratios are well in the range of $0.7<\Upsilon_{\ast}^H <1.3$ which is based on observed SDSS colours $0.1<(g-r)<1.1$ for 22,679 galaxies and 2MASS photometry \citep{bell03}. From  these three independent values we derive the error weighted mean of $\Upsilon_{\ast}^H=1.0\pm 0.4$.

\section{Are there any genuine young galaxies?}\label{s:nondetect}

Observations for  11 galaxies were not included in our photometric study either because the galaxy remained invisible in
the final mosaics despite our faint $H$-band surface brightness limit of $24-26$\,mag arcsec${}^{-2}$, or the galaxy was detected but foreground stars interfered with the analysis.  
In this section, we  discuss the four galaxies labelled  as ``no galaxy detected" in Table~\ref{badgals}: AM0717-571,  KK2000-04, KK2000-06 and NGC2784 DW1. While they remain as candidates for galaxies with a pure young stellar component, we show it is unlikely that this is the case. We also  include KK2000-03 which had a marginal detection, and
the galaxy  pair HIZOAJ1616-55 and SJK98 J1616-55 which were not detected but we note that the images had serious foreground contamination.
\begin{deluxetable*}{ccccccc}
\tablewidth{0pt}
\tabletypesize{\small}
\tablecaption{Galaxies not analysed further. \label{badgals}}
\tablecolumns{7}
\tablehead{\colhead{} & \colhead{R.A.} &\colhead{Decl.} &
  \colhead{}& \colhead{$m_H$} &\colhead{$M_{H,0}$} &\colhead{$\log_{10} ({\mathcal{M}}_{\ast})$}\\ 
\colhead{Name} &\colhead{(J2000.0)} & \colhead{(J2000.0)} &
\colhead{Reason} & \colhead{(mag)} &\colhead{(mag)} & \colhead{$\log_{10} ({\mathcal{M}}_{\ast})$}\\ 
\colhead{(1)} & \colhead{(2)} & \colhead{(3)} &\colhead{(4)} & \colhead{(5)}& \colhead{(6)}& \colhead{(7)}} 
\startdata
KK2000-03& 02:24:44.58 &-73:30:49.2& marginal detection &$-$&$-$&$-$\\
KK2000-04&  03:12:46.14  &-66:16:12.5&no galaxy detected &$>11.8$& $>-16.5$& $<7.0$\\
KK2000-06& 03:14:26.14  &-66:23:27.9& no galaxy detected &$>11.8$& $>-16.5$& $<7.9$\\
ESO490-G017 & 06:37:57.09 & -26:00:03.1 &foreground stars& $-$& $-$& $-$\\
HIZSS003 & 07:00:29.3   & -04:12:30 & foreground stars & $-$&$-$& $-$\\
ESO558-PN011& 07:06:56.80 &-22:02:26.0 &foreground stars& $-$& $-$& $-$\\
AM0717-571 &07:18:37.90  & -57:24:46.5& no galaxy detected&$>11.8$& $>-18.7$&$<8.8$\\
NGC2784 DW1 &09:12:18.5   &-24:12:41 & no galaxy detected &$>11.8$&$>-18.3$&$<8.7$\\
SJK98 J1616-55 & 16:16:49.0    &-55:44:57 & Galactic plane& $-$& $-$& $-$\\
HIZOAJ1616-55 &16:18:46     &-55:37:30& Galactic plane& $-$& $-$& $-$\\
ESO594-G004& 19:29:58.97  &-17:40:41.3& foreground stars&$-$&$-$&$-$\\
\enddata
\end{deluxetable*}

A lower bound of the total apparent magnitude for these galaxies can be calculated.  For that purpose, we consider a hypothetical galaxy with a constant  star density  equivalent  to the survey's mean surface brightness limit of $\langle\mu_{0,lim}\rangle=25$\,mag arcsec${}^{-2}$ out to a cutoff radius, $r_{cut}$, at which point the stellar density drops to zero. We set the cutoff radius, $r_{cut}$, to be 250\,arcsec which is equivalent to the size of the largest galaxies in our sample. This yields the brightest apparent magnitude an undetected galaxy could possibly have: 
\begin{equation}
m_{tot}> \mu_{0,lim} -2.5\log_{10} (\pi r_{cut}^2)=11.8\,{\rm mag}.
\end{equation}
This lower bound is applicable to AM0717-571,  KK2000-04, KK2000-06 and NGC2784 DW1 but not KK2000-03 which has a foreground star located directly in front of the galaxy (see discussion below), nor  HIZOAJ1616-55 and SJK98 J1616-55 which have serious foreground contamination. A lower bound on the absolute magnitude and an upper bound on the stellar mass is calculated using distances from the literature (see Table~\ref{badgals} ).

\subsection*{AM0717-571}
Although AM0717-571 has appeared in several lists of nearby galaxies~\cite[eg.][]{karachentseva98, whiting02, whiting07}, this object was actually never followed-up in the optical or near-IR and consequently was never confirmed to be a galaxy. The only optical ($B$-band) image
available comes from the Digital Sky Survey (DSS) and shows an object with a morphology resembling closely that of a Galactic nebula.  An H\,{\sc i} signal at the position of AM0717-571 was reported  by HIPASS  and  included in the HIPASS Bright Galaxy Catalog~\citep{koribalski04}. However, to understand this apparent detection one has to know that AM0717-571 has two neighboring galaxies, ESO123-G001 and ESO162-G017, located at an angular distance  of only 20.7 and 22.4 arcmin respectively. These two galaxies have heliocentric velocities of 1160\kms and 1098\kms similar to and bracketing AM0717-571's listed velocity of 1148 \kms. The HIPASS spectra at the $RA,DEC$ positions of ESO162-G017 and AM0717-571 are almost identical in velocity width and peak flux. The fact that our deep $H$-band image did not reveal any galaxy and considering
the relatively large uncertainty of HIPASS coordinates due to the 15 armin beam size of the Parkes Radio Telescope, it is conceivable that the 21cm emission at 1148\,km\,s$^{-1}$ detected by HIPASS is coming from the extended H\,{\sc i} halos of ESO162-G017 and/or
ESO123-G001. It has been pointed out by~\cite{koribalski04} that larger offsets between the H\,{\sc i} and the optical positions usually occur when multiple galaxies contribute to the signal or when the H\,{\sc i} distribution is asymmetric or peculiar.
A clarification of the true nature of AM0717-571 will require further investigations.

\subsection*{HIZOAJ1616-55 and SJK98 J1616-55}

HIZOAJ1616-55 (Juraszek et al.~2000; listed as HIZOAJ1618-55 in NED) and  SJK98 J1616-55 have almost identical catalog positions and are close to the Galactic plane  ($b=-3.8^\circ$). Staveley-Smith et al.~(1998) concluded from deep ATCA  21\,cm mapping that these objects are likely part of a single or interacting pair of low-mass H\,{\sc i} galaxies with a total H\,{\sc i} mass of $8\times 10^{7} {\mathcal{M}}_{\odot}$. Their heliocentric velocities (402\,km\,s$^{-1}$ and 430\,km\,s$^{-1}$) are very similar to that of the Circinus galaxy (439\,km\,s$^{-1}$; Jones et al.~1998) suggesting a possible physical connection to Circinus and the nearby Cen\,A group. However, no optical counterparts have been found to date and hence it is plausible that the H\,{\sc i} detection originates from a compact high velocity cloud~\citep{putman02}. To search for more evidence for either scenario we included HIZOAJ1616-55 and  SJK98 J1616-55 in our imaging survey. The Galactic extinction in the $H$-band ($A_H=0.4$\,mag) is much lower than in the optical (eg. $A_B=2.7$\,mag) and hence increases the chances to detect the stellar components of these objects at low Galactic latitude. However, despite our deep imaging, no stellar counterpart to the H\,{\sc i} was found and consequently the picture of high velocity cloud(s) seems more plausible.

\subsection*{KK2000-03}
KK2000-03, also known as PGC 9140, is well away from the Galactic plane 
at $l=294.2$, $b=-42.0$ but remained almost  invisible on our 1800\,sec $H$-band image. \cite{whiting07} list the galaxy and report an $R$-band surface brightness of $23.7\pm0.2$\,mag\,arcsec${}^{-2}$. They calculate this  as the average surface brightness for an area of roughly 1\,arcmin in diameter, located such that it contains the brightest parts of the galaxy, and  excludes stars (where possible). Assuming an $R-H$ color of  $1.10$ for the sun \citep[calculated using results from][]{colina96} implies KK2000-03 should have an $H$-band surface brightness of 22.6\,mag\,arcsec$^{-2}$, well above our detection limit. The DSS image shows that KK2000-03 is located directly behind a foreground star. Despite this contamination, KK2000-03 must have an unusual blue stellar population for it not to be detected more prominently in our survey.

\subsection*{KK2000-04 and KK2000-06}
Little is known about these two extremely low surface brightness irregular galaxies  which were first mentioned in a catalogue of dwarf galaxy candiates by~\cite{karachentseva00}.  The authors estimated their total $B$-band magnitudes from photographic plates as 17.8 and 17.0, respectively, and speculated that they may be companions of the barred spiral NGC\,1313 ($v_\odot=475$\,km\,s$^{-1}$), another of our sample galaxies. We did not detect either KK2000-04 or KK2000-06  despite obtaining deep imaging data of fields with only low levels of foreground contamination. While KK2000-04 also remains undetected at 21\,cm,  the H\,{\sc i} spectrum for KK2000-06 measured with the Effelsberg 100-m radio telescope \citep{huchtmeier01} suggests a heliocentric velocity of  $\sim$2250\,km\,s$^{-1}$. Therefore it is possible that KK2000-04 is  a plate flaw and KK2000-06 is a distant galaxy unrelated to NGC1313.

\subsection*{NGC2784 DW1}
The extremely low surface brightness dwarf galaxy NGC2784 DW1 was first detected by~\cite{parodi02}. It is located between the S0 galaxy NGC2784 and  the nucleated early-type dwarf KK98-73  on the sky and due to its morphology (dE), its size and location, \cite{parodi02} suggest it is likely to be a satellite of NGC2784.  \cite{karachentsev04}  give NGC2784 DW1 a membership distance and include it in the census of galaxies within 10 Mpc. However, it is important to note that no independent distance measurement has been obtained to date. Our deep near-IR observation detected the faint  galaxy KK98-73, while NGC2784 DW1 is barely visible as expected from the recorded mean effective surface brightness of $\langle \mu\rangle_{eff}\approx25$\,mag\,arcsec$^{-2}$ and $r_{eff}\approx 20$\,arcsec in the $B$-band.

\section{Results}\label{s:results}

\subsection{LSI survey versus 2MASS photometry}

\begin{figure}[!hbt]
\centering
\begin{tabular}{cc}
 \mbox{\includegraphics[width=0.45\linewidth]{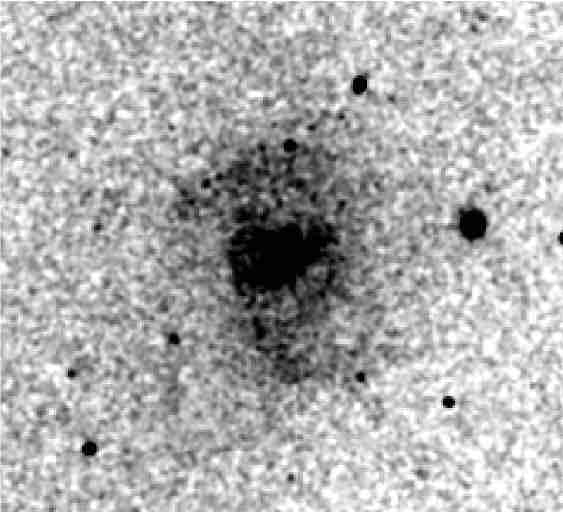}}&
 \mbox{\includegraphics[width=0.45\linewidth]{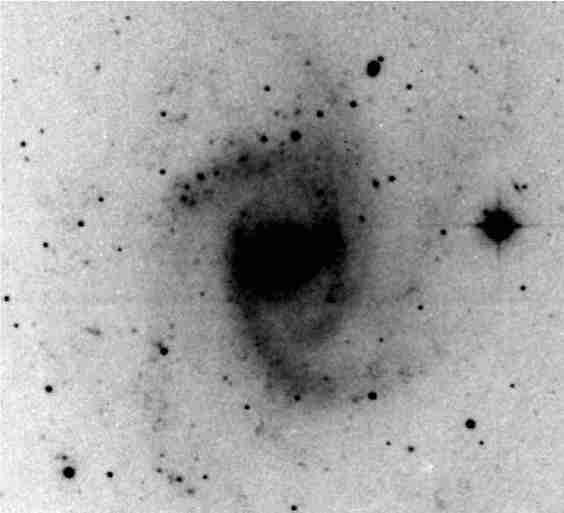}}\\
 \mbox{\includegraphics[width=0.45\linewidth]{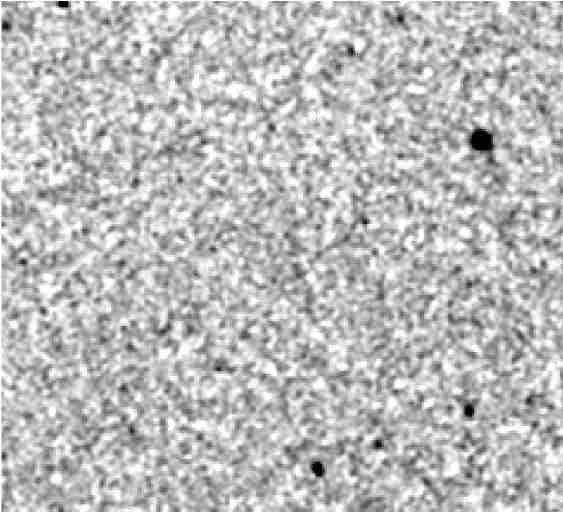}}&
 \mbox{\includegraphics[width=0.45\linewidth]{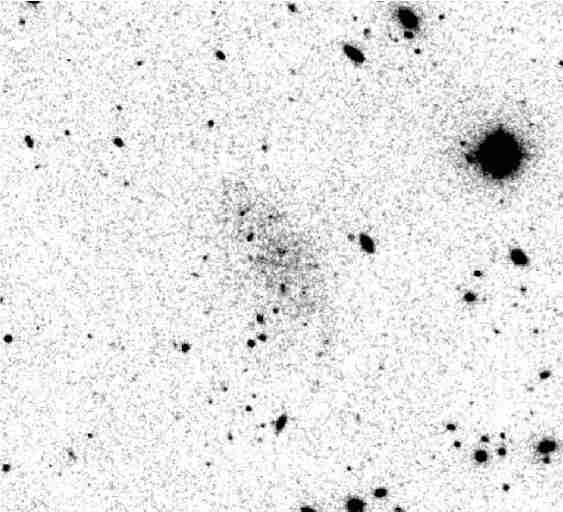}}\\
 \end{tabular}
\caption{The 2MASS (left panel) and the LSI (right panel) $H$-band images of the spiral galaxy NGC2835 ($V_{helio}=886$\,km\,s$^{-1}$) and the dwarf irregular galaxy ESO473-G024 ($V_{helio}=541$\,km\,s$^{-1}$) are  shown in the upper and lower panels, respectively. The new LSI images probe to surface brightness levels $\approx$4\,mag\,arcsec${}^{-2}$ fainter than 2MASS. A complex morphology and additional spiral arms are detected in the case of NGC2835 whereas only the deeper LSI image detects a stellar component in the case of ESO473-G024. Higher resolution version available at  \url{http://www.mso.anu.edu.au/~emma/KirbyHband.pdf}}
\label{fig:2mass}
\end{figure}

It is instructive to see how galaxies  can change  their appearance when going from 2MASS to the deeper LSI observations. For example, our image of the  barred Sc galaxy NGC2835 (Fig.~\ref{fig:2mass}) reveals the rich near-IR morphology and extent of this spiral galaxy  for the first time. The almost face-on view presents a well-ordered 4 or 5-arm spiral pattern outlined by the star-dominated Population II disk, which closely traces the gas-dominated Population I disk morphology observed in the $B$-band \citep{sandage94}. The 2MASS  image for the irregular Sculptor group galaxy ESO473-G024 shows  qualitatively the limitation of that survey  to study  dwarf  galaxies.   With a central $H$-band surface brightness of $\approx 20.5$\,mag\,arcsec$^{-2}$, ESO473-G024 remains effectively undetected in 2MASS (Fig.~\ref{fig:2mass}). Our image uncovers a smooth, dE-like morphology with little evidence of irregularity. This stands in stark contrast to the $B$-band image  that is dominated by a number of prominent H\,{\sc ii} regions and dust features.

\begin{figure}[!hbt]
\centering
\includegraphics[width=1\linewidth]{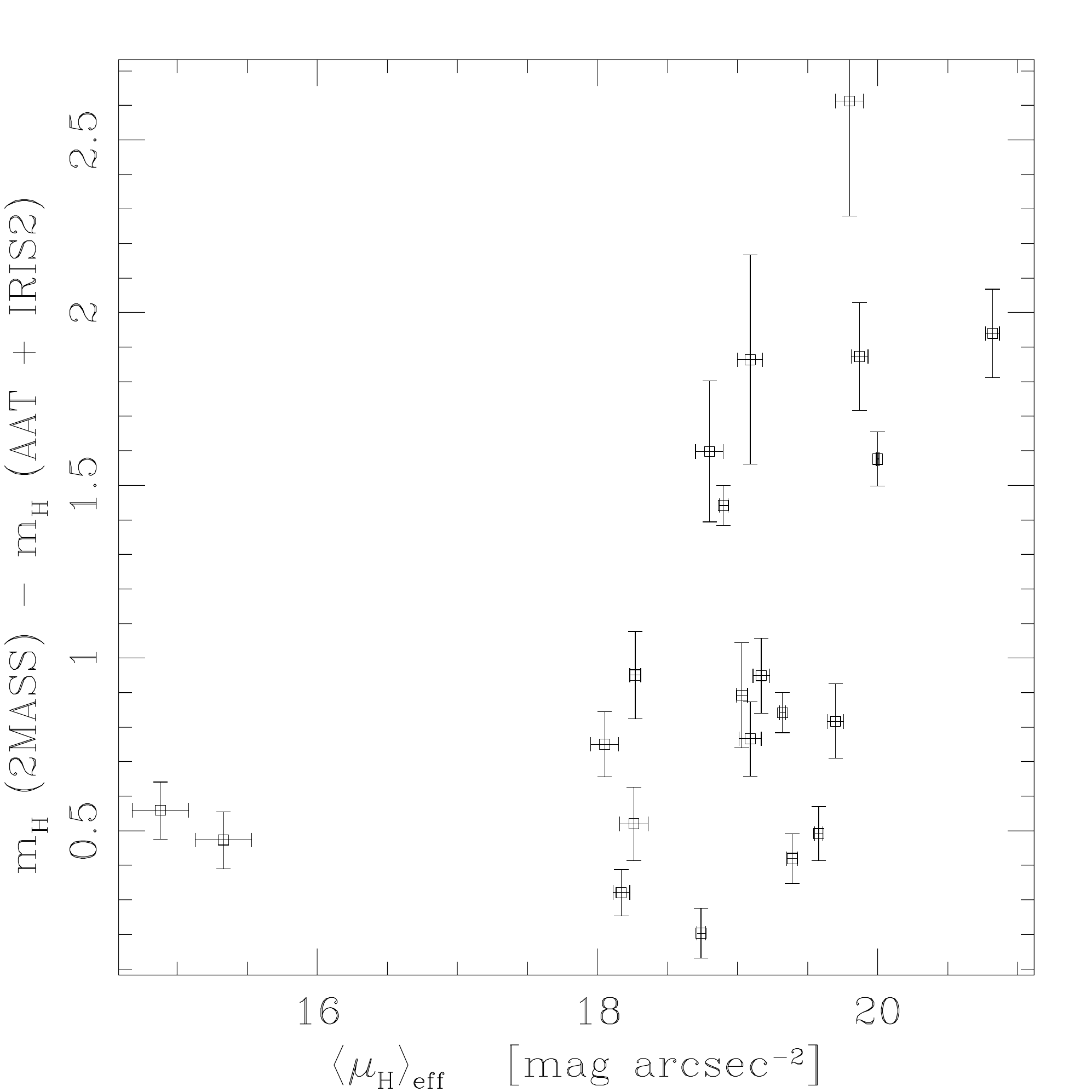}
\caption{2MASS versus LSI magnitudes showing the systematic underestimation of galaxy fluxes by 2MASS for the galaxies that it did detect.}
\label{fig:comp}
\end{figure}

It has been previously pointed out  by \cite{andreon02} that the short integration time of 2MASS failed to detect most  of the lower surface brightness (dwarf) galaxies and that, if they were  detected, fluxes were underestimated by as much as 70 percent. To investigate  this issue further, we plot in Figure~\ref{fig:comp} the difference between  our total extrapolated apparent magnitudes ($m_{H,obs}-\Delta m$; Table 3, col.~2 and Table 4, col~6) and the total magnitudes from the 2MASS All-Sky Extended Source Catalog for 21 galaxies we have in common, as a function of mean effective surface brightness ($\langle\mu_H \rangle_{eff}$; Table 3, col.~4). 
LSI galaxies with a surface brightness fainter  than $\mu_{H}=18$ mag\,arcsec${}^{-2}$ are affected at different levels with the missing flux in the range between 0.2 and 2.5\,mag. Even in the cases of the luminous galaxies NGC2784 and NGC3115 our analysis finds that their 2MASS $H$-band magnitudes are 0.5\,mag too faint.

 To demonstrate that the disparity between the 2MASS and our $H$-band magnitudes  is not caused by differences in the measuring procedure we analysed 2MASS images using our method. The photometric parameters listed in the 2MASS  Extended Source Catalog, as well as the surface brightness profiles, were reproduced within the quoted uncertainties. It should be noted that the 2MASS Large Galaxy Atlas~\citep{jarrett03} has employed two different methods for recovering the flux below the sky background noise and that we have made the comparison to the magnitudes which were obtained by extrapolating the surface brightness profile.  The 2MASS magnitudes which were obtained using \cite{kron80}  apertures are systematically fainter (see Figure 9 of \citealt{jarrett03}) and therefore the difference between our total apparent magnitudes and the 2MASS magnitudes obtained using Kron apertures is also larger.

The Extended Source Catalog is contaminated by a small (1\% to 5\%) number of artefacts which can significantly affect the photometry of real sources \citep{jarrett00}. Each of our galaxy images were visually inspected for artefacts which were effectively removed (unless documented otherwise). This is obviously an impractical approach for the much larger 2MASS dataset and  \cite{jarrett00} notes that the pipeline is not 100\% effective. It is conceivable that this accounts for a fraction of the discrepancy.

In summary, our finding is in good agreement with that of \cite{andreon02} and emphasizes again that the 2MASS magnitudes are significantly fainter than those obtained from deeper near-IR imaging. As the mean effective surface brightness correlates with the luminosity of a galaxy (see \S\ref{ss:lumsb}, Figure~\ref{fig:lsp}), serious selection biases must be expected, for instance, for the 2MASS-based $H$-band galaxy luminosity function at magnitudes fainter than M$_{H}=-20$\,mag.

\subsection{Optical - Near Infrared magnitude transformation} \label{s:photometric}

\begin{figure}
\centering
\includegraphics[width=1\linewidth]{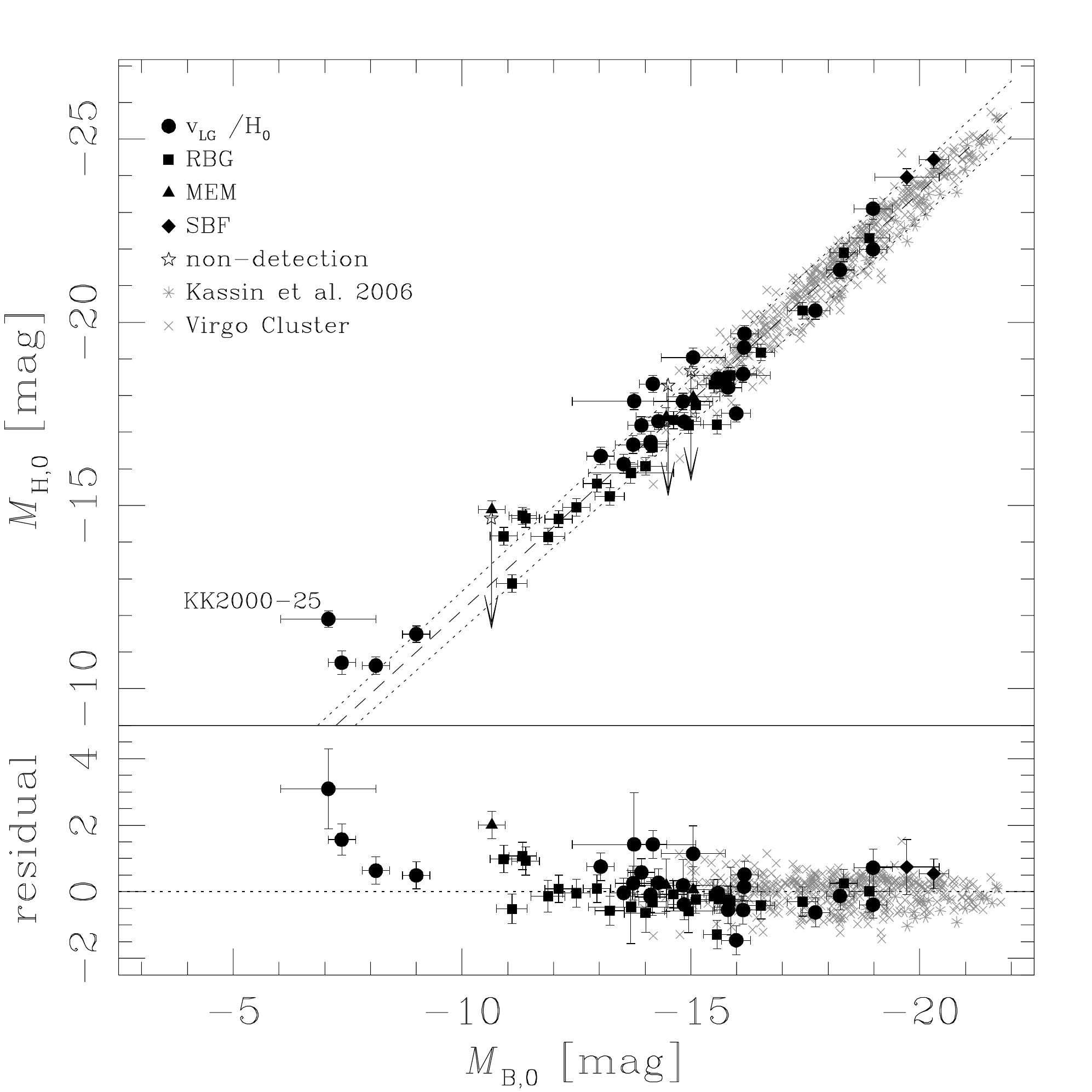}
\caption{Comparison of the integrated absolute $B$- and $H$-band magnitude. Plotted here is our new data with an indicator of the distance estimate method.  Also plotted is the data of \cite{kassin06} and the Virgo cluster data from  the Goldmine database \citep{gavazzi03}. Marked is   the galaxy KK2000-25  which appears to be significantly under-luminous  in the $B$-band. The dashed line is the $B - H$ mean colour for all galaxies in the Virgo Cluster sample and the dotted lines indicates data within 1 $\sigma$. The galaxies which were not detected have been given a lower bound magnitude estimate.} 
  \label{fig:BvsH}
\end{figure}

\begin{figure}
\centering
\includegraphics[width=1\linewidth]{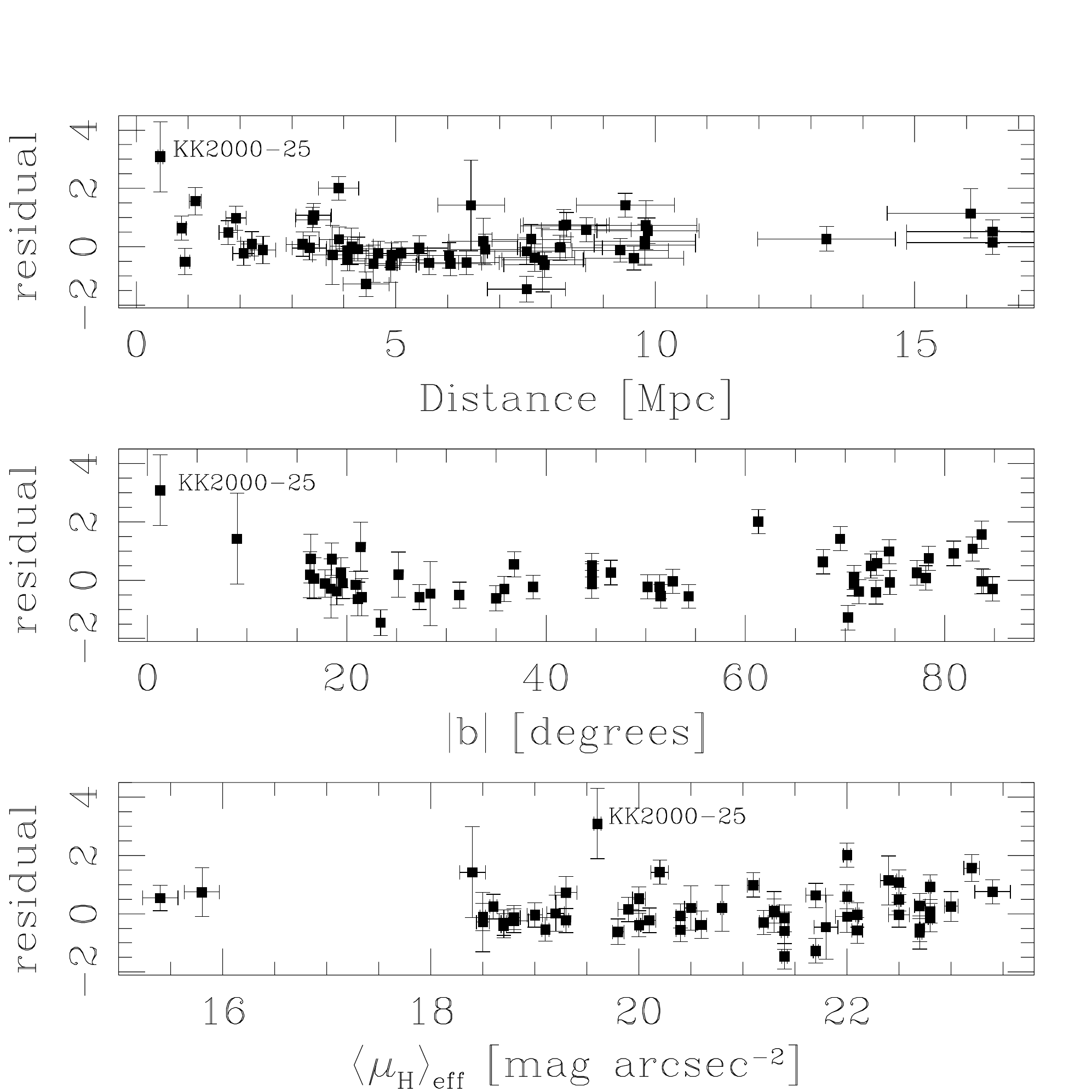}
\caption{The deviation of individual LSI galaxies from equation~\ref{eq:bh} as a function of distance [upper panel], Galactic latitude [middle panel] and mean effective surface brightness [lower panel]. The error in the distance is taken to be 10\%. KK2000-25 is located at low Galactic latitude and significantly deviates from the relationship.} 
  \label{fig:reslat}
\end{figure}

 The $B - H$ colour of each galaxy is an indicator of the ratio of the population II to population I stars, as modulated by the effects of dust. A comparison of the absolute  $B$- and $H$-band magnitude, corrected for Galactic extinction,  is shown in Figure~\ref{fig:BvsH} for our sample galaxies.
We also plot the Virgo cluster data from the Goldmine database \citep{gavazzi03}. The data of the Virgo galaxies  were  extinction-corrected  using  $A_B=0.13$\,mag and $A_H=0.01$\,mag.   We have adopted the mean cluster distance of 15.8\,Mpc \citep{jerjen04} based on surface brightness fluctuation measurements of early-type galaxies.  In addition to the Virgo cluster data, we have also included  the data for  30 bright spiral galaxies from  \cite{kassin06} which were corrected for extinction  \citep[using][]{schlegel98}. By including the two  additional data sets  we are able to  investigate  the $B - H$ colour for  late-type galaxies over a range of 15 magnitudes. Figure~\ref{fig:BvsH} shows that there is a tight  correlation between the $B$- and $H$-band luminosity of a galaxy. This linear relation is 
\begin{equation}\label{eq:bh}
M_{H,0}=(1.14\pm 0.02)M_{B,0}-(0.74\pm 0.32)
\end{equation}
which is  a least squares fit to the Virgo cluster data. While the more luminous galaxies in our sample obey the relation closely,  the residual plot suggests that the scatter marginally increases in the dwarf regime and possibly has a slight  upwards trend to redder colours (average residual $\approx1$ mag). The most deviant galaxy in our sample, KK2000-25,  appears under-luminous in the $B$-band by 3 magnitudes. While we cannot exclude the possibility that this galaxy has had an unusual star formation history, we need to point out two things. Firstly, KK2000-25 is located almost in the Galactic plane  ($b=1.28^\circ$) and thus has a large 
$B$-band extinction uncertainty. Secondly,  the $B$-band magnitude for KK2000-25 was estimated visually from a photographic film \citep{huchtmeier01}.  Consequently, the deviation from the line of best fit could be entirely due to a large uncertainty in the $B$-band magnitude.  By excluding KK2000-25, there is no correlation between the residual in the $B-H$ plot and the galaxy distance, Galactic latitude or mean effective surface brightness (see Figure~\ref{fig:reslat}).  

It is worth noticing that the least squares fit  deviates from a line of unity slope. The gradient of $1.14\pm 0.02$ implies  that dwarf galaxies are in general bluer than the more luminous galaxies.  It is well known that galaxy colour correlates with luminosity~\citep[eg,][]{tully98,hogg02, blanton01, blanton03}.

Figure~\ref{fig:BvsH} gives a useful indication of the stellar population of galaxies. Galaxies which lie well below the line are bluer  than most galaxies which suggests that they have a relatively young stellar population. Conversely, galaxies which lie well above the line are  redder than expected indicating  a larger old stellar population.  The tight correlation (correlation coefficient = 0.97) between the $B$- and $H$-band luminosities comes somewhat as a surprise. A $B$-band light profile of a galaxy can be significantly attenuated and distorted by dust. Moreover, short-lived giant O and B stars contribute to the  $B$-band emission  and hence the profile can be distorted by transient star-formation events. The stellar mass of most galaxies is dominated by the older, low luminosity stellar population whose energy output peaks at near-IR wavelengths~\citep{gavazzi96}. Hence it has been argued that the near-IR is the optimal wavelength regime for investigations of structural properties~\citep{driver04}. The tight correlation between the $B$- and $H$-bands, however, suggests that the advantages of the $H$-band may not be as significant as previously thought, at least for late-type giant galaxies. A detailed comparison of the observed scatter with the predictions from population synthesis models \citep[eg,][]{bruzual93, bruzual03, maraston98, li08} is beyond the scope of this paper because of the wide range of stellar compositions and star formation histories represented
by our galaxy sample. 

\begin{figure}
\centering
\includegraphics[width=1\linewidth]{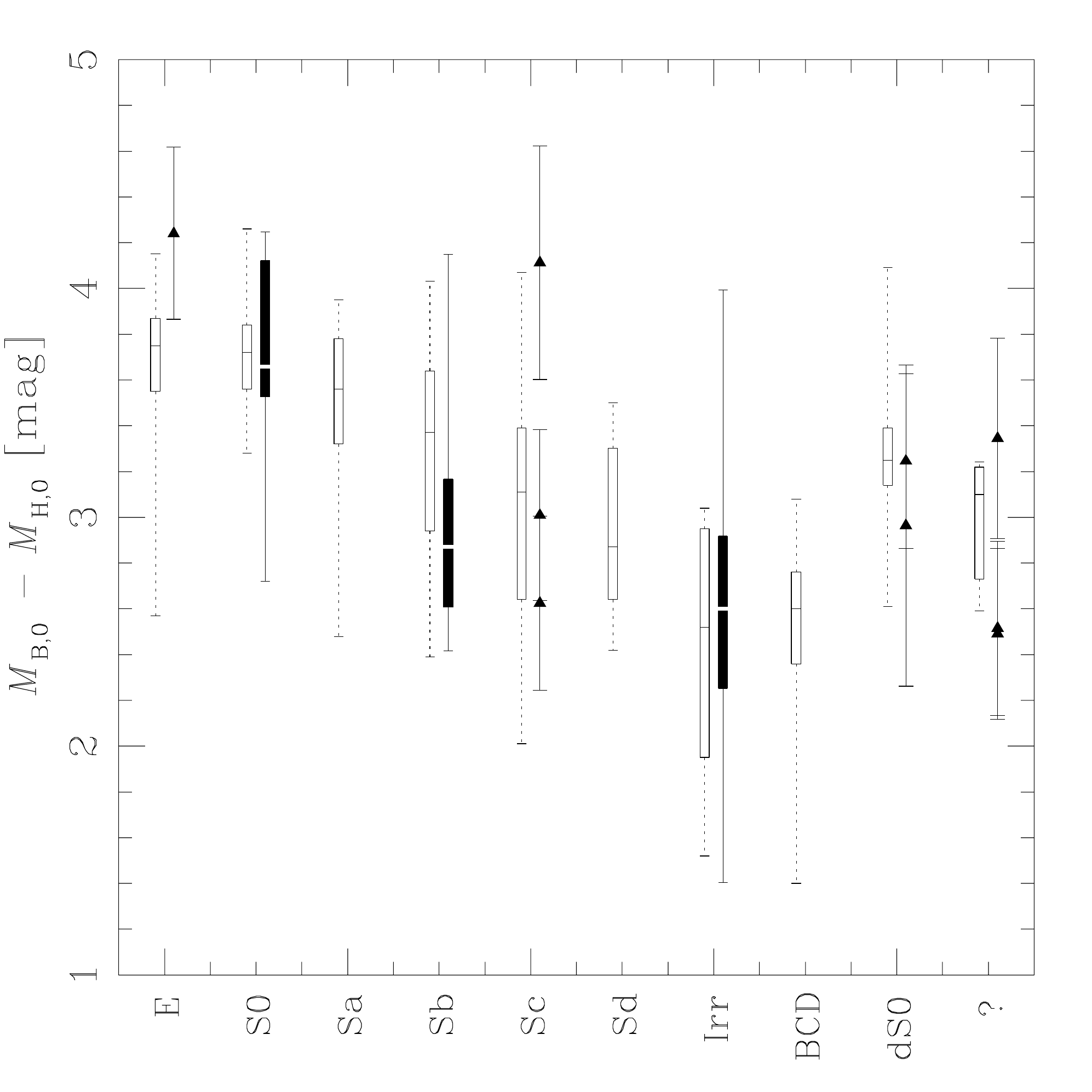}
\caption{Comparison of the integrated apparent $B - H$ colour and the morphological type. The  black boxes show our data and the white boxes show the combined samples of \cite{kassin06} and the Virgo cluster data listed in the Goldmine database \citep{gavazzi03}. The boxes show the median values as well as the quartile range. The errorbars on the boxes show the maximum and minimum $B-H$ colour for each morphological type. For morphologies with insufficient data, the individual data points are displayed with their associated uncertainty. } 
\label{fig:BmHvstype}
\end{figure}

The $B - H$ colour for each galaxy can be compared with the morphological type (Figure~\ref{fig:BmHvstype}). There we include the combined samples of \cite{kassin06} and the Virgo cluster data in the  white boxes and the new LSI data is shown by the black boxes. 
 The combined Virgo cluster and \cite{kassin06} sample is ten times the size of our sample but is dominated by giant, luminous galaxies. Our sample,   in contrast, contains four times as many irregular  dwarf galaxies. Therefore, when interpreting the color vs morphology plot, it must be noted that our sample dominates the morphological bin of irregular galaxies and the literature data dominates the larger galaxies.

The comparison of morphology to the $B - H$ colour shows that the early-type galaxies are redder than the late-type, irregular and dwarf galaxies. A similiar study by \cite{jarrett03} for galaxies in the 2MASS Large Galaxy Atlas also showed this  trend.

\subsection{Luminosity - Surface Brightness Relation} \label{ss:lumsb}
\begin{figure}
\centering
\includegraphics[width=1\linewidth]{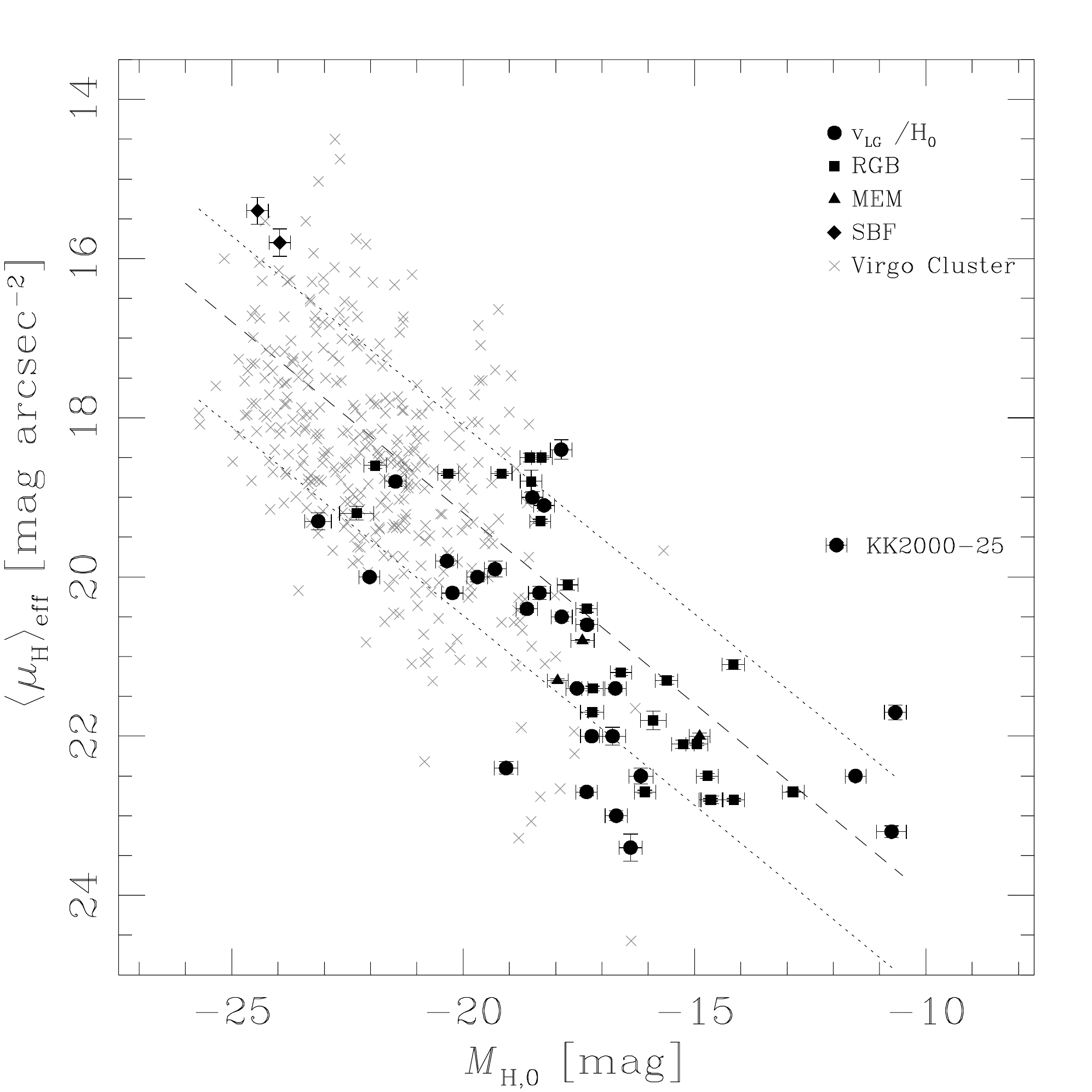}
\caption{Comparison of the mean effective surface brightness and the absolute magnitude for our sample galaxies. Symbol shapes  indicate the distance estimation method used. Also plotted are 560 late-type galaxies in the Virgo cluster~\citep{gavazzi03}. 
Marked is the galaxy KK2000-25 which deviates from the general trend probably caused by an incorrect distance estimate. The dashed line is the line of best fit for all galaxies and the dotted lines indicates data within 1$\sigma$. 
\label{fig:lsp}}
\end{figure}

In Figure~\ref{fig:lsp} we plot the mean effective surface brightness of our sample galaxies as a function of absolute magnitude. In addition to our $H$-band data, we include 560 late-type Virgo cluster galaxies (obtained from the Goldmine database; \citealt{gavazzi03}). The mean effective surface brightness for Virgo cluster galaxies was calculated as
$$
\langle\mu_H\rangle_{eff}=M_H + 2.5\log_{10}(\pi r_{eff}^2)
$$
and  the data was corrected for extinction using $A_H=0.01$\,mag. The morphologies of Virgo cluster galaxies included in the sample range from  S0 to Sd, Irr and BCD (listed as types 1 to 18 in the Goldmine database).

As previously discussed in \cite{dejong00}, the relationship between the two photometric parameters provides an important link to the underlying physical parameters of a galaxy, namely its total mass ${\cal M}_{tot}$ and total angular momentum. The total angular momentum of a galaxy, expressed as the dimensionless spin parameter $\lambda=J|E|^{1/2}{\cal M}_{tot}^{-5/2}G^{-1}$ \citep{peebles69},  is related to the scale length of its disk \citep{fall80, dalcanton97,mao98}. \cite{dejong00} 
showed that $\lambda$ can be transformed into observable quantities. The authors presented a model of a singular isothermal sphere with $E\propto {\cal M}_{tot}V_c^2$ from the virial theorem and a perfect exponential disk with angular momentum 
$J_{disk}\propto {\cal M}_{disk}r_{eff}V_c$ (assuming $V_{disk}=V_c$). They showed that if $J_{disk}/{\cal M}_{disk}\propto J/{\cal M}_{tot}$ and ${\cal M}_{disk}\propto {\cal M}_{tot}$ then $\lambda\propto r_{eff}V_c^2/{\cal M}_{disk}$.  Furthermore, using the relation ${\cal M}_{tot}\propto V_c^3$ predicted for dark matter halos, de Jong and Lacey showed that $\lambda\propto r_{eff} L^{\Upsilon /3}$, where $\Upsilon$ is the mass-to-light ratio for the disk. This can be transformed into an expression between surface brightness and total magnitude  by invoking
$$
\Sigma_{eff}=L(2\pi r_{eff}^2)^{-1}
$$
which results in:
$$
\lambda\propto \Sigma_{eff}^{-1/2} L^{-\Upsilon/3 + 1/2}
$$
Using the two identities $-2.5\log_{10}\Sigma_{eff}=\langle\mu_H\rangle_{eff}+c$ and $-2.5\log_{10}L=M+c'$ ($c$ and $c'$ being constants) we finally arrive at the theoretical luminosity --surface brightness relation:
\begin{equation}\label{eq:muefftheo}
\langle\mu\rangle_{eff} = (1-2\Upsilon/3) M +5\log_{10}\lambda+C
\end{equation}
Here, we can see that the gradient of the luminosity -- surface brightness relation is a function of the galaxy mass-to-light ratio and that the dispersion in the empirical relation reflects the distribution of the spin parameter.

The empirical relation between the two observational quantities in Figure~\ref{fig:lsp} for the late-type galaxies of the Virgo cluster sample is analytically best described by the linear equation:
\begin{equation}\label{eq:mueff}
\langle\mu_H\rangle_{eff}=a\cdot M_{H,0} +b
\end{equation}
where $a$ is $0.47\pm 0.08$, $0.44\pm 0.06$ and $0.48\pm0.05$ for our LSI sample, the Virgo cluster sample and the combined LSI + Virgo samples respectively, and $b$ is $29.0\pm 1.4$, $28.0\pm 1.3$ and $28.8\pm 1.1$. The uncertainty in $a$ and $b$ is the formal uncertainty in the linear fit plus the uncertainty due to the robustness of the sample obtained using bootstrap resampling. In Figure~\ref{fig:lsp}, the empirical relation between the mean effective surface brightness and the absolute magnitude is plotted for the combined LSI + Virgo cluster samples.

A comparison of the empirical result (see equation~\ref{eq:mueff}) with the theoretical prediction (equation~\ref{eq:muefftheo}) allows to estimate $\Upsilon_{\ast}^H$: $a=(1-2\Upsilon/3)\Rightarrow\Upsilon_{\ast}^H=0.80\pm0.12$, $0.84\pm 0.09$ and $0.78\pm0.08$ for our LSI sample, the Virgo cluster sample and the combined data sets respectively. These results are in excellent agreement with the value of $1.0\pm 0.4$ adopted in section~\ref{ss:mtol}.

\section{Interesting Galaxies}\label{s:interesting}

\subsection*{KKS2000-09}
 
KKS2000-09  has a  morphological classification of  a spiral ``S". However, close inspection of our  $H$-band image reveals that KKS2000-09  has a double nucleus and thus should be classified as a peculiar galaxy.

\subsection*{HIZSS003}

HIZSS003 is a H\,{\sc i} source originally detected at 21\,cm by the 
Dwingeloo obscured galaxies survey \citep{henning98}. It is
located behind the zone of avoidance ($b=0.09^\circ$) in the 
outskirts of the LG at 1.7\,Mpc \linebreak~\citep{silva05}.  Our image is highly contaminated with foreground stars preventing further analysis. \cite{begum05} show that HIZSS003 is actually a galaxy pair  (HIZSS003A and HIZSS003B). Our imaging shows that there is a 
resolved stellar overdensity associated with the H\,{\sc i} peak (labelled HIZSS003A using the~\citealt{begum05} convention) as well as a stellar component associated with the peak of the narrowband $H_{\alpha}$ emission first detected by~\cite{massey03} (labelled HIZSS003B by~\citealt{begum05}).

\subsection*{KK2000-25}
The morphology of KK2000-25 is listed in NED as Irr. Our deep imaging (see Figure~\ref{fig:galimages3}, third row, right panel) shows distinct spiral arms and that KK2000-25 should have a morphological classification of  ``Sb''. This morphology is not consistent with its angular size given the listed distance of 0.5\,Mpc. This distance estimate was obtained using a H\,{\sc i} profile observed by \cite{huchtmeier01} and the spectrum given does not show a clear galaxy detection. Based on the updated morphological classification, KK2000-25 must be more distant than the estimate given. 

The galaxy KK2000-25, although prominent when observed in the $H$-band, appears to be under-luminous in the $B$-band (see Figure~\ref{fig:BvsH}). This galaxy, however, is located at a low Galactic latitude and hence the extinction correction will not be very reliable. An inaccurate extinction correction will affect the $B$-band magnitude significantly but will have minimal affect on the $H$-band magnitude. Hence it is possible that this galaxy does not obey the $H$-band to $B$-band transformation (equation~\ref{eq:bh}) simply because the absolute $B$-band magnitude is incorrect.  The discrepancy between the absolute $B$- and $H$-band for KK2000-25 is not due to the incorrect distance estimate.

\section{Summary}\label{s:summary}

We have presented the deepest $H$-band images available to date for 57 galaxies in the Local Sphere of Influence ($D<10$\,Mpc), obtained using the near-IR camera IRIS2 at the 3.9m Anglo-Australian Telescope. Of the 68 targets, 11 remained undetected or could not be usefully analysed due to contamination by foreground stars. The surface brightness limit reaches down 
to $\mu_{lim}<26$\,mag\,arcsec, 4 magnitudes fainter than 2MASS.

The images, cleaned from Galactic foreground contamination, reveal the morphology and extent of many of the galaxies for the first time. For 56 galaxies, we derive radial luminosity profiles, ellipticities, and position angles, together with global parameters such as  total magnitude, mean effective surface brightness, half-light radius, S\'ersic parameters, and stellar mass.

No genuine young galaxies have been found in this survey. Some sample galaxies were previously identified on $B$-band photographic plates but remain undetected in the near-IR. In each case there is a plausible alternative explanation for the non-detection:
\begin{itemize}
\item AM0717-571: DSS $B_J$-band morphology resembles that of a Galactic nebula, but true nature still remains unclear.
\item HIZOAJ1616-55 and SJK98 J1616-55: possibly one or two high velocity clouds.
\item KK2000-03: Superimposed star hampers analysis however the marginal detection in the H-band suggests an unusual blue galaxy.
\item KK2000-04: Originally assumed to be a companion of NGC1313 however possibly a photographic plate flaw.
\item KK2000-06: Originally assumed to be a companion of NGC1313. More likely a background galaxy at $\approx$2250\,km\,s$^{-1}$.
\item NGC2784 DW1: intrinsic extreme low surface brightness dwarf satellite
of NGC2784.
\end{itemize}

We also detected a double nucleus in KKS2000-09 and propose to reclassify this system as a peculiar galaxy. KKS2000-25 was shown to have distinct spiral arms in the $H$-band and thus should be classified as ``Sb". Morphology and angular size strongly suggest that this is a background galaxy beyond 10\,Mpc.

We found compelling evidence that the short integration time of 2MASS resulted in serious underestimation of a galaxy's luminosity. The magnitudes of galaxies, with  $H$-band surface brightnesses fainter than 18\,mag\,arcsec${}^{-2}$, obtained in our study are up to 2.5\,mag brighter than those obtained by 2MASS. As the mean effective surface brightness correlates with the luminosity of a galaxy, we expect serious selection biases for a 2MASS-based $H$-band galaxy luminosity function fainter than M$_{H}=-20$\,mag.

There is a tight correlation (correlation coefficient = 0.97) between the $B$- and $H$-band magnitudes of a galaxy and this correlation has been demonstrated over a range of 15 magnitudes. The linear transformation between the $B$- and $H$-bands has a small scatter (0.3 mag) for bright galaxies. In the dwarf regime, there is a marginal  increase in scatter and possibly a slight trend for galaxies to be redder (by approximately 1 magnitude) than indicated by the transformation found for bright galaxies. 

The galaxy luminosity -- mean effective surface brightness  relation has been analysed to derive a semi-empirical stellar mass-to-light ratio of $\Upsilon_{\ast}^H=0.78\pm0.08$ in the $H$-band.

All raw and reduced $H$-band images of the 57 program galaxies in this near-IR survey will be made publicly available and can be obtained via email request.

\section{Acknowledgments}
We thank the referee for the useful comments. The authors acknowledge financial support from the Australian Research Council Discovery Project Grant DP0451426. This paper is based on data obtained with the Anglo-Australian Telescope. The study made use of data products from the Two Micron All Sky Survey (2MASS), which is a joint project of the University of Massachusetts and the Infrared Processing and Analysis Center/California Institute of Technology, funded by the National Aeronautics and Space Administration and the National Science Foundation. Support for IRIS2 data reduction within ORAC-DR is provided by the Joint Astronomy Centre. This research has made use of the GOLD Mine Database. This research has made use of the NASA/IPAC Extragalactic Database (NED) which is operated by the Jet Propulsion Laboratory, California Institute of Technology, under contract with the National Aeronautics and Space Administration. This research has made use of NASA's Astrophysics Data System.

%

\bibliographystyle{aa}
\bibliography{Kirby}

{\clearpage
\LongTables
\begin{landscape}	
\begin{deluxetable*}{cccccccccccccc}
\tablewidth{0pt}
\tabletypesize{\scriptsize}
\tablecaption{Basic properties of sample galaxies and observing log \label{basicdata}}
\tablecolumns{14}
\tablehead{\colhead{} & \colhead{} & \colhead{R.A.} &\colhead{Decl.} &  \colhead{$v_\odot$} & \colhead{$v_{LG}$} & \colhead{$D$}   & \colhead{}  &\colhead{$m_B$}& \colhead{$E(B-V)$}& \colhead{Obs date} & \colhead{} & \colhead{$t_{\rm tot}$} & \colhead{Seeing}\\ 
\colhead{Name} & \colhead{Type} & \colhead{(J2000.0)} & \colhead{(J2000.0)} & \colhead{(\kms)} & \colhead{(\kms)} &\colhead{(Mpc)}   & \colhead{Method}& \colhead{(mag)} &\colhead{(mag)} &\colhead{YYYY-MM-DD} & \colhead{Method} & \colhead{(sec)} & \colhead{(arcsec)}\\ 
\colhead{(1)} & \colhead{(2)} & \colhead{(3)} & \colhead{(4)}  & \colhead{(5)} & \colhead{(6)} & \colhead{(7)}  & \colhead{(8)}& \colhead{(9)}&\colhead{(10)}&\colhead{(11)}&\colhead{(12)}&\colhead{(13)}&\colhead{(14)}  }
\startdata
        SC18 &    LSB & 00:00:59.12 & -41:09:19.6 &   151 & 129 &   1.8 &  H   &    17.30\tablenotemark{j} & 0.013& 2004-10-22 & JSF & 1620 & 1.2\\
 ESO349-G031 &    Irr & 00:08:13.36 & -34:34:42.0 &   207 & 216 &    3.21 & TRGB\tablenotemark{b} &  15.48\tablenotemark{a}& 0.012& 2005-01-11 & JSF & 1620 & 1.5\\
 ESO294-G010 &    dS0 & 00:26:33.37 & -41:51:19.0 &    $-$&    $-$&    1.92 & TRGB\tablenotemark{a}  &    15.53\tablenotemark{a} & 0.006& 2006-01-02 & JSF & 1620 & 1.3\\
 ESO473-G024 &    Irr & 00:31:22.51 & -22:45:57.5 &   541 & 596 &   8.2 &  H   &  16.11\tablenotemark{e}& 0.019& 2004-10-23 & JSF & 1620 & 1.3\\
        SC24 &    LSB & 00:36:38.31 & -32:34:25.2 &    79 &  83 &   1.1 &  H   &   17.98\tablenotemark{j} & 0.015& 2006-01-02 & JSF & 1620 & 1.3\\
      IC1574 &    Irr & 00:43:03.82 & -22:14:48.7 &   363 & 413 &    4.92 & TRGB\tablenotemark{a}  &   14.36\tablenotemark{a} & 0.015& 2005-01-11 & JSF & 1350 & 1.9\\
  ESO540-G030 &    Irr & 00:49:20.96 & -18:04:31.5 &    $-$&    $-$&     3.4 & TRGB\tablenotemark{a}  &     16.37\tablenotemark{a} & 0.023& 2005-01-10 & JSF & 1728 & 1.6\\
      UGCA15 &    Irr & 00:49:49.20 & -21:00:54.0 &   294 &   346 &    3.34 & TRGB\tablenotemark{a}  &   15.19\tablenotemark{a} & 0.017  & 2004-10-22 & JSF & 1620 & 1.3\\
 ESO540-G032 &    Irr & 00:50:24.32 & -19:54:24.2 &    $-$&    $-$&    3.42 & TRGB\tablenotemark{a}  &      16.44\tablenotemark{a} &0.020& 2006-01-02 & JSF & 1620 & 1.4\\
  AM0106-382 &    dIrr & 01:08:21.93 & -38:12:34.5 &   645 &   605 &   8.3 &  H   & 16.61\tablenotemark{f} &0.012& 2004-10-22 & JSF & 1350 & 1.1\\
     NGC0625 &     Sb & 01:35:04.63 & -41:26:10.3 &   396 &   309 &    4.07 & TRGB\tablenotemark{a}  &    11.59\tablenotemark{a}& 0.016& 2006-01-02 & JSF & 1620 & 1.3\\
        SC42 &    LSB & 01:39:15.92 & -47:17:51.4 &   162 &    64 &  0.9 &  H   &   16.66\tablenotemark{j} & 0.015& 2004-10-22 & JSF & 1620 & 1.1\\
 ESO245-G005 &    Irr & 01:45:03.74 & -43:35:52.9 &   394 & 308 &    4.43 & TRGB\tablenotemark{a} &    12.73\tablenotemark{a}& 0.016& 2005-12-29 & JSF & 1350 & 1.4 \\
KK2000-03${}^{\ast \ast}$& dE & 02:24:44.58 &-73:30:49.20&$-$ &$-$ &4.1 & MEM\tablenotemark{a} & 16.0\tablenotemark{a} & 0.051& 2006-01-01& JSF& 1620&1.9\\
 ESO115-G021 &     Sc & 02:37:48.10 & -61:20:18.0 &   513 & 337 &    4.66 & TRGB\tablenotemark{a} &  13.34\tablenotemark{a} & 0.026& 2005-12-29 & JSF & 1350 & 1.4 \\
 ESO154-G023 &     Sb & 02:56:50.38 & -54:34:17.1 &   578 &   412 &   5.6 &  H  &    12.69\tablenotemark{a} & 0.017& 2004-10-23 & CSJ & 1800 & 1.1\\
KK2000-04${}^{\ast \ast}$& Irr& 03:12:46.14  &-66:16:12.5& $-$ &$-$ &4.2${}^{\ast\ast\ast}$& MEM\tablenotemark{i} & 17.8\tablenotemark{i}&0.032& 2006-01-01 & JSF & 810 & 1.8\\ 
KK2000-06${}^{\ast \ast}$& Irr& 03:14:26.14  &-66:23:27.9& $-$ &$-$ & 4.2${}^{\ast\ast\ast}$ &MEM\tablenotemark{i}&17.0\tablenotemark{i} & 0.055 & 2006-01-01 &JSF& 810 & 1.8\\
     NGC1313 &     Sd & 03:18:16.05 & -66:29:53.7 &   475 & 270 &    4.15 & TRGB\tablenotemark{a}  &     9.66\tablenotemark{a} & 0.109& 2005-12-30 & CSJ &  750 & 1.4 \\
     NGC1311 &     Sb & 03:20:06.96 & -52:11:07.9 &   570 & 398 &   5.5 &  H   &   13.18\tablenotemark{a}& 0.021& 2005-12-30 & JSF &  540 & 1.5\\
  AM0319-662 &    dIrr & 03:21:02.40 & -66:19:09.0 &   232 &  26 &    4.07 & TRGB\tablenotemark{a}  &  16.5\tablenotemark{a}& 0.077& 2005-12-30 & JSF &  810 & 1.3 \\
      IC1959 &     Sb & 03:33:12.59 & -50:24:51.3 &   639 & 464 &   6.4 &  H   &   13.26\tablenotemark{a} & 0.011& 2005-12-31 & JSF & 1620 & 1.3\\
  AM0333-611 &    Irr & 03:34:15.34 & -61:05:47.6 &  1172 & 971 &   13.3 &  H   &   16.47\tablenotemark{j}& 0.032& 2004-10-24 & JSF & 1080 & 1.2 \\
      IC2038 &     Sd & 04:08:53.75 & -55:59:22.4 &   712 &   505 &   16.5 &  MEM\tablenotemark{d}   &     14.98\tablenotemark{a} & 0.011& 2005-12-31 & JSF & 1620 & 1.3\\
      IC2039 &     S0 & 04:09:02.37 & -56:00:42.1 &   857 &   649 &   16.5 &  MEM\tablenotemark{d}   &      14.97\tablenotemark{e} &0.012& 2005-12-31 & JSF & 1620 & 1.3 \\
     NGC1705 &     S0 & 04:54:13.50 & -53:21:39.8 &   627 & 400 &     5.1 & TRGB\tablenotemark{a}  &     12.76\tablenotemark{a}& 0.008& 2004-10-21 & JSF & 1620 & 1.1\\
     NGC1744 &     Sb & 04:59:57.80 & -26:01:20.0 &   741 & 574 &   7.9 &  H   &    11.94\tablenotemark{e}& 0.041& 2006-01-02 & JSF & 1620 & 1.2\\
  AM0521-343 &    dIrr & 05:23:23.72&  -34:34:29.5 &   963 & 756 &    10.4 &  H   &     15.74\tablenotemark{g}& 0.028& 2004-10-21 & JSF & 1620 & 1.1 \\
  KKS2000-55${}^{\ast \ast}$ &     Sb & 05:50:17.71 & -10:17:51.6 &   901 &   736 &     10.1  & H&  $-$&0.827& 2005-12-31 & JSF & 1620 & 1.4 \\
 ESO364-G029 &    Irr & 06:05:45.22 & -33:04:51.0 &   787 &   549 &   7.5 &  H   &    13.58 \tablenotemark{a}& 0.044& 2005-12-29 & JSF & 1620 & 1.2\\
 ESO121-G020 &    Irr & 06:15:54.19 & -57:43:31.6 &   575 &   311 &    6.05 & TRGB\tablenotemark{b} & 15.85\tablenotemark{a} & 0.040& 2004-10-21 & JSF & 1620 & 1.1  \\
 ESO490-G017&Irr & 06:37:57.09 & -26:00:03.10 & 503 & 264 &4.23 & TRGB\tablenotemark{b} &  14.01\tablenotemark{a} &0.078& 2004-10-22& JSF & 1080& 1.3\\
 ESO308-G022 &    dIrr & 06:39:32.70 & -40:43:15.0 &   821 & 556 &   7.6 &  H  &    16.05\tablenotemark{a} & 0.089& 2005-12-30 & JSF &  810 & 1.2 \\
   KKS2000-09${}^{\ast \ast}$ &    S & 06:46:56.63  & -17:56:27.2  &   693 &   471 &   6.5 &  H   &   17.2\tablenotemark{a} & 0.443& 2006-01-01 & JSF & 1620 & 1.2\\
HIZSS003 & $-$ & 07:00:29.3   & -04:12:30 & 280&  101 & 1.4 & H&18\tablenotemark{a} &1.032& 2005-12-29 & JSF & 1620 & 1.4\\
        Argo &     dIrr & 07:05:18.80 & -58:31:13.0 &   564 &   284 &     4.9 & TRGB\tablenotemark{a}  &      14.95\tablenotemark{a}&0.119& 2005-12-30 & JSF &  810 & 1.2 \\
 ESO558-PN011& Irr &07:06:56.80 &-22:02:26.0 & 731 &489 & 6.7& H &14.43\tablenotemark{a} &0.372&2004-10-23 & JSF & 1620 & 1.2\\
  AM0717-571 & Irr& 07:18:37.90  & -57:24:46.5& 1148 &865 &11.8 & H & $-$ & 0.165&2006-01-02 & JSF& 1620& 1.2\\
 ESO059-G001 &    Irr & 07:31:18.20 & -68:11:16.8 &   530 &   255 &    4.57 & TRGB\tablenotemark{b} &   13.98\tablenotemark{a} & 0.147& 2005-12-31 & JSF & 1620 & 1.3\\
  AM0737-691 &    Irr &  07:37:12.6 &   -69:20:31 &  1456 &  1174 &   16.1 &  H   &  16.82\tablenotemark{h}&0.213& 2004-10-24 & JSF & 1080 & 1.1 \\
   KK2000-25${}^{\ast \ast}$ &    Irr & 07:56:38.48 & -26:15:01.9 &   241 &  -34 &  0.5${}^{\ast\ast\ast}$ &  H  &  17.7\tablenotemark{a}& 0.335& 2006-01-02 & JSF &  810 & 1.0\\
 ESO006-G001 &    Sb & 08:19:22.14 & -85:08:35.9 &   738 &   488 &   6.7 &  H  &     15.13\tablenotemark{a}& 0.193& 2006-01-01 & JSF & 1620 & 1.8\\
     UGCA148 &    Irr & 09:09:46.54 & -23:00:33.0 &   725 & 439 &     9.8 & MEM\tablenotemark{a}  & 15.63\tablenotemark{a} & 0.167& 2006-01-02 & JSF & 1620 & 1.1\\
     NGC2784 DW1 & dE &  09:12:18.5   &-24:12:41 &$-$&$-$&$9.8$&MEM\tablenotemark{a}&17.27\tablenotemark{a} & 0.206& 2006-01-01 & CSJ &  900 & 1.3\\ 
     NGC2784 &     S0 & 09:12:19.50 & -24:10:21.4 &   697 &   402 &    9.82 & SBF\tablenotemark{k} &   11.17\tablenotemark{a} &0.214& 2006-01-01 & CSJ &  900 & 1.3\\
     KK98-73 &    dE & 09:12:29.30 & -24:14:28.0 &    $-$&    $-$&     9.8 & MEM\tablenotemark{a}  &   16.35\tablenotemark{a} & 0.197& 2006-01-01 & CSJ &  900 & 1.3\\
     UGCA153 &    Irr & 09:13:12.08 & -19:24:31.0 &   768 & 491 &   6.7 &  H   &  15.40\tablenotemark{a} & 0.088& 2006-01-02 & JSF & 1620 & 1.1\\
     NGC2835 &     Sc & 09:17:52.91 & -22:21:16.8 &   886 & 601 &   8.2 &  H &  11.03\tablenotemark{e}&0.101& 2006-01-01 & CSJ &  600 & 1.3\\
     UGCA162 &    Irr & 09:21:28.07 & -22:30:06.8 &   846 & 560 &   7.7 &  H   & 14.87\tablenotemark{e}& 0.067& 2006-01-01 & JSF & 1620 & 1.2 \\
 ESO565-G003 &    Irr & 09:23:10.00 & -20:10:03.2 &   829 &   549 &   7.5 &  H  &    15.53\tablenotemark{a} & 0.062& 2006-01-02 & JSF & 1620 & 1.2\\
     NGC2915 &    Irr & 09:26:11.53 & -76:37:34.8 &   468 &   192 &    3.78 & TRGB\tablenotemark{a}  &   13.20\tablenotemark{a} & 0.275& 2005-01-10 & JSF & 1728 & 1.5\\
     NGC3115 &     S0 & 10:05:13.98 & -07:43:06.9 &   720 &   478 &    9.86 & SBF\tablenotemark{k}  &    9.86\tablenotemark{a} & 0.047& 2006-01-01 & CSJ &  900 & 1.4\\
        SJK98 J1616-55& $-$ & 16:16:49.0    &-55:44:57 & 421 &247& 3.4 & H& $-$ & 0.633& 2006-05-14& JSF & 810 & 1.4\\
 HIZOAJ1616-55${}^{\ast}$ & $-$&16:18:46     &-55:37:30& 409 & 236 & 3.2 & H & $-$ &0.616& 2006-05-04 & JSF & 810 & 1.5\\
      IC4662 &    Irr & 17:47:08.86 & -64:38:30.3 &   302 &   153 &    2.44 & TRGB\tablenotemark{b} &     11.74\tablenotemark{a} & 0.070& 2006-05-14 & JSF & 1620 & 1.3\\
 ESO594-G004& Irr& 19:29:58.97  &-17:40:41.3& -79 &24 &1.04&TRGB\tablenotemark{a} & 14.12\tablenotemark{a} &0.070& 2006-05-14&JSF & 1620& 1.3\\
 ESO461-G036 &    dIrr & 20:03:57.38 & -31:40:53.8 &   427 &   470 &    7.83 & TRGB\tablenotemark{b} &     17.06\tablenotemark{a}& 0.296& 2006-09-01 & JSF & 1620 & 1.2\\
     DDO210 &    Irr & 20:46:51.80 & -12:50:52.5 &  -141 &     3 &    0.94 & TRGB\tablenotemark{a}  &     14.0\tablenotemark{a}& 0.051& 2006-05-14 & JSF & 1620 & 1.3\\
      IC5052 &     Sb & 20:52:01.63 & -69:11:35.9 &   584 &   445 &    6.03 & TRGB\tablenotemark{c} &    11.68\tablenotemark{a} & 0.051& 2006-09-01 & CSJ & 450 & 1.1\\
      IC5152 &    Irr & 22:02:41.51 & -51:17:47.2 &   122 &    69 &    2.07 & TRGB\tablenotemark{a}  &  11.06\tablenotemark{a} & 0.025& 2006-09-01 & CSJ &750 & 1.2\\
 ESO468-G020 &    E3& 22:40:43.93 & -30:48:00.2 &    $-$&    $-$&     3.9 & MEM\tablenotemark{a}  &    17.36\tablenotemark{a} & 0.013& 2006-09-02 & JSF & 1620 &  1.0\\
     UGCA438 &    Irr & 23:26:27.52 & -32:23:19.5 &    62 &  99 &    2.23 & TRGB\tablenotemark{a}  &    13.86\tablenotemark{a} & 0.015& 2004-10-21 & JSF & 1620 & 1.1\\
 ESO347-G017 &     Sb & 23:26:56.21 & -37:20:48.9 &   692 &   688 &   9.4 &  H   &   15.77\tablenotemark{e} & 0.017& 2006-09-02 & JSF & 1620 & 1.0\\
      IC5332 &     Sa & 23:34:27.49 & -36:06:03.9 &   701 &   700 &   9.6 &  H   &  11.00\tablenotemark{e}& 0.017& 2006-09-01 & CSJ & 900 & 1.3\\
     NGC7713 &     Sb & 23:36:14.99 & -37:56:17.1 &   692 &   681 &   9.3 &  H   &    11.66\tablenotemark{e}& 0.017& 2006-09-04 &CSJ & 900 & 1.5\\
     UGCA442 &     Sb & 23:43:45.55 & -31:57:24.4 &   267 &   282 &    4.27 & TRGB\tablenotemark{a}  &    13.58\tablenotemark{a} & 0.017& 2006-09-04 & CSJ & 900 & 1.7\\
 ESO348-G009 &    Irr & 23:49:23.47 & -37:46:18.9 &   648 &   633 &   8.7 &  H   & 15.83\tablenotemark{j}&0.013 & 2006-09-03 & JSF & 1620 & 2.9\\
     NGC7793 &     S0 & 23:57:49.83 & -32:35:27.7 &   229 & 252 &    3.91 & TRGB\tablenotemark{a}  &    9.70\tablenotemark{a} & 0.019& 2004-10-24 & CSJ &  900 & 2.3\\
\enddata
\tablerefs{ (a) \citealt{karachentsev04};
		(b) \citealt{karachentsev06};
		(c) \citealt{seth05};
		(d) \citealt{carrasco01};
		(e) ESO-LV catalogue,~\citealt{lauberts89};
		(f) \citealt{makarova05}; 
		(g) \citealt{parodi02};
		(h) \citealt{vadar94};
		(i) \citealt{karachentseva00};
		(j) \citealt{maddox90};
		(k) \citealt{tonry01}.\\
($\ast$) This galaxy is listed as HIZOAJ1618-55 in the NASA Extragalactic Database (NED). We use the name of HIZOAJ1616-55 which is the original name listed in \cite{juraszek00}.\\
($\ast \ast$) We use KK2000 to indicate that the original listing of this galaxy was in the \cite{karachentseva00} paper and KKS2000 to indicate that it was originally listed in the \cite{karachentsev00} paper. This is consistent with the names listed in NED.\\
($\ast \ast \ast$) The distance estimate for KK2000-04 and KK2000-06 is based on a group membership listed by  \cite{karachentseva00}. These values are discussed in detail in \S\ref{s:nondetect}. The distance estimate for KK2000-25 is based on a H\,{\sc i} spectra obtained by \cite{huchtmeier01}. Our new data clearly shows that this value is incorrect.}
\end{deluxetable*}
\clearpage
\end{landscape}}

\begin{deluxetable*}{cccccc}
\tablewidth{0pt}
\tabletypesize{\scriptsize}
\tablecaption{Galaxy Parameters: Measured \label{tab:phot1}}
\tablecolumns{6}
\tablehead{\colhead{} & \colhead{$m_{H,obs}$} & \colhead{$r_{eff}$} &\colhead{$\langle\mu_H\rangle_{eff}$} &\colhead{} & \colhead{} \\ 
\colhead{Name} & \colhead{(mag)} & \colhead{(arcsec)} & \colhead{(mag arcsec${}^{-2}$)}& \colhead{e} & \colhead{PA}\\
\colhead{(1)} & \colhead{(2)} & \colhead{(3)} & \colhead{(4)}  & \colhead{(5)}  & \colhead{(6)}}
\startdata
        SC18 & 14.94  $\pm$   0.05 & 12.9  $\pm$   0.5 &  22.50  $\pm$   0.02 &   0.5 &    50\\
 ESO349-G031 & 12.96  $\pm$  0.06 & 29.4  $\pm$   1.6 & 22.29  $\pm$  0.04 &  0.05 &     0\\
 ESO294-G010 &  12.4  $\pm$  0.1 & 19.7  $\pm$  1.4 & 20.84  $\pm$   0.06 &   0.3 &     5\\
 ESO473-G024 & 13.7  $\pm$   0.1 &    24.0  $\pm$  2.7 & 22.6  $\pm$    0.1 &   0.5 &    30\\
        SC24 & 14.8  $\pm$  0.2 & 16.8  $\pm$   2.6 & 22.88  $\pm$  0.08 &   0.5 &    -5\\
      IC1574 & 11.89  $\pm$   0.08 & 30.1  $\pm$   1.7 & 21.28  $\pm$   0.04 &   0.6 &   -10\\
 ESO540-G030 & 13.0  $\pm$  0.1 & 30.5  $\pm$   1.7 & 22.40  $\pm$  0.02 &  0.05 &     0\\
      UGCA15 & 12.73  $\pm$   0.09 & 32.2  $\pm$   1.6 & 22.26  $\pm$  0.02 &   0.7 &    30\\
 ESO540-G032 &  13.10  $\pm$   0.08 & 28.8  $\pm$  1.4 & 22.39  $\pm$   0.03 &   0.4 &   -45\\
  AM0106-382 & 13.4  $\pm$   0.1 & 25.0  $\pm$  3.5 & 22.4  $\pm$   0.2 &   0.2 &    45\\
     NGC0625 &  8.94  $\pm$   0.04 & 36.3  $\pm$   1.1 & 18.74  $\pm$  0.03 &  0.58 &   -89\\
        SC42 & 14.13  $\pm$   0.08 & 10.4  $\pm$    0.9 &  21.2  $\pm$  0.1 &   0.5 &    -5\\
 ESO245-G005 & 11.1  $\pm$  0.1 &  49.7  $\pm$  3.4 & 21.55  $\pm$  0.03 &  0.38 &   -53\\
 ESO115-G021 & 10.71  $\pm$   0.06 & 33.6  $\pm$   2.0 & 20.34  $\pm$  0.07 &  0.7 &  42\\
ESO154-G023 & 10.37  $\pm$   0.07 & 50.9  $\pm$  2.7 &  20.90  $\pm$  0.05 &  0.77 &    39\\
     NGC1313 &  6.7  $\pm$  0.3 & 101.6  $\pm$  15.5 & 18.74  $\pm$  0.09 &   0.3 &    15\\
     NGC1311 & 10.28  $\pm$  0.08 & 23.0  $\pm$   1.6 & 19.09  $\pm$  0.08 &  0.62 &    39\\
  AM0319-662 & 14.00  $\pm$  0.05 & 21.8  $\pm$  0.7 &  22.70  $\pm$   0.02 &   0.1 &   -50\\
      IC1959 & 10.84  $\pm$  0.05 & 20.5  $\pm$   0.7 & 19.39  $\pm$  0.03 &   0.7 &   -30\\
  AM0333-611 & 13.65  $\pm$   0.08 & 20.9  $\pm$  1.3 & 22.26  $\pm$   0.05 &  0.25 &     5\\
      IC2038 & 11.8  $\pm$  0.09 & 15.8  $\pm$  1.3 &  19.8  $\pm$    0.1 &  0.65 &   -28\\
      IC2039 & 11.43  $\pm$   0.04 &  14.1  $\pm$  0.6 & 19.17  $\pm$   0.06 &   0.2 &   -57\\
     NGC1705 & 10.16  $\pm$  0.07 & 15.1  $\pm$  1.5 & 18.1  $\pm$   0.1 &   0.2 &    60\\
     NGC1744 &  9.31  $\pm$   0.09 & 47.9  $\pm$  3.0 &  19.70 $\pm$  0.06 &   0.6 &   -10\\
  AM0521-343 & 14.2  $\pm$  0.4 &  7.1  $\pm$  0.2 & 20.5  $\pm$   0.4 &   0.3 &   -45\\
  KKS2000-55 & 10.44  $\pm$  0.05 & 32.5  $\pm$   0.7 &    20  $\pm$  0     &  0.55 &   -60\\
 ESO364-G029 & 11.92  $\pm$  0.08 & 35.1  $\pm$   1.8 & 21.64  $\pm$  0.02 &   0.6 &    57\\
ESO121-G020 & 13.87  $\pm$   0.09 & 17.6  $\pm$   1.2 &  22.10  $\pm$   0.05 &  0.25 &    45\\
 ESO308-G022 & 13.4  $\pm$  0.1 & 25.0  $\pm$  1.9 & 22.32  $\pm$  0.07 &   0.2 &   -50\\ 
   KKS2000-09 & 11.48  $\pm$  0.07 &  8.4 $\pm$  0.9 & 18.3  $\pm$  0.1 &  0.35 &    25\\
       Argo & 12.72  $\pm$  0.08 & 37.3  $\pm$  1.7 & 22.57  $\pm$   0.02 &   0.5 &    45\\
 ESO059-G001 & 11.28  $\pm$   0.06 & 34.4  $\pm$  1.4 & 20.96  $\pm$   0.03 &   0.3 &   -20\\
  AM0737-691 & 12.2  $\pm$  0.1 & 32.1  $\pm$  3.3 & 21.94  $\pm$   0.08 &   0.2 &   -30\\
   KK2000-25 & 11.66  $\pm$  0.05 & 11.9  $\pm$  0.5 & 19.03  $\pm$  0.03 &  0.15 &    60\\
 ESO006-G001 & 11.37  $\pm$   0.06 & 20.0  $\pm$  1.1 & 19.87  $\pm$   0.06 &  0.12 & -13\\
     UGCA148 & 12.14  $\pm$   0.05 & 19.0  $\pm$  0.6 & 20.53  $\pm$   0.02 &  0.35 &    60\\
     NGC2784 &  6.16  $\pm$   0.07 & 27.2  $\pm$   2.9 & 15.3  $\pm$   0.2 &  0.56 &    73\\
     KK98-73 & 13.0  $\pm$  0.1 & 13.0  $\pm$  0.8 & 20.57  $\pm$  0.01 &   0.4 &    45\\
     UGCA153 & 12.6  $\pm$   0.2 & 30.9  $\pm$   3.8 &    22.0  $\pm$   0.1 &   0.6 &   -48\\
     NGC2835 &  7.1  $\pm$  0.2 & 88.0  $\pm$  10.2 &  18.8  $\pm$  0.1 &   0.3 &   -20\\
     UGCA162 & 12.49  $\pm$  0.09 & 20.8  $\pm$   1.4 & 21.08  $\pm$  0.06 &   0.8 &    30\\
 ESO565-G003 & 12.79  $\pm$   0.07 & 16.4  $\pm$   0.8 & 20.85  $\pm$  0.05 &  0.25 &    30\\
     NGC2915 &  9.53  $\pm$   0.05 & 21.3 $\pm$   1.1 & 18.17  $\pm$   0.06 &   0.4 &   -53\\
     NGC3115 &   5.70  $\pm$   0.07 & 27.3  $\pm$  2.8 & 14.9  $\pm$   0.2 &   0.6 &    44\\
      IC4662 &  8.71  $\pm$  0.03 & 33.1  $\pm$  0.7 & 18.31  $\pm$  0.03 &  0.34 &   -80\\
 ESO461-G036 & 13.9  $\pm$  0.2 &  14.7  $\pm$   2.2 & 21.7  $\pm$  0.1 &   0.4 &    25\\
       DDO210 &  12.30  $\pm$  0.09 & 47.0  $\pm$  1.7 & 22.65  $\pm$  0.01 &   0.4 &   -80\\
      IC5052 &  8.89  $\pm$  0.05 & 54.9  $\pm$   1.8 & 19.6  $\pm$  0.03 &  0.85 &   -40\\
      IC5152 &  8.26  $\pm$   0.03 & 53.5  $\pm$  1.5 &  18.90  $\pm$  0.03 &  0.35 &   -85\\
 ESO468-G020 &  13.2  $\pm$  0.1& 21.4  $\pm$   1.0 & 21.85  $\pm$  0.04 &   0.4 &    25\\
     UGCA438 & 11.2  $\pm$    0.1 & 36.1  $\pm$  2.5 & 20.96  $\pm$   0.05 &   0.2 &   -40\\
      IC5332 &  8.14  $\pm$   0.03 & 68.62  $\pm$   1.5 & 19.32  $\pm$   0.02 &  0.05 &     0\\
 ESO347-G017 & 11.78  $\pm$  0.08 & 21.8  $\pm$  1.6 & 20.47  $\pm$  0.09 &  0.65 &   -85\\
     NGC7713 &  8.45  $\pm$   0.06 & 44.0  $\pm$  2.3 & 18.66  $\pm$   0.06 &  0.55 &   -10\\
     UGCA442 & 11.27  $\pm$   0.07 &  32.5  $\pm$  1.8 & 20.82  $\pm$  0.05 &  0.75 &    53\\
 ESO348-G009 & 12.66  $\pm$  0.09 & 30.5  $\pm$  1.7 & 22.08  $\pm$  0.03 &   0.6 &    75\\
     NGC7793 &  6.5  $\pm$   0.1 & 88.9  $\pm$   5.5 & 18.27  $\pm$  0.04 &   0.4 &   -80\\
\enddata
\end{deluxetable*}

{\clearpage
\LongTables
\begin{landscape}	
\begin{deluxetable*}{cccccccccc}
\tablewidth{0pt}
\tabletypesize{\scriptsize}
\tablecaption{Galaxy Parameters: Derived \label{tab:phot2}}
\tablecolumns{10}
\tablehead{\colhead{}& \colhead{$r_{eff}$} & \colhead{$\mu_{0}$} & \colhead{} & \colhead{$\alpha$} & \colhead{$\Delta m$}& \colhead{$R_{eff}$}& \colhead{$R_{eff}$} & \colhead{$M_{H,0}$} & \colhead{log${}_{10}({\mathcal{M}}_{\ast})$}\\
\colhead{Name}& \colhead{(kpc)} & \colhead{(mag arcsec${}^{-2}$)} & \colhead{$n$} & \colhead{(arcsec)} & \colhead{(mag)} & \colhead{(arcsec)} & \colhead{(kpc)} & \colhead{(mag)} & \colhead{(log${}_{10}({\mathcal{M}}_{\odot}$))}\\
\colhead{(1)}& \colhead{(2)} & \colhead{(3)} & \colhead{(4)} & \colhead{(5)} & \colhead{(6)} & \colhead{(7)}&\colhead{(8)}&\colhead{(9)}&\colhead{(10)}}
\startdata
SC18             & 0.11 $\pm$ 0.01 &  22.11 $\pm$ 0.09  &   2.08 $\pm$ 0.3     &  20.15 $\pm$ 1.0   & 0.18 & 16.1 $\pm$ 1.3   &   0.14 $\pm$ 0.01 & -11.5 $\pm$ 0.2&   5.9  $\pm$ 0.2\\
ESO349-G031 & 0.46 $\pm$ 0.03 &  21.57 $\pm$ 0.4   &   1.21 $\pm$ 0.2   &  26.80 $\pm$ 4.0  & 0.05 & 33.7 $\pm$ 6.6    &   0.52 $\pm$ 0.09  & -14.6 $\pm$ 0.2&   7.2  $\pm$ 0.2\\
ESO294-G010 & 0.18 $\pm$ 0.02 &  20.11 $\pm$ 0.4   &   1.10 $\pm$ 0.5    &  18.20 $\pm$ 5.8  & 0.14 & 26.1 $\pm$ 13.1 &   0.24 $\pm$ 0.09 & -14.2 $\pm$ 0.2&   7.0  $\pm$ 0.2\\
ESO473-G024 & 1.0 $\pm$ 0.1      &  21.63 $\pm$ 0.3   &   1.17 $\pm$ 0.3   &  23.41 $\pm$ 6.5  & 0.28 & 30.8 $\pm$ 11.0 &   1.2 $\pm$ 0.3       & -16.1 $\pm$ 0.3&   7.8  $\pm$ 0.2\\
SC24             & 0.09 $\pm$ 0.02 &  21.58 $\pm$ 0.4    &   0.98 $\pm$ 0.2     &  12.97 $\pm$ 4.6   & 0.18 & 22.4 $\pm$ 9.6   &   0.12 $\pm$ 0.04 & -10.7 $\pm$ 0.3&   5.6  $\pm$ 0.2\\
IC1574              & 0.72 $\pm$ 0.04 &  20.54 $\pm$ 0.08 &   1.50 $\pm$ 0.1   &  42.91 $\pm$ 2.3  & 0.01 & 42.9 $\pm$ 3.3   &   1.02 $\pm$ 0.06   & -16.6 $\pm$ 0.2&   8.0  $\pm$ 0.2\\
ESO540-G030 & 0.50 $\pm$ 0.03 &  21.70 $\pm$ 0.4   &   1.17 $\pm$ 0.4   &  26.77 $\pm$ 9.3  & 0.01 & 35.2 $\pm$ 16.3 &   0.6 $\pm$ 0.2       & -14.7 $\pm$ 0.2&   7.2  $\pm$ 0.2\\
UGCA15       & 0.52 $\pm$ 0.03 &  21.51 $\pm$ 0.03 &   1.49 $\pm$ 0.07    &  54.51 $\pm$ 1.2   & 0.05 & 54.9 $\pm$ 2.1  &   0.89 $\pm$ 0.03 & -15.0 $\pm$ 0.2&   7.3  $\pm$ 0.2\\
ESO540-G032 & 0.48 $\pm$ 0.03 &  20.89 $\pm$ 3.1   &   0.90 $\pm$ 1.7   &  17.73 $\pm$ 19.2& 0.15 & 36.0 $\pm$ 84.9 &   0.6 $\pm$ 1.1       & -14.7 $\pm$ 0.2&   7.2  $\pm$ 0.2\\
AM0106-382  & 1.0 $\pm$ 0.1     &  19.91 $\pm$ 0.4   &   0.56 $\pm$ 0.1    &   3.33 $\pm$ 2.3     & 0.13 &  27.2 $\pm$ 21.0&   1.1 $\pm$ 0.8      & -16.4 $\pm$ 0.2&   7.9  $\pm$ 0.2\\
NGC0625     & 0.72 $\pm$ 0.03  &  17.20 $\pm$ 0.04  &   0.82 $\pm$ 0.1     &  23.38 $\pm$ 0.8  & 0.05 & 58.0 $\pm$ 10.5 &   1.1 $\pm$ 0.1     & -19.2 $\pm$ 0.2&   9.0  $\pm$ 0.2\\
SC42             & 0.04 $\pm$ 0.01 &  18.21 $\pm$ 0.6    &   0.43 $\pm$ 0.05   &   0.75 $\pm$ 0.5    & 0.04 & 22.5 $\pm$ 16.8 &   0.10 $\pm$ 0.03 & -10.6 $\pm$ 0.2&   5.6  $\pm$ 0.2\\
ESO245-G005 & 1.07 $\pm$ 0.08 &  20.95 $\pm$ 0.04 &   1.46 $\pm$ 0.09 &  61.74 $\pm$ 2.3  & 0.04 &63.3 $\pm$ 3.6    &   1.36 $\pm$ 0.07 & -17.2 $\pm$ 0.2&   8.2  $\pm$ 0.2\\
ESO115-G021 & 0.76 $\pm$ 0.05 &  18.81 $\pm$ 0.1   &   0.95 $\pm$ 0.08 &  27.52 $\pm$ 3.6  & 0.09 & 50.3 $\pm$ 7.9  &   1.1 $\pm$ 0.1       & -17.7 $\pm$ 0.2&   8.4  $\pm$ 0.2\\
ESO154-G023 & 1.39 $\pm$ 0.08 &  19.85 $\pm$ 0.03 &   1.20 $\pm$ 0.03 &  85.02 $\pm$ 2.5  & 0.19 & 108.1 $\pm$ 3.9 &   3.00 $\pm$ 0.06 & -18.6 $\pm$ 0.2&   8.8  $\pm$ 0.2\\
NGC1313     & 2.0 $\pm$ 0.3      &  16.41 $\pm$ 0.2   &   0.46 $\pm$ 0.03    &  12.69 $\pm$ 3.6   & 0.86 & 260.6 $\pm$ 83.2&   5.2 $\pm$ 0.7    & -22.3 $\pm$ 0.4&   10.3  $\pm$ 0.2\\
NGC1311     & 0.61 $\pm$ 0.04  &  17.70 $\pm$ 0.06 &   0.93 $\pm$ 0.03   &  19.30 $\pm$ 1.2   & 0.05 & 36.7 $\pm$ 2.6   &   0.97 $\pm$ 0.04 & -18.5 $\pm$ 0.2&   8.7  $\pm$ 0.2\\
AM0319-662  & 0.43 $\pm$ 0.02 &  22.19 $\pm$ 0.4   &   1.77 $\pm$ 0.8    &  29.05 $\pm$ 6.8   & 0.06 & 25.6 $\pm$ 8.7   &   0.5 $\pm$ 0.1      & -14.2 $\pm$ 0.2&  7.0  $\pm$ 0.2\\
IC1959              & 0.63 $\pm$ 0.02 &  18.46 $\pm$ 0.08 &   1.42 $\pm$ 0.06 &  32.04 $\pm$ 1.6  & 0.04 & 33.7 $\pm$ 1.9   &   1.04 $\pm$ 0.04  & -18.2 $\pm$ 0.2&   8.6  $\pm$ 0.2\\
AM0333-611  & 1.35 $\pm$ 0.08 &  20.69 $\pm$ 1.5   &   0.73 $\pm$ 0.3    &   8.58 $\pm$ 6.9    & 0.31 & 28.6 $\pm$ 28.4 &   1.8 $\pm$ 1.4      & -17.3 $\pm$ 0.2&   8.3  $\pm$ 0.2\\
IC2038              & 1.3 $\pm$ 0.1      &  17.62 $\pm$ 0.2   &   0.70 $\pm$ 0.05 &   6.32 $\pm$ 1.2   & 0.03 & 23.7 $\pm$ 5.0   &   1.9 $\pm$ 0.3       & -19.3 $\pm$ 0.2&   9.1  $\pm$ 0.2\\
IC2039              & 1.13 $\pm$ 0.05 &  15.04 $\pm$ 0.6   &   0.37 $\pm$ 0.03 &   0.25 $\pm$ 0.2   & 0.03 & 20.1 $\pm$ 12.9 &   1.6 $\pm$ 0.7       & -19.7 $\pm$ 0.2&   9.2  $\pm$ 0.2\\
NGC1705     & 0.37 $\pm$ 0.04 &  15.33 $\pm$ 0.4   &   0.44 $\pm$ 0.04    &   0.90 $\pm$ 0.4    & 0.15 & 23.7 $\pm$ 11.6  &   0.6 $\pm$ 0.2     & -18.5 $\pm$ 0.2&   8.8  $\pm$ 0.2\\
NGC1744     & 1.8 $\pm$ 0.1     &  17.86 $\pm$ 0.1    &   0.69 $\pm$ 0.05    &  21.97 $\pm$ 3.6  & 0.13 & 86.1 $\pm$ 16.7  &   3.3 $\pm$ 0.4     & -20.3 $\pm$ 0.2&   9.5  $\pm$ 0.2\\
AM0521-343  & 0.36 $\pm$ 0.01 & 			    $-$   &   $-$                         &   $-$                         & $-$   & $-$                        &   $-$                         & -15.9 $\pm$ 0.5&   7.7 $\pm$ 0.3\\
KKS2000-55  & 1.59 $\pm$ 0.04 &  18.05 $\pm$ 0.4    &   0.63 $\pm$ 0.09  &  11.36 $\pm$ 4.5  & 0.15 & 59.6 $\pm$ 27.1 &   2.9 $\pm$ 0.7     & -20.2 $\pm$ 0.2&   9.4 $\pm$ 0.2\\
ESO364-G029 & 1.30 $\pm$ 0.07 &  21.11 $\pm$ 0.05 &   1.78 $\pm$ 0.1   &  60.81 $\pm$ 2.2  & 0.02 & 53.3 $\pm$ 2.7    &   1.94 $\pm$ 0.07  & -17.5 $\pm$ 0.2&   8.3 $\pm$ 0.2\\
ESO121-G020 & 0.52 $\pm$ 0.04 &  20.82 $\pm$ 0.8   &   0.95 $\pm$ 0.4    &  12.22 $\pm$ 5.9  & 0.19 & 22.3 $\pm$ 14.0&   0.7 $\pm$ 0.3      & -15.3 $\pm$ 0.2&   7.4 $\pm$ 0.2\\
ESO308-G022 & 0.92 $\pm$ 0.07 &  20.87 $\pm$ 0.3   &   0.71 $\pm$ 0.2    &  10.76 $\pm$ 5.2  & 0.53 & 38.8 $\pm$ 22.8 &   1.4 $\pm$ 0.5      & -16.7 $\pm$ 0.2&   8.0 $\pm$ 0.2\\
KKS2000-09    & 0.26 $\pm$ 0.03 &  14.12 $\pm$ 4.6   &   0.42 $\pm$ 0.2    &   0.28 $\pm$ 1.8  & 0.06 &  9.7 $\pm$ 63.4  &   0.3 $\pm$ 1.7       & -17.9 $\pm$ 0.2&   8.5 $\pm$ 0.2\\
Argo                  & 0.89 $\pm$ 0.04 &  21.87 $\pm$ 0.1   &   1.32 $\pm$ 0.2    &  48.77 $\pm$ 4.6  & 0.27 & 55.4 $\pm$ 7.7   &   1.3 $\pm$ 0.1      & -16.1 $\pm$ 0.2&   7.8 $\pm$ 0.2\\
ESO059-G001 & 0.76 $\pm$ 0.04 &  20.23 $\pm$ 0.2   &   1.33 $\pm$ 0.2   &  36.68 $\pm$ 6.2  & 0.1   & 41.3 $\pm$ 8.3   &   0.9 $\pm$ 0.2       & -17.2 $\pm$ 0.2&   8.2 $\pm$ 0.2\\
AM0737-691  & 2.5 $\pm$ 0.3      &  20.67 $\pm$ 0.2   &   1.07 $\pm$ 0.2    &  21.41 $\pm$ 3.9   & 0.1   & 32.0  $\pm$ 7.1  &   2.5 $\pm$ 0.6      & -19.0 $\pm$ 0.3&    9.0 $\pm$ 0.2\\
KK2000-25      & 0.03 $\pm$ 0.01 &  18.13 $\pm$ 0.08 &   1.12 $\pm$ 0.03  &   9.41 $\pm$ 0.5  & 0.03 & 13.1 $\pm$ 0.7   &   0.03 $\pm$ 0.01  & -11.9 $\pm$ 0.2&   6.1 $\pm$ 0.2\\
ESO006-G001 & 0.65 $\pm$ 0.04 &  19.20 $\pm$ 0.5   &   1.07 $\pm$ 0.2   &  16.11 $\pm$ 5.2  & 0.08 & 24.1 $\pm$ 8.9   &   0.8 $\pm$ 0.2       & -17.8 $\pm$ 0.2&   8.5 $\pm$ 0.2\\
UGCA148     & 0.90 $\pm$ 0.03 &  19.85 $\pm$ 0.07 &   1.37 $\pm$ 0.09   &  22.15 $\pm$ 1.3   & 0.05 & 24.2 $\pm$ 1.9   &   1.15 $\pm$ 0.07 & -18.0 $\pm$ 0.2&   8.5 $\pm$ 0.2\\
NGC2784     & 1.3 $\pm$ 0.1     &  15.26 $\pm$ 0.4    &   0.82 $\pm$ 0.1      &  26.85 $\pm$ 9.0  & 0.04 & 66.6 $\pm$ 24.4  &   3.2 $\pm$ 0.5     & -24.0 $\pm$ 0.2&   10.9 $\pm$ 0.2\\
KK98-73            & 0.62 $\pm$ 0.04 &  19.48 $\pm$ 0.3   &   0.95 $\pm$ 0.1   &  10.74 $\pm$ 2.2 & 0.36 & 19.6 $\pm$ 4.8   &   0.9 $\pm$ 0.1       & -17.4 $\pm$ 0.2&   8.3 $\pm$ 0.2\\
UGCA153     & 1.0 $\pm$ 0.1      &  20.93 $\pm$ 0.2   &   1.15 $\pm$ 0.1     &  33.31 $\pm$ 5.2   & 0.11 & 44.8 $\pm$ 8.5   &   1.5 $\pm$ 0.2      & -16.7 $\pm$ 0.3&   8.0 $\pm$ 0.2\\
NGC2835     & 3.5 $\pm$ 0.4     &  16.50 $\pm$ 0.04  &   0.52 $\pm$ 0.01   &  14.27 $\pm$ 0.8  & 0.54 & 160.0 $\pm$ 10.7&   6.4 $\pm$ 0.2     & -23.1 $\pm$ 0.3&   10.6 $\pm$ 0.2\\
UGCA162     & 0.77 $\pm$ 0.05 &  20.22 $\pm$ 0.02 &   1.35 $\pm$ 0.02   &  47.70 $\pm$ 0.7   & 0.32 & 52.9 $\pm$ 1.0  &   2.00 $\pm$ 0.02  & -17.3 $\pm$ 0.2&    8.3 $\pm$ 0.2\\
ESO565-G003 & 0.60 $\pm$ 0.03 &  19.97 $\pm$ 0.5   &   1.01 $\pm$ 0.2   &  12.69 $\pm$ 4.1  & 0.05 & 20.8 $\pm$ 7.6   &   0.8 $\pm$ 0.2        & -16.7 $\pm$ 0.2&   8.0 $\pm$ 0.2\\
NGC2915     & 0.39 $\pm$ 0.02 &  15.97 $\pm$ 0.2    &   0.63 $\pm$ 0.03   &   5.21 $\pm$ 0.8   & 0.03 & 27.4 $\pm$ 4.8    &   0.50 $\pm$ 0.07 & -18.6 $\pm$ 0.2&   8.8 $\pm$ 0.2\\
NGC3115     & 1.3 $\pm$ 0.1      &  10.20 $\pm$ 0.4    &   0.31 $\pm$ 0.02   &   0.18 $\pm$ 0.1   & 0.14 & 62.1 $\pm$ 43.4 &   3.0 $\pm$ 0.9      & -24.4 $\pm$ 0.2&   11.1 $\pm$ 0.2\\
IC4662              & 0.39 $\pm$ 0.02 &  17.27 $\pm$ 0.2   &   0.94 $\pm$ 0.05 &  22.68 $\pm$ 2.9  & 0.03 & 42.3 $\pm$ 5.8   &   0.50 $\pm$ 0.06  & -18.3 $\pm$ 0.2&   8.7 $\pm$ 0.2\\
ESO461-G036 & 0.56 $\pm$ 0.08 &  21.16 $\pm$ 0.2   &   1.80 $\pm$ 0.4   &  21.07 $\pm$ 2.8  & 0.14 & 18.3 $\pm$ 3.4    &   0.7 $\pm$ 0.1       & -15.9 $\pm$ 0.3&   7.7 $\pm$  0.2\\
DDO210           & 0.21 $\pm$ 0.02 &  22.07 $\pm$ 0.08 &   1.52 $\pm$ 0.2   &  64.49 $\pm$ 4.7  & 0.27 & 63.8 $\pm$ 6.4   &   0.29 $\pm$ 0.03  & -12.9 $\pm$ 0.2&   6.5 $\pm$ 0.2\\
IC5052              & 1.61 $\pm$ 0.06 &  18.07 $\pm$ 0.02 &   1.27 $\pm$ 0.02 &  88.82 $\pm$ 1.4 & 0.28 & 105.4 $\pm$ 2.1 &   3.08 $\pm$ 0.04  & -20.3 $\pm$ 0.2&   9.5 $\pm$ 0.2\\
IC5152              & 0.54 $\pm$ 0.03 &  17.65 $\pm$ 0.08 &   0.96 $\pm$ 0.04 &  35.44 $\pm$ 2.7 & 0.01 & 63.5 $\pm$ 5.5   &   0.64 $\pm$ 0.05  & -18.3 $\pm$ 0.2&   8.7 $\pm$ 0.2\\
ESO468-G020 & 0.41 $\pm$ 0.02 &  21.13 $\pm$ 0.07 &   1.28 $\pm$ 0.1   &  24.93 $\pm$ 1.9  & 0.13 & 29.3 $\pm$ 3.0   &   0.55 $\pm$ 0.04   & -14.9 $\pm$ 0.2&   7.3  $\pm$ 0.2\\
UGCA438     & 0.39 $\pm$ 0.03 &  20.49 $\pm$ 0.2   &   1.31 $\pm$ 0.08   &  38.34 $\pm$ 3.4   & 0.03 & 43.9 $\pm$ 4.4  &   0.47 $\pm$ 0.04  & -15.6 $\pm$ 0.2&    7.6  $\pm$ 0.2\\
IC5332              & 3.19 $\pm$ 0.08 &  16.92 $\pm$ 0.2   &   0.52 $\pm$ 0.04 &   8.51 $\pm$ 2.4   & 0.21 & 95.4 $\pm$ 30.5 &   4.4 $\pm$ 1.0      & -22.0 $\pm$ 0.2&   10.1  $\pm$ 0.2\\
ESO347-G017 & 1.00 $\pm$ 0.07 &  18.19 $\pm$ 0.4   &   0.74 $\pm$ 0.08 &   9.02 $\pm$ 2.9    & 0.22 & 29.0 $\pm$ 10.3 &   1.3 $\pm$ 0.4      & -18.3 $\pm$ 0.2&   8.7  $\pm$ 0.2\\
NGC7713     & 2.0 $\pm$ 0.1     &  16.98 $\pm$ 0.2     &   0.83 $\pm$ 0.06   &  24.46 $\pm$ 3.8  & 0.02 & 59.0 $\pm$ 10.6 &   2.7 $\pm$ 0.4      & -21.4 $\pm$ 0.2&   9.9  $\pm$ 0.2\\
UGCA442     & 0.67 $\pm$ 0.04 &  19.94 $\pm$ 0.03 &   1.42 $\pm$ 0.04   &  57.38 $\pm$ 1.5   & 0.44 & 60.4 $\pm$ 2.0  &   1.25 $\pm$ 0.03  & -17.3 $\pm$ 0.2&   8.3  $\pm$ 0.2\\
ESO348-G009 & 1.28 $\pm$ 0.07 &  21.26 $\pm$ 0.05 &   1.31 $\pm$ 0.07 &  44.50 $\pm$ 2.0  & 0.15 & 51.0 $\pm$ 3.1   &   2.14 $\pm$ 0.08  & -17.2 $\pm$ 0.2&   8.2  $\pm$ 0.2\\
NGC7793     & 1.7 $\pm$ 0.1      &  16.47 $\pm$ 0.08  &   0.62 $\pm$ 0.02   &  28.72 $\pm$ 3.0  & 0.46 & 159.4 $\pm$ 18.7&   3.0 $\pm$ 0.2    & -21.9 $\pm$ 0.2&    10.1  $\pm$ 0.2\\
\enddata
\end{deluxetable*}
\clearpage
\end{landscape}}

\end{document}